\documentclass[11pt]{article}
\usepackage{jheppub}
\usepackage{graphicx}
\usepackage{subfigure}
\usepackage{dcolumn}
\usepackage{bm}
\usepackage{slashed}
\usepackage{extarrows}
\usepackage{multirow,array}
\usepackage{amsmath,amssymb,physics}
\usepackage{slashed}
\usepackage{orcidlink}
\usepackage{url}
\usepackage{hyperref}
\usepackage{xcolor}
\definecolor{nicered}{rgb}{0.5,0.,0.}
\definecolor{nicegreen}{rgb}{0.,0.5,0.}
\definecolor{niceblue}{rgb}{0.,0.,0.5}
\definecolor{darkmagenta}{rgb}{0.4,0,0.4}
\hypersetup{colorlinks,citecolor=nicegreen,linkcolor=nicered,urlcolor=niceblue}
\usepackage{braket}
\usepackage{colortbl}
\usepackage{gensymb}
\usepackage{comment}
\usepackage{enumitem}
\setlist{nolistsep}
\usepackage{setspace}
\usepackage{wasysym} 

\renewenvironment{eqnarray}{\begin{equation}\begin{aligned}}{\end{aligned}\end{equation}}

\newcommand{\GeV}{\textrm{GeV}}
\newcommand{\TeV}{\textrm{TeV}}
\newcommand{\MeV}{\textrm{MeV}}
\newcommand{\PDF}{\textrm{PDF}}

\newcommand{\calO}{\mathcal{O}}
\newcommand{\msbar}{\overline{\textrm{MS}}}
\newcommand{\el}{\textrm{el}}

\newcommand{\HT}{\textrm{HT}}

\newcommand{\github}[1]{\href{https://github.com/#1}{\includegraphics[width=10pt]{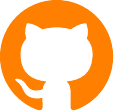}}}

\usepackage{tikz}
\usetikzlibrary{positioning,fit,shapes.misc,shapes.geometric}
\tikzstyle{startstop} = [rectangle, rounded corners, 
minimum width=3cm, 
minimum height=1cm,
text centered, 
draw=black, 
fill=red!30]

\tikzstyle{io} = [trapezium, 
trapezium stretches=true, 
trapezium left angle=70, 
trapezium right angle=110, 
minimum width=3cm, 
minimum height=1cm, text centered, 
draw=black, fill=blue!30]

\tikzstyle{process} = [rectangle, 
minimum width=6cm, 
minimum height=1cm,
text width=6cm,
text centered, 
draw=black, 
fill=orange!30]

\tikzstyle{decision} = [diamond, 
minimum width=3cm, 
minimum height=0.5cm, 
aspect=4,
text centered, 
draw=black, 
fill=green!30]
\tikzstyle{arrow} = [thick,->,>=stealth]

\author[a,b,\orcidlink{0000-0003-4261-3393}]{Keping Xie,}
\affiliation[a]{Pittsburgh Particle Physics, Astrophysics, and Cosmology Center, \\
	Department of Physics and Astronomy, University of Pittsburgh, Pittsburgh, PA 15260, USA}
\affiliation[b]{Department of Physics and Astronomy, Michigan State University, East Lansing, MI 48824, USA}
\emailAdd{xiekeping@pitt.edu}

\author[c,\orcidlink{0000-0003-1600-8835}]{Bei Zhou,}
\affiliation[c]{William H. Miller III Department of Physics and Astronomy, \\
	Johns Hopkins University, Baltimore, MD 21218, USA}
\emailAdd{beizhou@fnal.gov}

\author[d,\orcidlink{0000-0002-2729-0015}]{and T.~J.~Hobbs}
\affiliation[d]{High Energy Physics Division, Argonne National Laboratory, Argonne, Illinois 60439, USA}
\emailAdd{tim@anl.gov}
\preprint{ANL-182626, MSUHEP-24-004, PITT-PACC-2314}
\collaboration{CTEQ-TEA Collaboration}

\title{The Photon Content of the Neutron}

\abstract{
	In this work, we complete our CT18qed study with the neutron's photon parton distribution function (PDF), which is essential for the nucleus scattering phenomenology. 
	Two methods, CT18lux and CT18qed, based on the LUXqed formalism and the DGLAP evolution, respectively, to determine the neutron's photon PDF have been presented. Various low-$Q^2$ non-perturbative variations have been carefully examined, which are treated as additional uncertainties on top of those induced by quark and gluon PDFs. 
	The impacts of the momentum sum rule as well as isospin symmetry violation have been explored and turned out to be negligible.
	A detailed comparison with other neutron's photon PDF sets has been performed, which shows a great improvement in the precision and a reasonable uncertainty estimation.
	Finally, two phenomenological implications are demonstrated with photon-initiated processes: neutrino-nucleus $W$-boson production, which is important for the near-future TeV--PeV neutrino observations, and the axion-like particle production at a high-energy muon beam-dump experiment.
}

\begin{document}
	\date{\today}
	
	\maketitle
	
	\section{Introduction}
	\label{sec:intro}
	Similar to the proton as a composite hadron, the neutron also contains its internal structure, mostly consisting of quarks and gluons as partons. The corresponding parton distribution functions (PDFs) can describe its internal structure up to very high precision, based on the modern strong interaction theory, Quantum Chromodynamics (QCD). 
	A precise determination of the neutron's parton content plays a critical role in the hadron and particle physics frontier, which deepens our understanding of the fundamental parton dynamics~\cite{Kovarik:2019xvh}.
	
	Even as an electrically neutral particle, the neutron can still have photon content, due to its internal charged particles. Experimentally, the neutron's magnetic moment is measured to be $\mu_n = -1.913\mu_N$~\cite{ParticleDataGroup:2022pth}, where $\mu_N=e\hbar/2m_p$ is the nuclear magneton. In a more modern picture, the neutron's electromagnetic property can be described with the corresponding form factors, which are measured in various scattering experiments. See Ref.~\cite{Ye:2017gyb} for an overview of the most updated experiments and the global fitted results. 
	The photon content induced by the electromagnetic form factors can be ascribed to the elastic component, as the neutron remains intact in a scattering process.
	The neutron's elastic photon can be determined through the so-called equivalent photon approximation (EPA)~\cite{Fermi:1924tc,vonWeizsacker:1934nji,Williams:1934ad,Budnev:1974de}, which maps the electromagnetic form factors to the equivalent photon number or spectrum. 
	This approach was adopted to include the elastic component of the proton's \emph{inclusive} photon in the CT14qed PDF set~\cite{Schmidt:2015zda}, while the neutron's elastic photon was assumed to be zero regarding its zero electric charges.
	
	Meanwhile, the neutron also has an inelastic photon component, which corresponds to inelastic scattering processes. In a deep-inelastic scattering (DIS), the internal quarks and gluons of the neutron can be resolved by a deep virtual photon. In such a picture, the photon becomes completely a parton of a neutron, which co-evolves with other partons (quarks and gluon) in terms of the DGLAP equations~\cite{Dokshitzer:1977sg,Gribov:1972ri,Lipatov:1974qm,Altarelli:1977zs}. The first quantum electrodynamics (QED) PDF set to include the photon as a parton of neutron came out as the MRST2004qed~\cite{Martin:2004dh}.
	The initial photon PDF there is parameterized with the radiation from the ``primordial" up and down quarks, governed by the corresponding current or constituent quark masses.
	A small isospin violation at $\mu_0$ was introduced to the valence quarks, with the corresponding amount fixed by momentum conservation. In such a way, the photon and other partons of neutron and proton were determined through a global QCD+QED analysis.
	With the isospin symmetric assumption at $\mu_0$, NNPDF fitted proton and neutron's photon content simultaneously through a global analysis of DIS and Drell-Yan data and released the corresponding PDF set as NNPDF2.3qed~\cite{Ball:2013hta}. Contemporaneously in the CT14qed PDFs~\cite{Schmidt:2015zda}, the proton's initial inelastic photon was constructed with the radiation from the valence quarks, with the initial momentum fraction $\langle x\gamma_p\rangle(\mu_0^2)$ determined with the ZEUS data on the isolated photon production~\cite{ZEUS:2009onz}. The neutron's PDFs are related to the proton ones with an isospin symmetry violation determined through $\langle x\gamma_p\rangle(\mu_0^2)$.
	
	The idea of EPA was extended to determine the inelastic photon as well in terms of the inelastic structure functions (SFs)~\cite{Anlauf:1991wr,Blumlein:1993ef,Mukherjee:2003yh}.
	This approach was inherited by the LUXqed group~\cite{Manohar:2016nzj,Manohar:2017eqh}, who established a rigorous framework based on the collinear factorization beyond the leading order by including proper $\overline{\rm MS}$ matching terms.
	With the precisely measured elastic and inelastic structure functions, the proton's photon PDF was constrained up to a high precision at the level of 1--2\%. Afterward, this LUXqed formalism~\cite{Manohar:2016nzj,Manohar:2017eqh} was adopted to determine the proton's initial photon PDF at $\mu_0$ in the QCD+QED global analyses, including CT18qed~\cite{Xie:2021equ}, MMHT2015qed~\cite{Harland-Lang:2019pla}/MSHT20qed~\cite{Cridge:2021pxm}, and NNPDF3.1luxQED\footnote{The NNPDF3.1luxQED PDFs choose a high initialization scale $\mu_0=100~\GeV$, while others take a low one $\mu_0\sim\calO(1~\GeV)$.}~\cite{Bertone:2017bme}, which advanced our understanding of the proton's structure in the precision frontier~\cite{Amoroso:2022eow}.
	Based on the MRST2004qed proton-neutron isospin relation~\cite{Martin:2004dh}, both the MMHT2015qed~\cite{Harland-Lang:2019pla} and MSHT20qed~\cite{Cridge:2021pxm} analyses have released the neutron PDFs as well.
	
	In this work, we extend our recent CT18qed study~\cite{Xie:2021equ} to include the neutron PDFs as well, to provide a key input for many phenomenological applications related to nucleus scattering. Different from the CT14qed neutron PDFs~\cite{Schmidt:2015zda}, we will include the elastic photon component in terms of the electromagnetic form factors, similar to the treatment in the MMHT2015qed~\cite{Harland-Lang:2019pla} and MSHT20qed~\cite{Cridge:2021pxm} PDF sets.
	Following the CT18qed proton study~\cite{Xie:2021equ}, the neutron's photon PDF can be generated either through applying the LUXqed formalism at any scale or through evolving the LUXqed initialized photon up to higher scales, which we dub as CT18lux and CT18qed PDF sets, respectively. 
	In the lower energy regime, the structure functions receive various nonpertubative contributions, such as higher-twist and target-mass corrections, which will result in different photon PDFs. We ascribed these variations as a part of the corresponding PDF uncertainties.
	
	This paper is organized as follows. Sec.~\ref{sec:lux} lays out the neutron's photon PDF determination. More specifically, in Sec.~\ref{sec:el}, we discuss the neutron's elastic photon component, followed by the CT18lux inelastic component in Sec.~\ref{sec:inel}, using the LUXqed formalism. In Sec.~\ref{sec:qed}, we develop the DGLAP methodology to consistently address the momentum sum rule and isospin violation in the mixed QCD and QED PDF evolution. A comprehensive comparison between the first and second generations of the neutron's photon PDFs is presented in Sec.~\ref{sec:compare}.
	Two examples of the phenomenological implications are demonstrated in Sec.~\ref{sec:imp} and conclusions come afterward in Sec.~\ref{sec:conclude}.

	\section{The LUXqed photon PDF formalism}
	\label{sec:lux}
	As implied by the equivalent photon approximation~\cite{Budnev:1974de}, the nucleon's photon PDF is related to the corresponding structure functions. The elastic structure functions can be constructed in terms of the electromagnetic form factors as~\cite{Budnev:1974de}
	\begin{eqnarray}\label{eq:FEFM}
		&F_E(Q^2)=\frac{4m^2G_E^2(Q^2)+Q^2G_M^2(Q^2)}{4m^2+Q^2},\\
		&F_M(Q^2)=G_M^2(Q^2),
	\end{eqnarray}
	where $m$ is the nucleon mass. By mapping the equivalent photon spectrum to the elastic\footnote{The ``elastic" and ``inelastic" are referred as ``coherent" and ``incoherent", respectively, by the MRST group~\cite{Martin:2014nqa}.} photon PDF, we obtain~\cite{Budnev:1974de}
	\begin{equation}\label{eq:EPA}
		x\gamma^{\el}=\frac{\alpha}{\pi}
		\int\frac{\dd Q^2}{Q^2}\left[\frac{Q_\perp^2}{Q^2}F_E(Q^2)+\frac{x^2}{2}F_M(Q^2)\right].
	\end{equation}
	where the photon virtuality $Q^2$ can be related to the transverse momentum $Q_\perp$ as
	\begin{equation}
		Q^2=\frac{Q_\perp^2+x^2m^2}{1-x}.
	\end{equation}
	An improved form based on the $Q_\perp^2$ integration including the proper integral interval can be found in Refs.~\cite{Harland-Lang:2016apc,Harland-Lang:2016kog}. 
	
	This idea was applied to the inelastic photon as well in Refs.~\cite{Anlauf:1991wr,Blumlein:1993ef,Mukherjee:2003yh}.
	With a rigorous collinear factorization framework, the LUXqed group extended this approach beyond the leading order by including a proper $\msbar$ matching term~\cite{Manohar:2016nzj,Manohar:2017eqh}, which we dub as the LUXqed formalism
	\begin{eqnarray}\label{eq:LUXqed}
		x \gamma(x,\mu^2)
		&=\frac{1}{2\pi\alpha(\mu^2)}\int_{x}^{1}\frac{\dd z}{z}
		\Bigg\{\int_{\frac{x^{2}m^{2}}{1-z}}^{\frac{\mu^{2}}{1-z}}
		\frac{\dd Q^{2}}{Q^{2}}\alpha_{\rm ph}^{2}(-Q^2)
		\Bigg[\left(zp_{\gamma q}(z)+
		\frac{2x^{2}m^{2}}{Q^{2}}\right)F_{2}(x/z,Q^{2})\\
		&-z^{2}F_{L}(x/z,Q^{2})\Bigg]
		-\alpha^{2}(\mu^2)z^{2}F_{2}(x/z,\mu^2) 
		\Bigg\} + \mathcal{O}(\alpha^2, \alpha\alpha_s). 
	\end{eqnarray}
	Here $p_{\gamma q}(z)\equiv[1+(1-z)^2]/z$ corresponds to the leading order DGLAP splitting kernel.
	The $\alpha_{\rm ph}$ is the physical fine structure constant, defined as
	\begin{eqnarray}
		\alpha_{\rm ph}(q^2)=\frac{\alpha^2(\mu^2)}{1-\Pi(q^2,\mu^2)},
	\end{eqnarray}
	where $\Pi(q^2,\mu^2)$ is the vacuum polarization function, evaluated at $q^2=-Q^2$.
	The first term in Eq.~(\ref{eq:LUXqed}) with the square bracket is referred to as the physical factorization component, and the second one that only involves $F_2$ is the $\msbar$ conversion~\cite{Manohar:2017eqh}.
	To perform the integration in Eq.~(\ref{eq:LUXqed}), the neutron's structure functions $F_{2,L}^n$ in the complete $(x,Q^2)$ plane are needed.
	With a precise knowledge of the structure functions $F_{2,L}^n$, the neutron's photon PDF can be determined up to a high precision, which is the main goal of this work.
	
	\begin{figure}[h]
		\centering
		\includegraphics[width=0.99\textwidth]{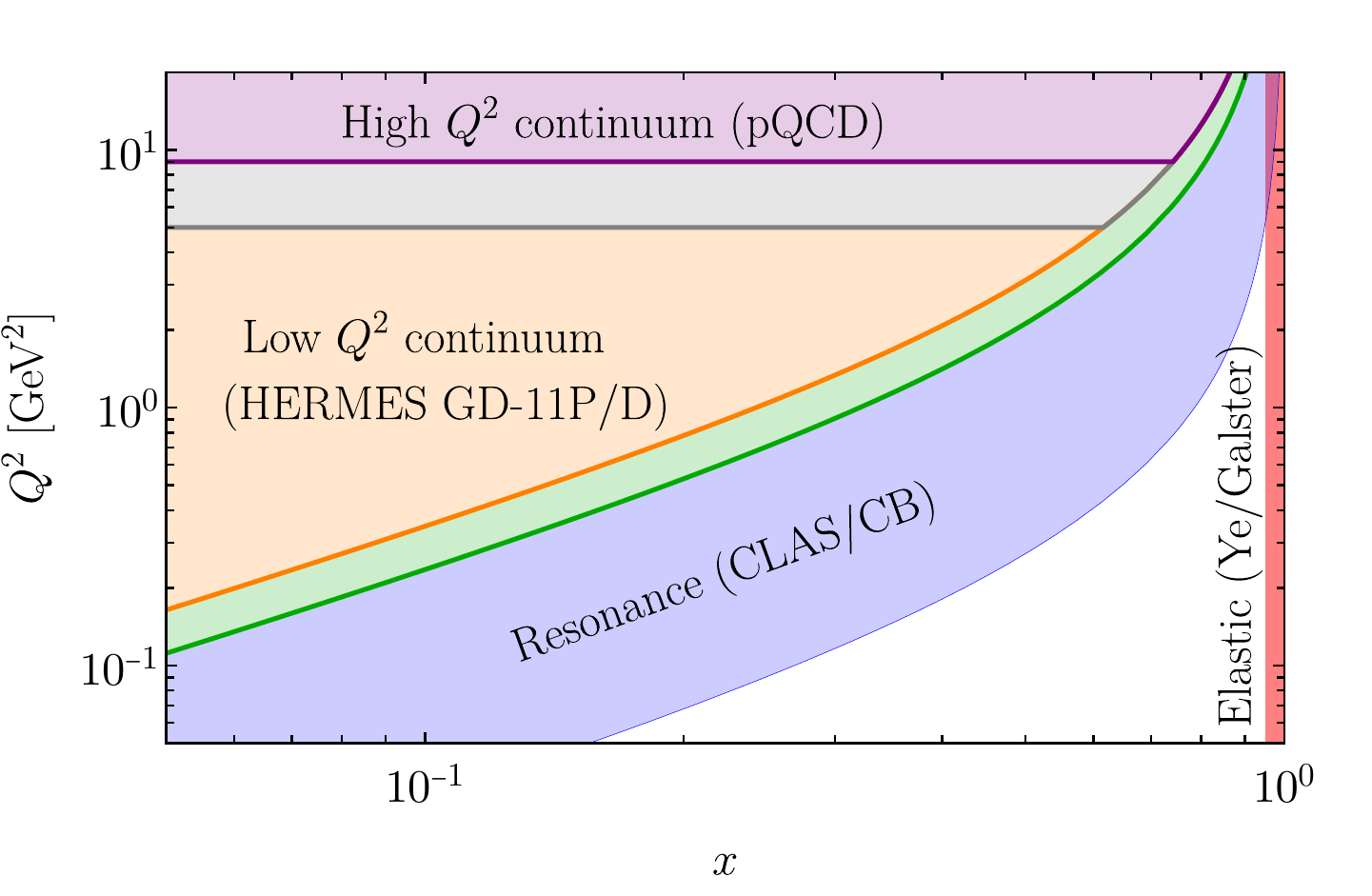}
		\caption{The breakup plane to determine the neutron structure functions, $F^{n}_{2,L}(x,Q^2)$.}
		\label{fig:breakup}
	\end{figure}
	
	Following the LUXqed treatment~\cite{Manohar:2016nzj,Manohar:2017eqh}, we divide the $(x,Q^2)$ plane mainly into three regions, with a smooth transition when crossing a boundary, illustrated in Fig.~\ref{fig:breakup}. In the high-$Q^2$ region when $Q^2>Q^2_{\rm PDF}$, perturbative QCD (pQCD) is well-established, thus we employ pQCD to calculate the structure functions. Here $Q^2_{\rm PDF}$ denotes the matching scale, which varies between 5 and 9 $~\GeV^2$ in this work, with the difference quantifying the corresponding uncertainty. When $Q^2$ decreases, we gradually enter the non-perturbative regions. The region with $Q^2<Q^2_{\rm PDF}$ and $W^2>W^2_{\rm res}$ with $W^2_{\rm res}\sim3-4~\GeV^2$ is termed as the low-$Q^2$ continuum region. In this region, we directly take the structure functions measured by the HERMES experiment with the GD-11P and GD-11D fits~\cite{HERMES:2011yno} with the ALLM parametrization~\cite{Abramowicz:1991xz,Abramowicz:1997ms}. 
	Considering the potential loss of reliability in the extrapolated low-$x$ region, $x\lesssim10^{-4}$, we take the pQCD structure functions as an alternative choice. In the resonance region $W^2<W^2_{\rm res}$, the nucleon structure functions show many resonance features. In this work, we take the CLAS~\cite{CLAS:2003iiq} and Christy-Bosted~\cite{Christy:2007ve,Bosted:2007xd,Christy:2021abc} fits to determine the resonance structure functions. Below the resonance threshold $W^2<(m+m_\pi)^2$, the nucleon inelastic structure functions vanish, which does not contribute. Finally, the elastic form factors are taken from the global fit of the elastic scattering data~\cite{Ye:2017gyb} or the Galster parameterization~\cite{Galster:1971kv}, shown as the red band around $x\sim1$ as the nucleon remains intact. A detailed description of the structure-function determination will come in the next two subsections.
	
	\subsection{The elastic photon PDF}
	\label{sec:el}
	In this subsection, we explore the nucleon's elastic photon, with the elastic form factors from fitting the global data~\cite{Ye:2017gyb}. We will also compare our results with the MMHT15qed~\cite{Harland-Lang:2019pla} and MSHT20qed~\cite{Cridge:2021pxm} ones, which uses the Galster parameterization~\cite{Galster:1971kv}.
	
	The LUXqed formalism, Eq.~(\ref{eq:LUXqed}), can be applied to the elastic photon as well. 
	The elastic structure functions in Eq.~(\ref{eq:FEFM}) can be rewritten as~\cite{Ricco:1998yr}
	\begin{eqnarray}\label{eq:Fel}
		F_2^{\el}(x,Q^2)&=\frac{G_E^2(Q^2)+G_M^2(Q^2)\tau}{1+\tau}\delta(1-x),\\
		F_L^{\el}(x,Q^2)&=\frac{G_E^2(Q^2)}{\tau}\delta(1-x),
	\end{eqnarray}
	where $\tau=Q^2/4m^2$. Here $\delta(1-x)$ factor indicates that the nucleon remains intact, which corresponds to the red narrow band in Fig.~\ref{fig:breakup}. Substituting Eq.~(\ref{eq:Fel}) into Eq.~(\ref{eq:LUXqed}) ends up with the LUXqed elastic photon as
	\begin{eqnarray}\label{eq:LUXqedel}
		x\gamma^{\el}(x,\mu^2)=\frac{1}{2\pi\alpha(\mu^2)}
		\int_{\frac{x^2m^2}{1-x}}^{\infty}\frac{\dd Q^2}{Q^2}\alpha_{\rm ph}(-Q^2)
		\bigg[
		&\left(1-\frac{x^2m^2}{Q^2(1-x)}\right)\frac{2(1-x)G_E^2(Q^2)}{1+\tau}\\
		+&\left(xp_{\gamma q}(x)+\frac{2x^2m^2}{Q^2}\right)
		\frac{G_M^2(Q^2)\tau}{1+\tau}\bigg].
	\end{eqnarray}
	Here this LUXqed form shares the same philosophy as the EPA in Eq.~(\ref{eq:EPA}), but improves with a proper matching, which facilitates the higher-order extension. Here the physical factorization and $\msbar$ terms are merged together, resulting in the $Q^2$ integration up to infinity.
	We remind the readers that this upper integration limit is different from the one in the MMHT15qed~\cite{Harland-Lang:2019pla} and MSHT20qed~\cite{Cridge:2021pxm} fits, which adopted the starting scale $\mu^2_0$, while the elastic photon at higher scale was obtained through a derived evolution equation. See Ref.~\cite{Harland-Lang:2019pla} for the details.

	As mentioned in Sec.~\ref{sec:intro}, the electromagnetic form factors $G_{E,M}(Q^2)$ can be directly extracted from the experimental measurements.
	In comparison with the proton form factors, the neutron ones are poorly known, as no free and stable neutron exists in a natural environment.
	Therefore, the neutron form factors have to be extracted from nuclei (such as Deuterium or Helium), by subtracting the dominant proton contribution after properly accounting for nuclear effects. See Ref.~\cite{Ye:2017gyb} for the details of the determination methodology. 
	
	For the neutron's electric form factor, MMHT2015qed~\cite{Harland-Lang:2019pla} and MSHT20qed~\cite{Cridge:2021pxm} PDF sets adopted a phenomenological parameterization by Galster et al.~\cite{Galster:1971kv},
	\begin{equation}
		G_{E}^{n}(Q^2)=\frac{A \tau}{1+B \tau} G_{D}\left(Q^{2}\right),
	\end{equation} 
	where the $G_D$ is the dipole form,
	\begin{equation}
		G_{D}(Q^2)=\frac{1}{(1+Q^2/\Lambda^2)^2},
	\end{equation}
	with $\Lambda^2=0.71~\GeV^2$. The parameters $A$ and $B$ are determined through the Deuterium ($^2_1$H) and Helium ($^3_2$He) scattering experiments~\cite{Kelly:2004hm},
	\begin{equation}
		A=1.70\pm0.04, ~B=3.30\pm0.32.
	\end{equation}
	The magnetic form factor was taken as a simple dipole approximation
	\begin{equation}
		G_M^{n}(Q^2)=\mu_n G_D(Q^2),
	\end{equation}
	where $\mu_n=-1.913$~\cite{ParticleDataGroup:2022pth}.
	
	\begin{figure}[h]
		\centering
		\includegraphics[width=0.49\textwidth]{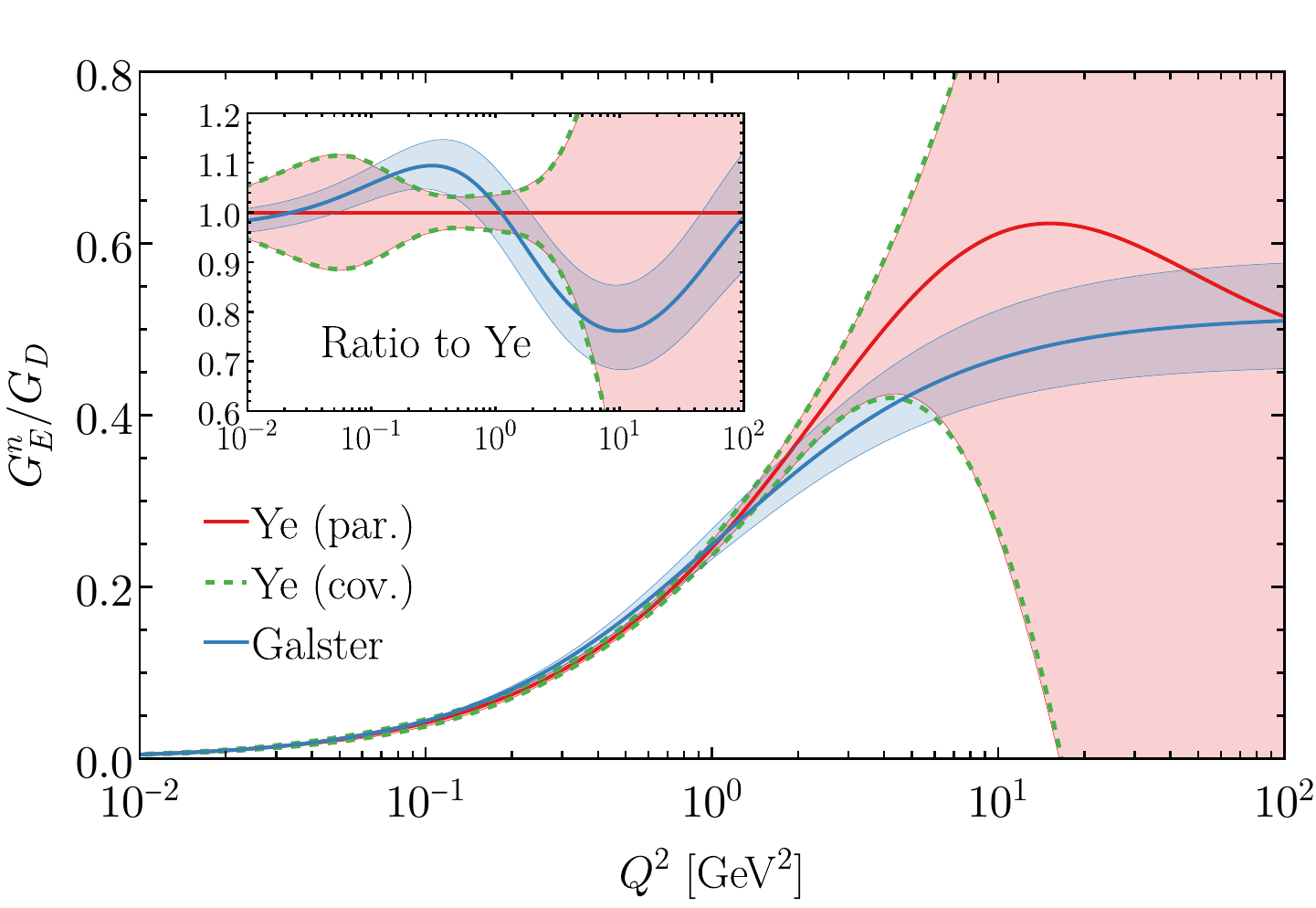}
		\includegraphics[width=0.49\textwidth]{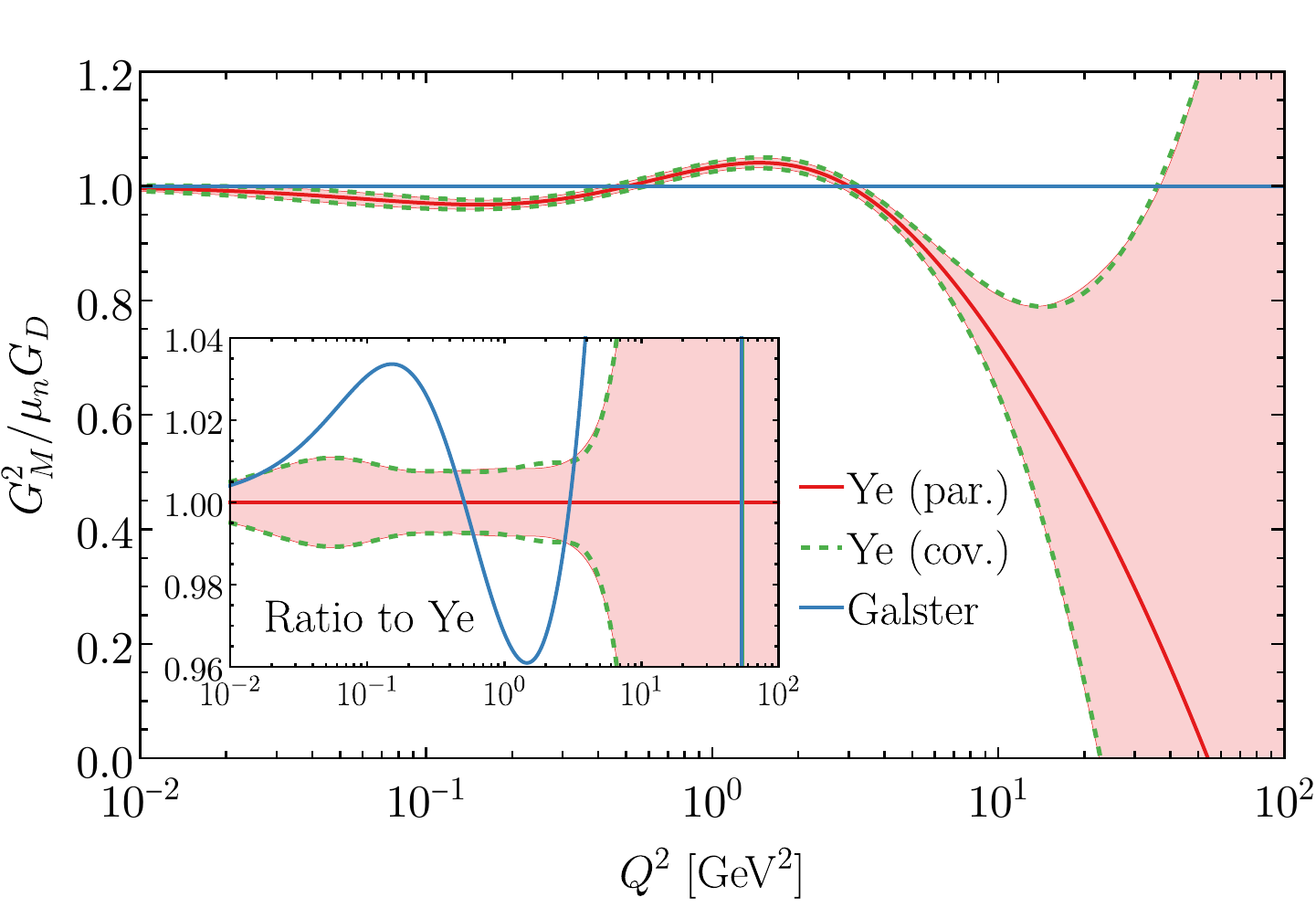}
		\caption{The recent world fit of neutron's elastic form factors, $G_{E,M}^{n}$, by Ye et al.~\cite{Ye:2017gyb}, compared with the Galster parameterization~\cite{Galster:1971kv} used in the MMHT2015qed~\cite{Harland-Lang:2019pla} and MSHT20qed~\cite{Cridge:2021pxm} PDF sets.
			The red thin bands (``par.") parameterize the corresponding uncertainties with analytical forms, while the green dashed ones (``cov.") are directly extracted from the covariance matrix of the fit~\cite{Ye:2017gyb}.}
		\label{fig:GEGM}
	\end{figure}
	
	Instead, we adopt the global fitted $G_{E,M}^n$ results by Ye et al.~\cite{Ye:2017gyb} to determine the neutron's elastic photon component in this work. The comparison between the two different elastic form factors is displayed in Fig.~\ref{fig:GEGM}.
	Here the global fitted $G_{E,M}^n$ are displayed as the red curves, and the red thin bands (``par.") indicate the corresponding uncertainties parameterized with analytical formulas. The green dashed curves (``cov.") refer to the same uncertainties obtained with the covariance matrix of the fit, which agree with the parameterized forms very well through the whole $Q^2$ range. 
	In practice, we adopt the analytical form in our calculation for convenience.
	In comparison with the Ye et al.'s fit, the Galster parameterization~\cite{Galster:1971kv,Kelly:2004hm} captures the general feature of the neutron's form factors, while with an underestimation of the uncertainties, especially at a high photon virtuality $Q^2$.
	
	\begin{figure}[h]
		\includegraphics[width=0.49\textwidth]{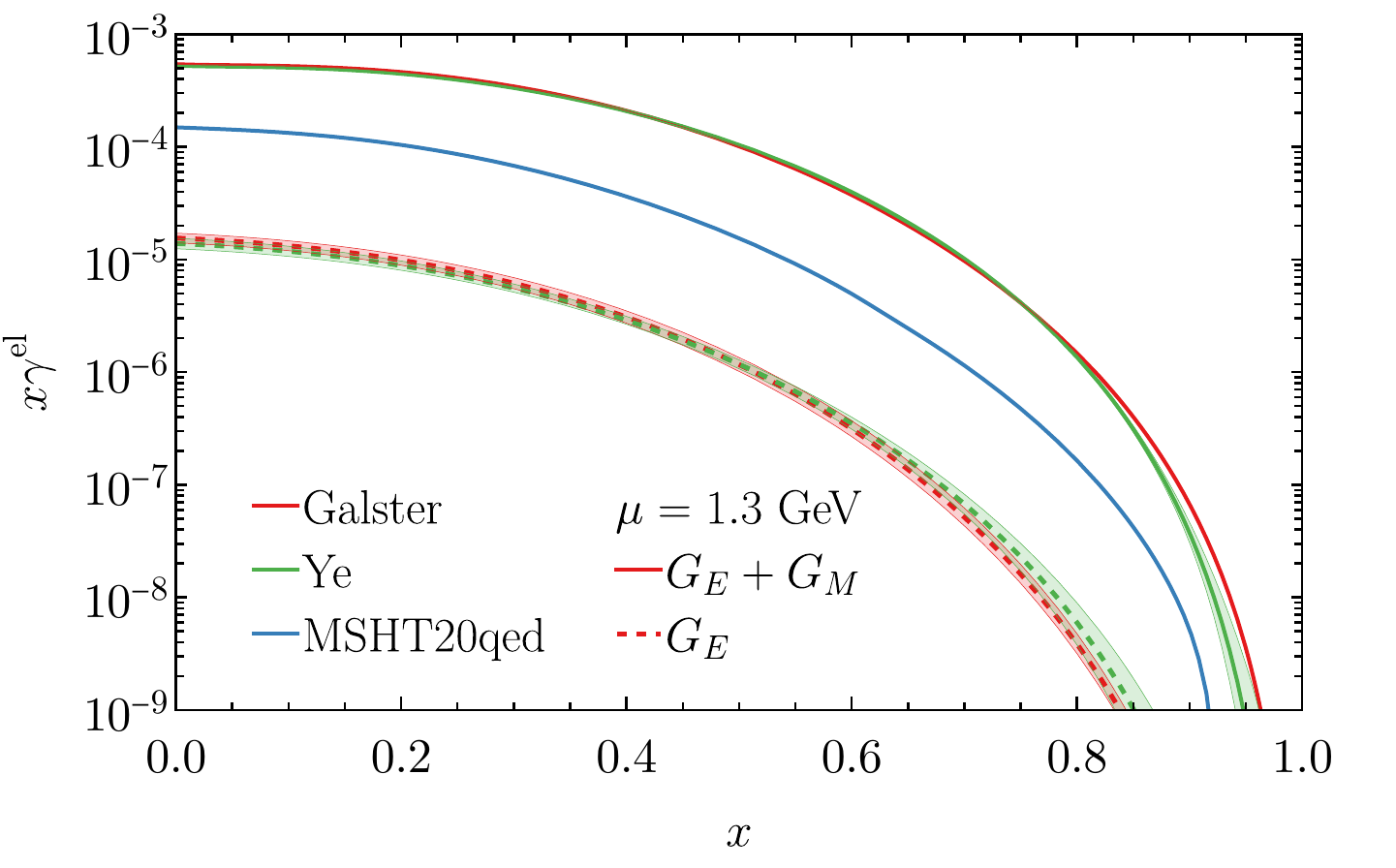}
		\includegraphics[width=0.49\textwidth]{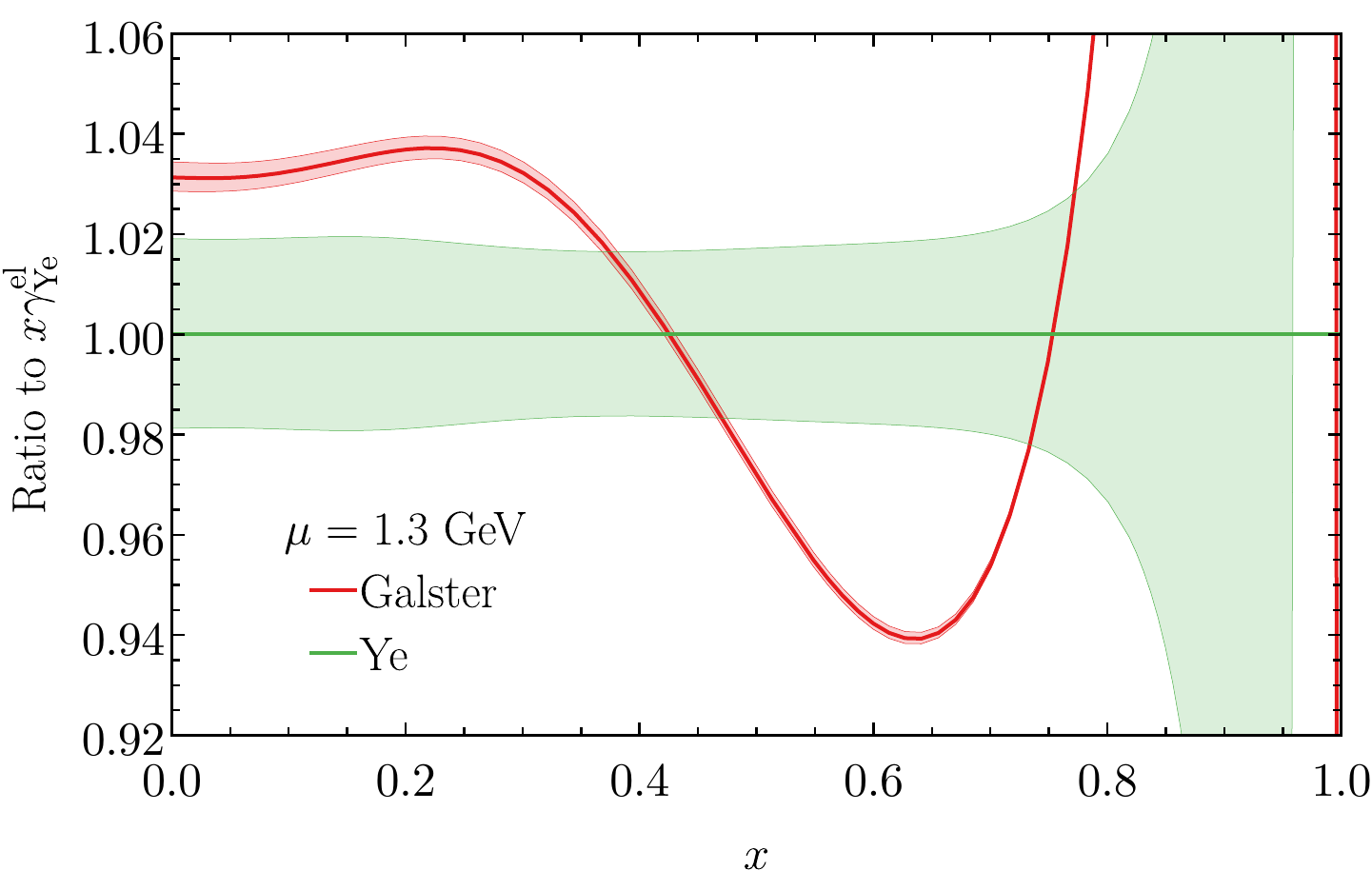}
		\includegraphics[width=0.49\textwidth]{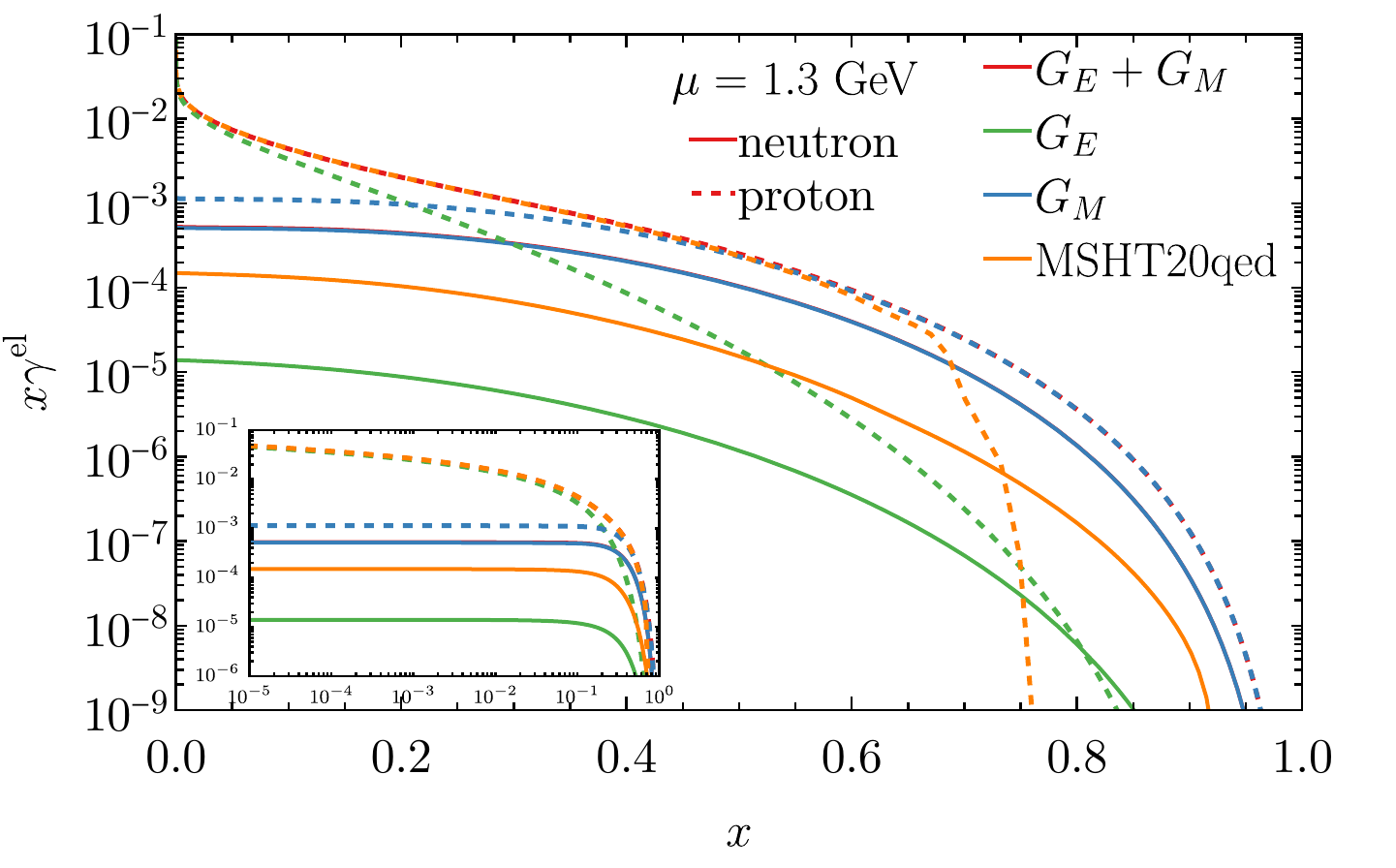}
		\includegraphics[width=0.49\textwidth]{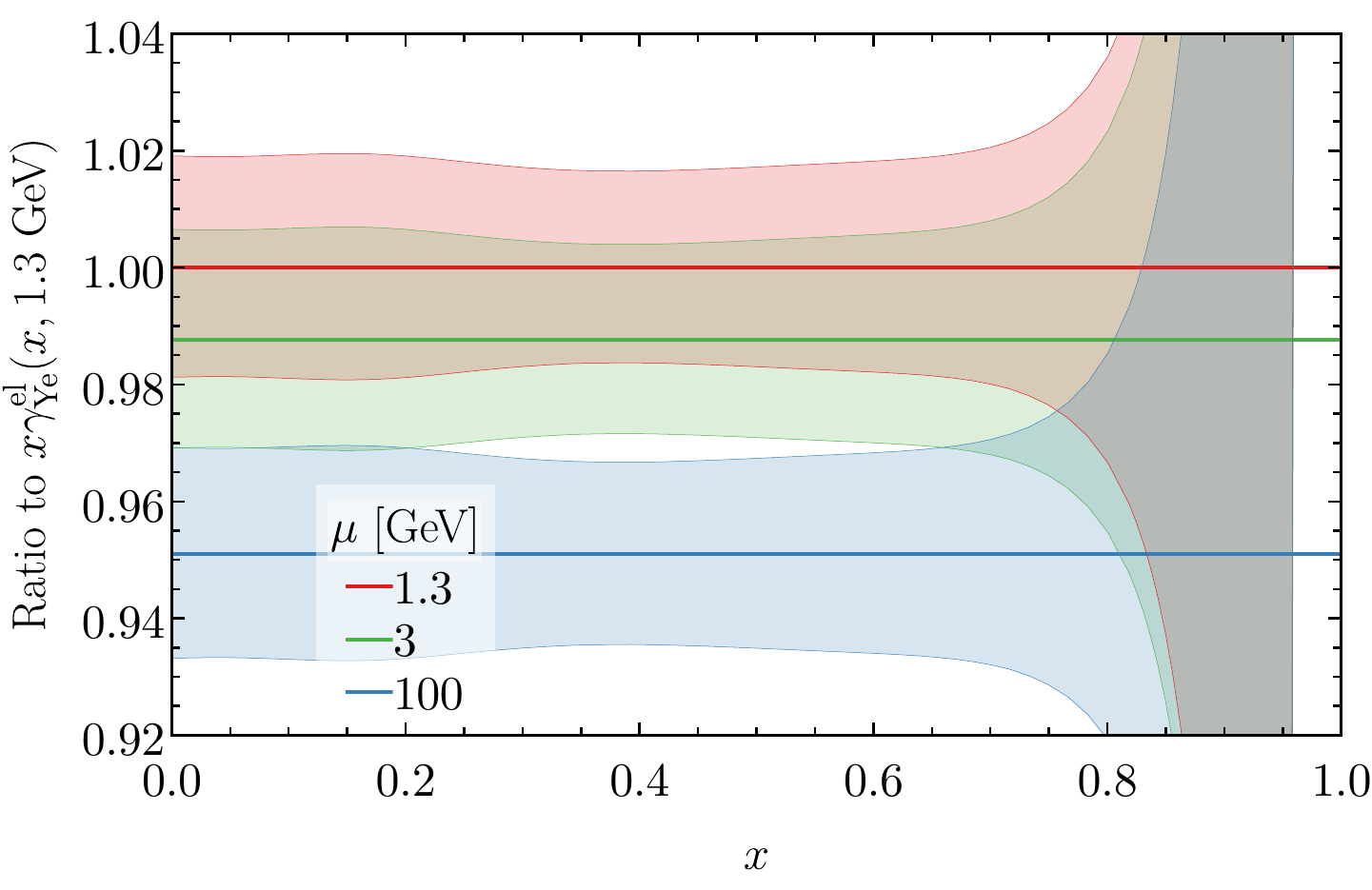}
		\caption{Upper: the neutron's elastic photon at scale $\mu=1.3~\GeV$, based on the Galster~\cite{Galster:1971kv} and Ye~\cite{Ye:2017gyb}'s electromagnetic form factors.
			Lower: the comparison between neutron and proton's elastic photon PDFs, generated with the Ye's form factors.
			We also include MSHT20qed result~\cite{Cridge:2021pxm} for a comparison.}
		\label{fig:el100GeV}
	\end{figure}
	
	In Fig.~\ref{fig:el100GeV}, we show the neutron's elastic photon PDF, $x\gamma^{\rm el}(x,\mu^2)$, at $\mu=1.3~\GeV$. The higher-scale photon PDFs behave similarly, with only a small variation resulting from the running of $\alpha(\mu^2)$ in Eq.~(\ref{eq:LUXqedel}), as shown in the lower right panel.
	First, we see that the neutron's elastic photon is dominated by the magnetic contribution, while the electric form factor only contributes a percent level to the total elastic photon. This can be easily understood in terms of its zero electric charge, which is different from the proton case, as shown in the lower left panel of Fig.~\ref{fig:el100GeV}. In the proton case, the electric (magnetic) contribution dominates the elastic photon at low (large) $x$, with equality around $x\sim0.2$. 
	Second, we also see that the neutron's total elastic photon PDF is generally much smaller than the proton's, by about two orders of magnitude at a small $x$. At a large $x$ ($x\gtrsim0.7$) where the magnetic contribution dominates, the neutron and proton elastic photons scale as the magnetic moment ratio $\mu_n/\mu_p$.
	Third, the neutron's elastic photon induced by the Galster parameterization is slightly larger than the one with Ye's form factors by about 3\% at a low $x$, 
	mainly due to the larger low-scale $G_M$ when $Q^2\lesssim0.6~\GeV^2$ as shown in Fig.~\ref{fig:GEGM}.
	As we observed before, the Galster parameterization underestimated the uncertainties, which resulted in smaller error bands.
	Meanwhile, we include the elastic photon from the MSHT20qed fit~\cite{Cridge:2021pxm} for a comparison, which is about 30\% of our result, mainly resulting from a much lower integration limit, as we mentioned above. 
	However, this difference of the proton's elastic photon between the one with Ye's form factors and the MSHT20qed result is much smaller, which only shows up when $x\gtrsim0.7$ at $\mu=1.3~\GeV$\footnote{We have also checked the difference at a higher scale, which occurs at a higher $x$ value, \emph{e.g.}, $x\gtrsim0.9$ at $\mu=100~\GeV$.
		We suspect that the smaller MSHT20qed proton's elastic photon at the large $x$ is caused by the LHAPDF interpolation of its $x$ grid.} as shown in Fig.~\ref{fig:el100GeV} lower left panel.  
	
	\subsection{The inelastic photon PDF with the LUXqed approach: CT18lux}
	\label{sec:inel}
	As discussed before, when applying the LUXqed formalism to the neutron's inelastic photon, we need to integrate Eq.~(\ref{eq:LUXqed}) throughout the complete $(x,Q^2)$ plane as shown in Fig.~\ref{fig:breakup}.
	Here we take similar breakups as the original LUXqed treatment~\cite{Manohar:2017eqh}.
	That is, we divide the $(x,Q^2)$ plane into regions discussed as follows.
	
\begin{figure}[h]
		\centering
		\includegraphics[width=0.49\textwidth]{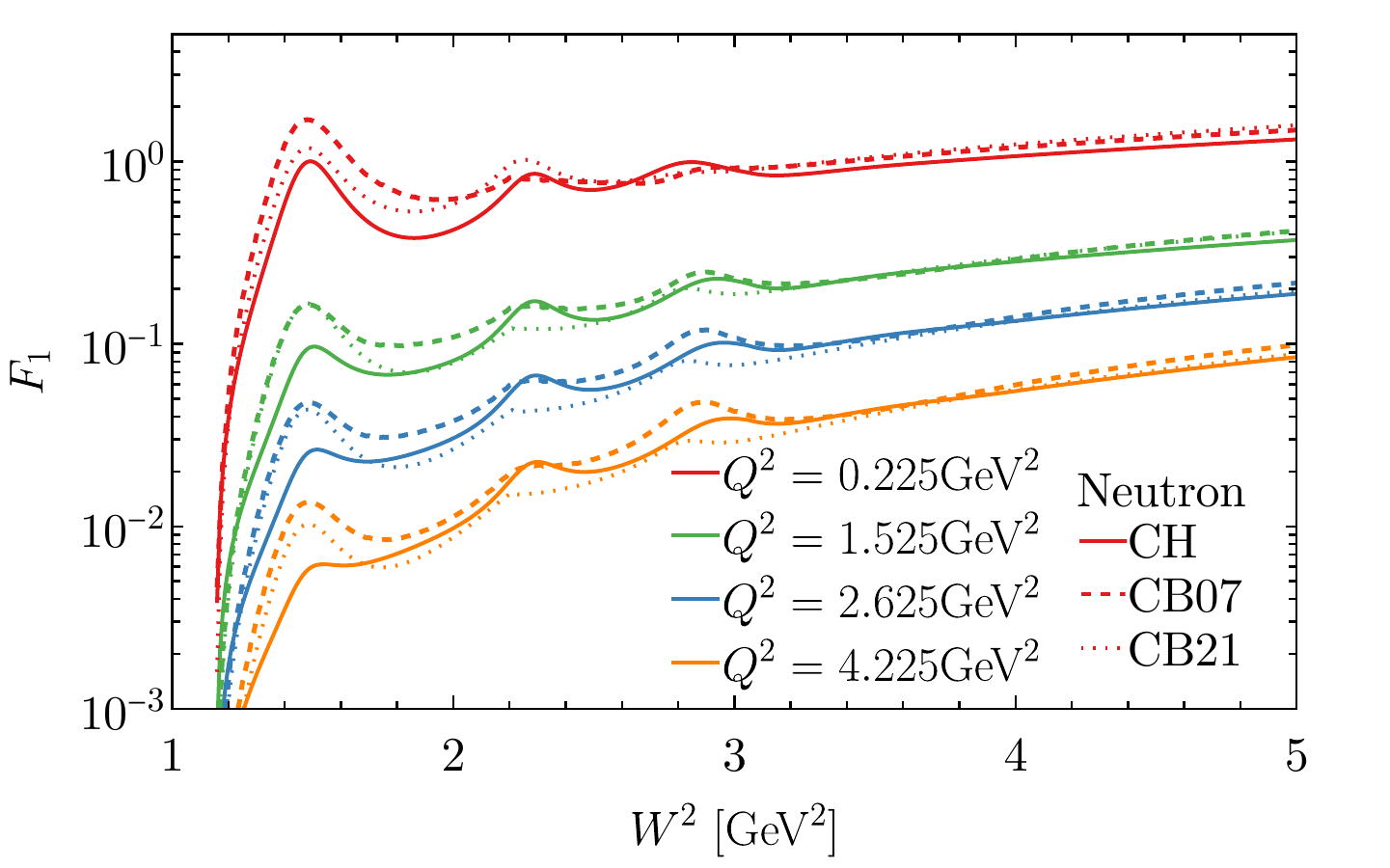}
		\includegraphics[width=0.49\textwidth]{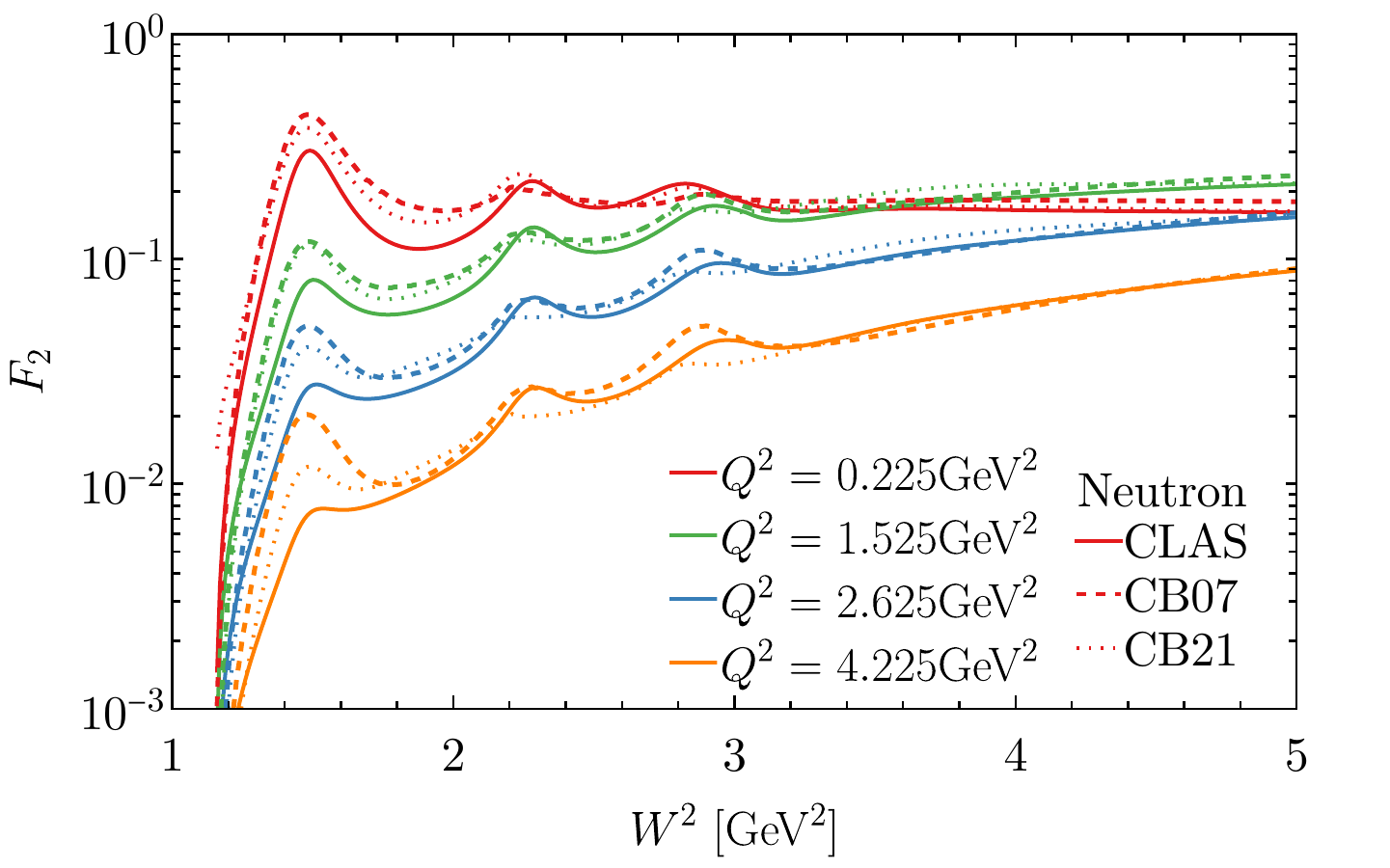}
		\caption{The neutron structure functions in the resonance region, taken from the CLAS experiment~\cite{CLAS:2003iiq} and the Christy-Bosted fit~\cite{Christy:2007ve,Bosted:2007xd,Christy:2021abc}, respectively. The ``CH" denotes the smooth transition from the CLAS to HERMES with Eq.~(\ref{eq:CH}).}
		\label{fig:SFres}
	\end{figure}

	\textbf{Resonance.} $W^2<W_{\rm lo}^2=3~\GeV^2$ is defined as the resonance region. We remind the readers that the hadronic inelastic kinematics is bounded from below with a threshold $W^2\geq(m+m_\pi)^2$, shown as the lower boundary in Fig.~\ref{fig:breakup}. Similar to the LUXqed treatment, we take the neutron's resonance structure functions either from the CLAS experiment~\cite{CLAS:2003iiq} or the Christy-Bosted fit in 2007 (CB07)~\cite{Christy:2007ve,Bosted:2007xd} as well as the updated one in 2021 (CB21)~\cite{Christy:2021abc}.
	With a few representative scales $Q^2$, we compare $F_{1,2}$ from these three resources in Fig.~\ref{fig:SFres}. 
	We see an overall agreement along the whole  $W^2$ range, even though the difference is slightly larger than the proton case, which can be found in our recent work~\cite{Xie:2021equ}. 
	We take the CLAS fit as our default choice, while the variation to CB21 fits quantifies the resonance uncertainty.
	
\begin{figure}[h]
		\centering
		\includegraphics[width=0.49\textwidth]{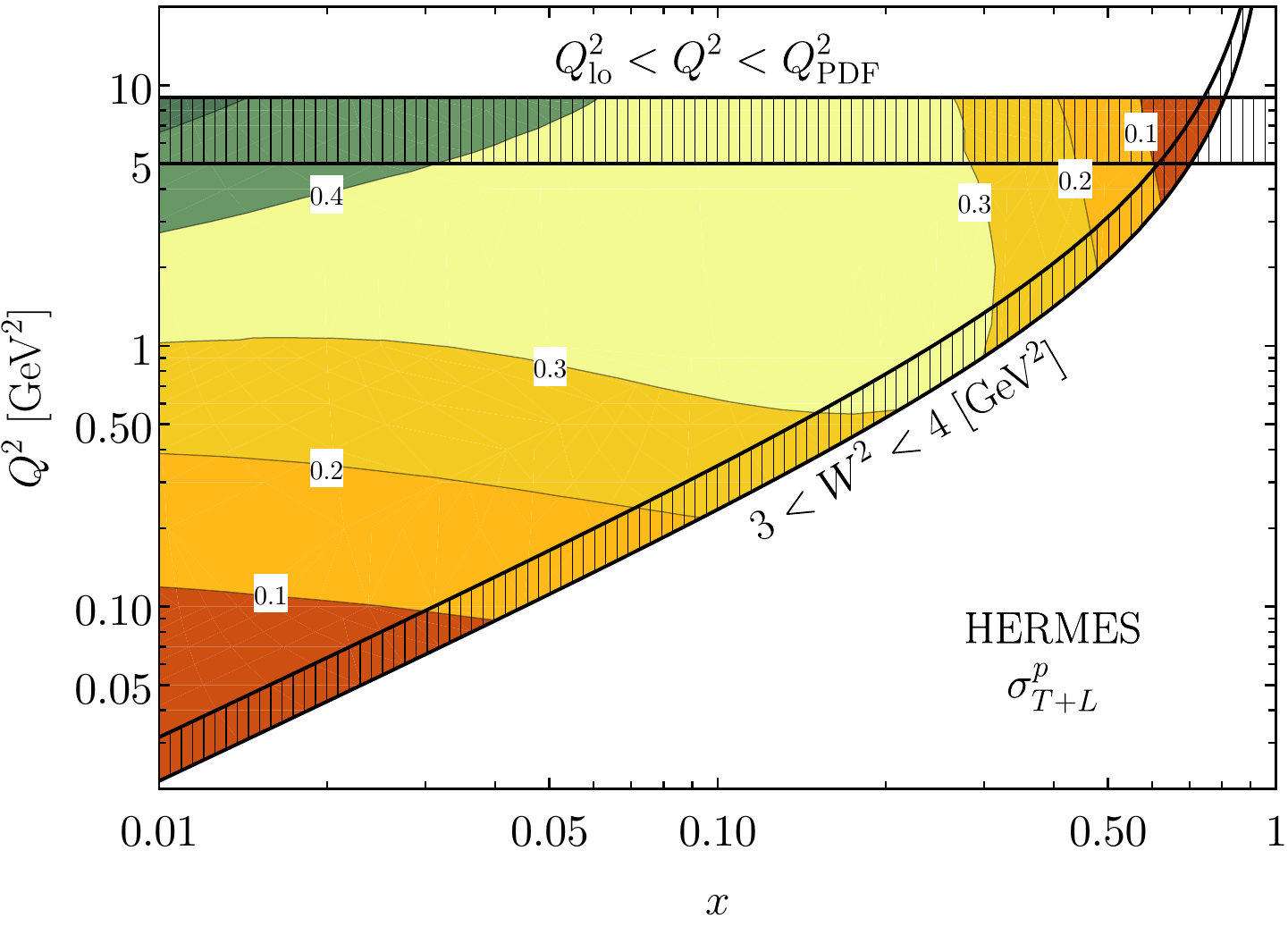}
		\includegraphics[width=0.49\textwidth]{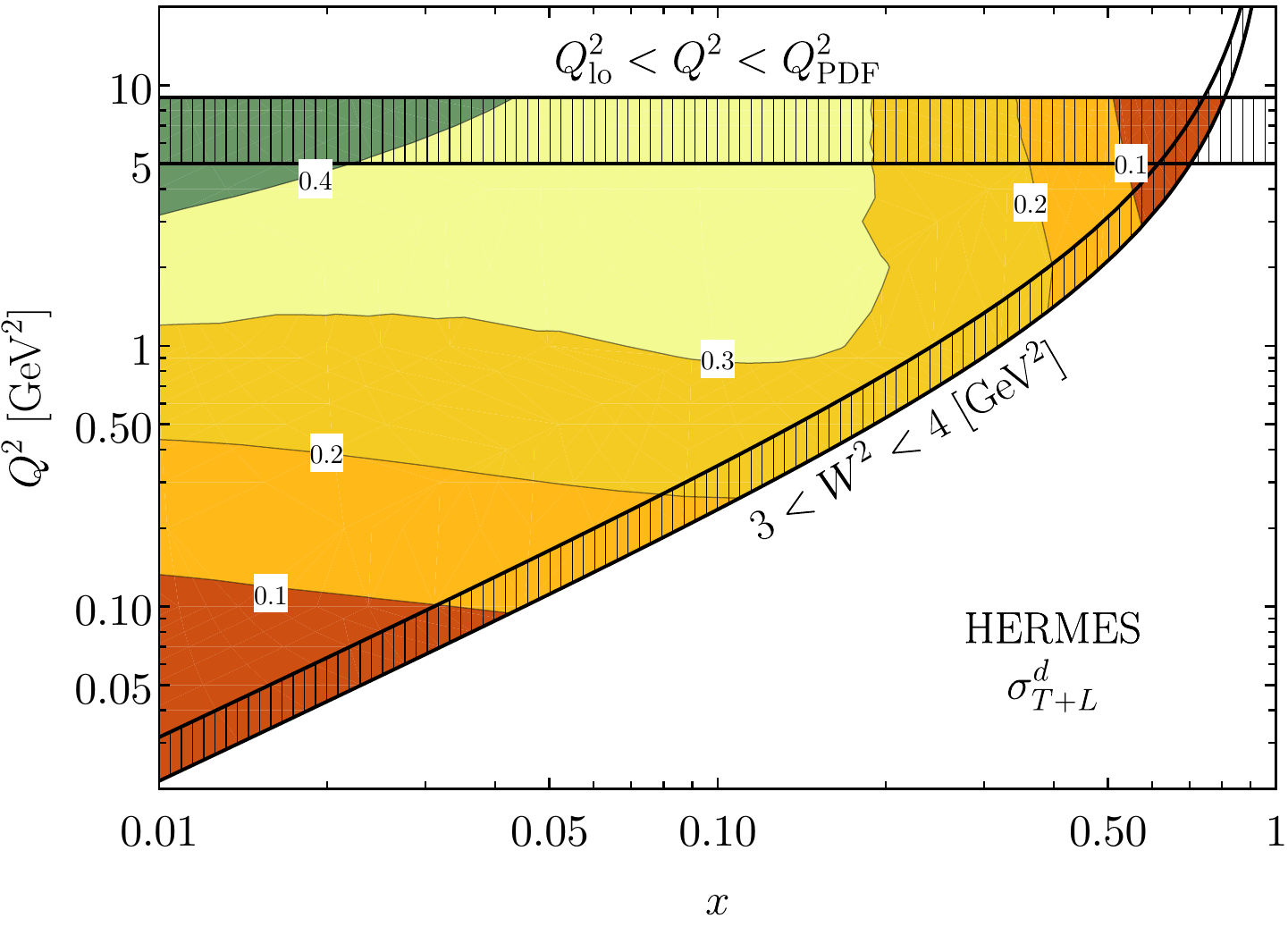}    
		\includegraphics[width=0.49\textwidth]{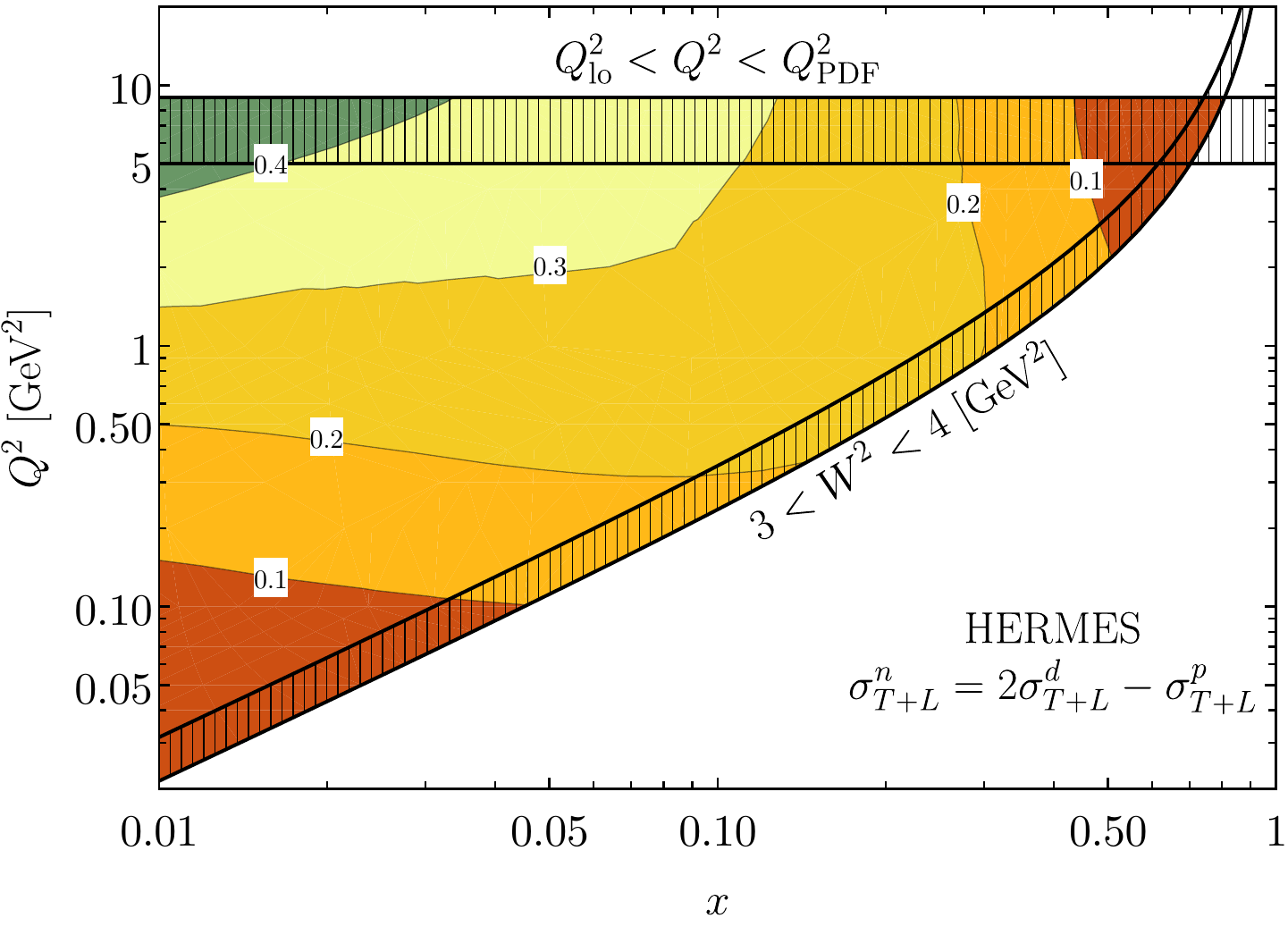}    
		\caption{Upper: the HERMES longitudinal-transverse cross-section sum $\sigma_{T+L}$~[barn] for the proton (left) and deuteron (right) in the low-$Q^2$ continuum region.
			Lower: the derived neutron cross-section sum (right) based on Eq.~(\ref{eq:neutron}), with nuclear corrections folded into the HERMES and $R^n_{L/T}$ ratio uncertainties. The shaded area denotes two smooth transition regions.}
		\label{fig:HERMES}
	\end{figure}

	\textbf{Low-$Q^2$ continuum.} $W^2>W_{\rm hi}^2=4~\GeV^2$ and $Q^2<Q_{\rm lo}^2$ is the low-$Q^2$ continuum region.
	We follow the LUXqed treatment~\cite{Manohar:2017eqh} and construct the structure functions based on the HERMES GD11-P and GD11-D fits~\cite{HERMES:2011yno}, which adopted a wide range of data and the ALLM functional form~\cite{Abramowicz:1991xz,Abramowicz:1997ms}. The HERMES Collaboration measured the inclusive DIS cross section from both proton and deuteron targets. In Fig.~\ref{fig:HERMES}, we show the sum of longitudinal and transverse photon-absorption cross sections $\sigma_{T+L}=\sigma_T+\sigma_L$, with which we can directly construct the $F_2$ structure function as
	\begin{equation}
		F_2(x,Q^2)=\frac{1}{4\pi^2\alpha}\frac{Q^2(1-x)}{1+4x^2m^2/Q^2}
		\sigma_{T+L}(x,Q^2).
	\end{equation}
	
	In order to get the $F_1$ or $F_L$, we need the longitudinal-to-transverse cross-section ratio 
	\begin{equation}
		R_{L/T}(x,Q^2)=\frac{\sigma_L(x,Q^2)}{\sigma_T(x,Q^2)}=\left(1+\frac{4x^2m^2}{Q^2}\right)\frac{F_2(x,Q^2)}{2xF_1(x,Q^2)}-1,
	\end{equation}
	which can be obtained from the R1998 fit~\cite{E143:1998nvx} or Ref.~\cite{Ricco:1998yr}.
	In such a way, the longitudinal structure function can be constructed as
	\begin{eqnarray}
		F_{L}(x, Q^{2})&=\left(1+\frac{4x^2m^2}{Q^2}\right)F_2(x,Q^2)-2xF_1(x,Q^2)\\
		&=\left(1+\frac{4 m^{2} x^{2}}{Q^{2}}\right) 
		F_{2}(x, Q^{2})\frac{R_{L/T}(x, Q^{2})}{1+R_{L/T}(x, Q^{2})}.
	\end{eqnarray}
	Here we always keep the target-mass factor $(1+4x^2m^2/Q^2)$, which will be explored in more detail for its corrections.
	In order to obtain the neutron's structure functions, we take the assumption that
	\begin{equation}\label{eq:neutron}
		2\sigma_{T+L}^{d}=\sigma_{T+L}^{p}+\sigma_{T+L}^{n},
	\end{equation}
	by ignoring the Deuterium's nuclear corrections, with the result shown in Fig.~\ref{fig:HERMES} lower panel. We remind that $\sigma^{d}_{T+L}$ is defined as a cross section per nucleon.
	The uncertainties induced by this low-$Q^2$ HERMES SFs are propagated from cross section quantified by 
	\begin{eqnarray}\label{eq:neutronunc}
		\delta\sigma^n_{T+L}=\sqrt{(2\sigma^d_{T+L})^2+(\sigma^p_{T+L})^2}.
	\end{eqnarray}
	
	The difference of $R_{L/T}$ ratio between Deuterium and proton $\Delta R=R^{d}-R^p$ is measured by many experiments, such as SLAC~\cite{Whitlow:1990gk,E140X:1995ims}, and NMC~\cite{NewMuon:1992olp,NewMuon:1996gam,NewMuon:1996uwk}.
	It is found that $\Delta R$  is compatible with zero within its uncertainty with size $\calO(\lesssim10\%)$, through a wide $x$ range, such as $0.002\sim0.4$~\cite{NewMuon:1996uwk} and even up to $x<0.8$~\cite{Whitlow:1990gk} .
	In this work, we take $R_{L/T}^{d}=R_{L/T}^{p}=R_{L/T}^{n}$, with the uncertainty captured by the variation of R1998~\cite{E143:1998nvx} by $\pm50\%$.
	
	\textbf{$W^2$ transition.} Following the LUXqed treatment~\cite{Manohar:2016nzj,Manohar:2017eqh}, the structure functions in the gap between the resonance and low-$Q^2$ continuum regions ($W^2_{\rm lo}<W^2<W^2_{\rm hi}$) are dealt with a smooth transition that
	\begin{equation}\label{eq:CH}
		F_{a}(x,Q^2)=(1-\rho)F_{a}^{\rm resonance}+\rho F_a^{\rm HERMES},
	\end{equation}
	where $a=2,L$ and
	\begin{equation}
		\rho=2\omega^2-\omega^4, ~
		\omega=\frac{W^2-W^2_{\rm lo}}{W^2_{\rm hi}-W^2_{\rm lo}}.
	\end{equation}
	
	\textbf{High-$Q^2$ continuum.} $Q^2>Q_{\PDF}^2=9~\GeV^2$ and $W^2>W^2_{\rm hi}=4~\GeV^2$ is dubbed as the high-$Q^2$ continuum region.
	In this region, the perturbative quantum chromodynamics (pQCD) is well established, which we can rely on to calculate the corresponding structure functions up to a high order~\cite{Guzzi:2011ew}.
	Based on the CT18lux methodology~\cite{Xie:2021equ}, we need to know the quark and gluon PDFs at the scale when $\mu>\mu_0$.
	In this work, we adopt the isospin symmetric approximation (ISA) to relate the proton and neutron quark PDFs,
	\begin{equation}\label{eq:IS}
		\overset{(-)}{d}_n=\overset{(-)}{u}_p,~\overset{(-)}{u}_n=\overset{(-)}{d}_p,
		~i_{n}=i_{p}~(i=g,\overset{(-)}{s},\overset{(-)}{c},\cdots).
	\end{equation}
	Due to the QED evolution effect, a small isospin violation will be generated through the $\gamma\to q\bar{q}$ splitting~\cite{Martin:2004dh,Ball:2013hta,Schmidt:2015zda}, which will be properly addressed later with the CT18qed approach~\cite{Xie:2021equ}. 
	Similar to the LUXqed treatment~\cite{Manohar:2017eqh}, a variation as $Q_{\PDF}^2=5~\GeV^2$ is introduced to explore the corresponding PDF matching uncertainty.
	
\begin{figure}[h]
		\centering
		\includegraphics[width=0.475\textwidth]{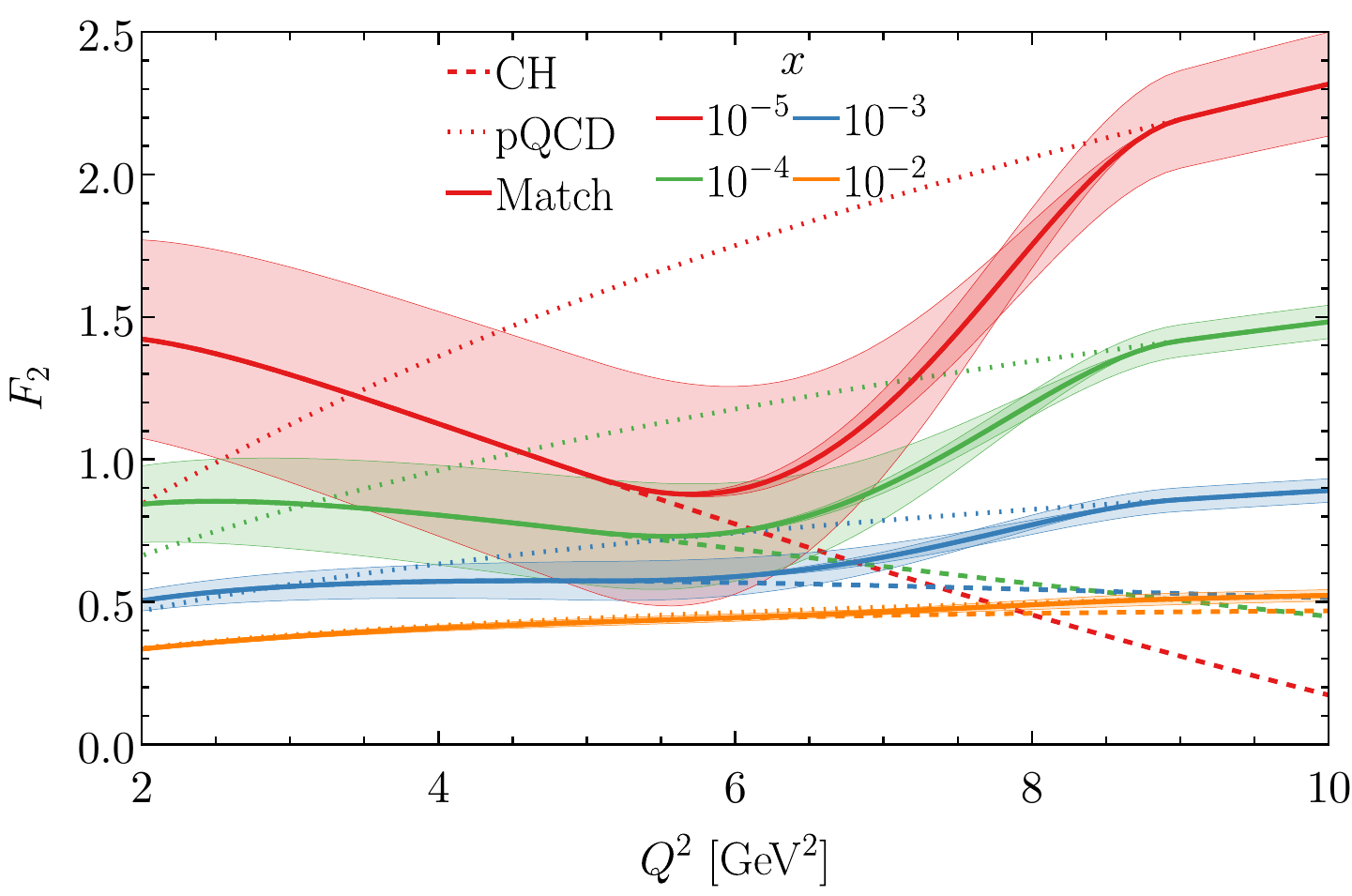}
		\includegraphics[width=0.515\textwidth]{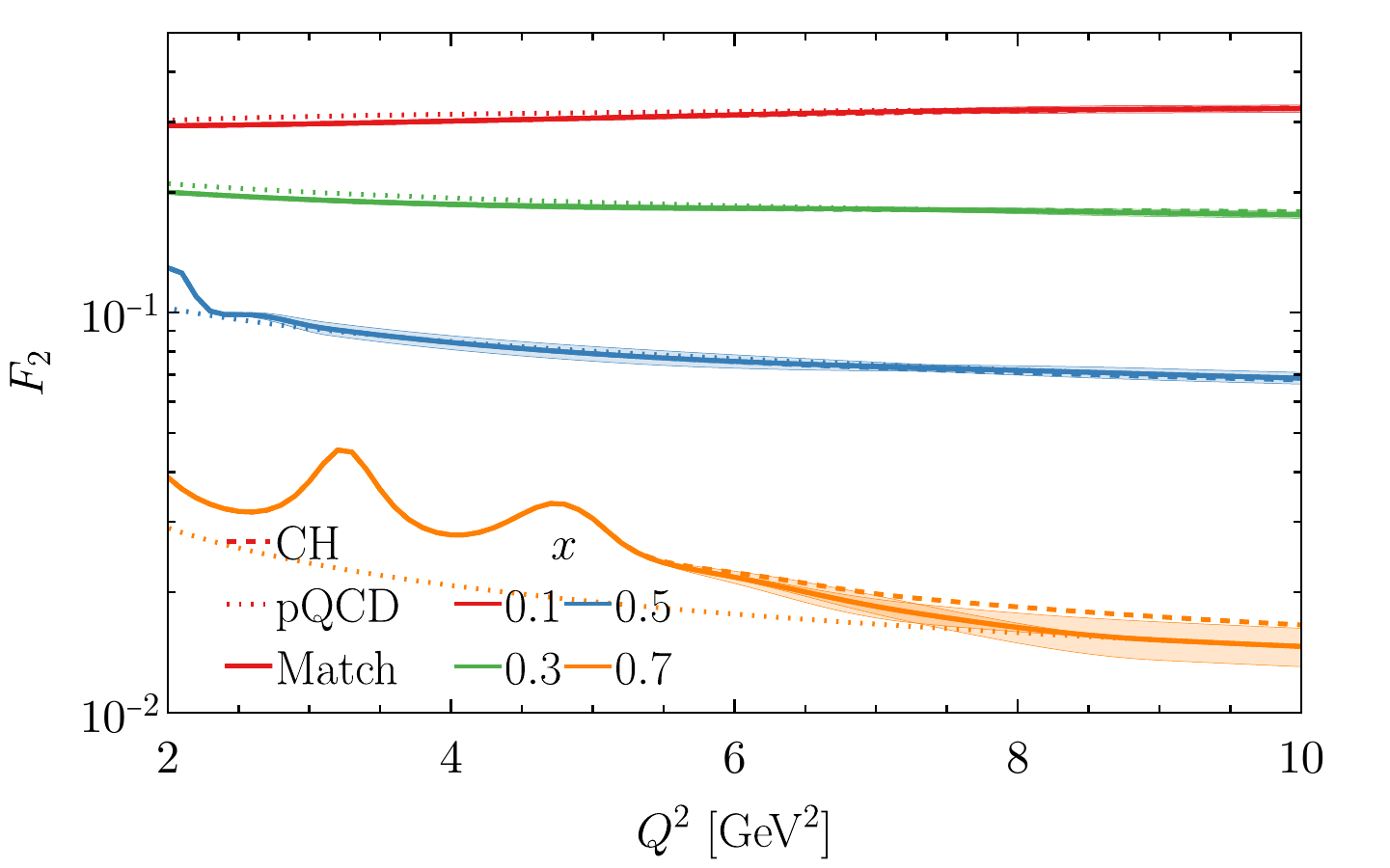}
		\includegraphics[width=0.475\textwidth]{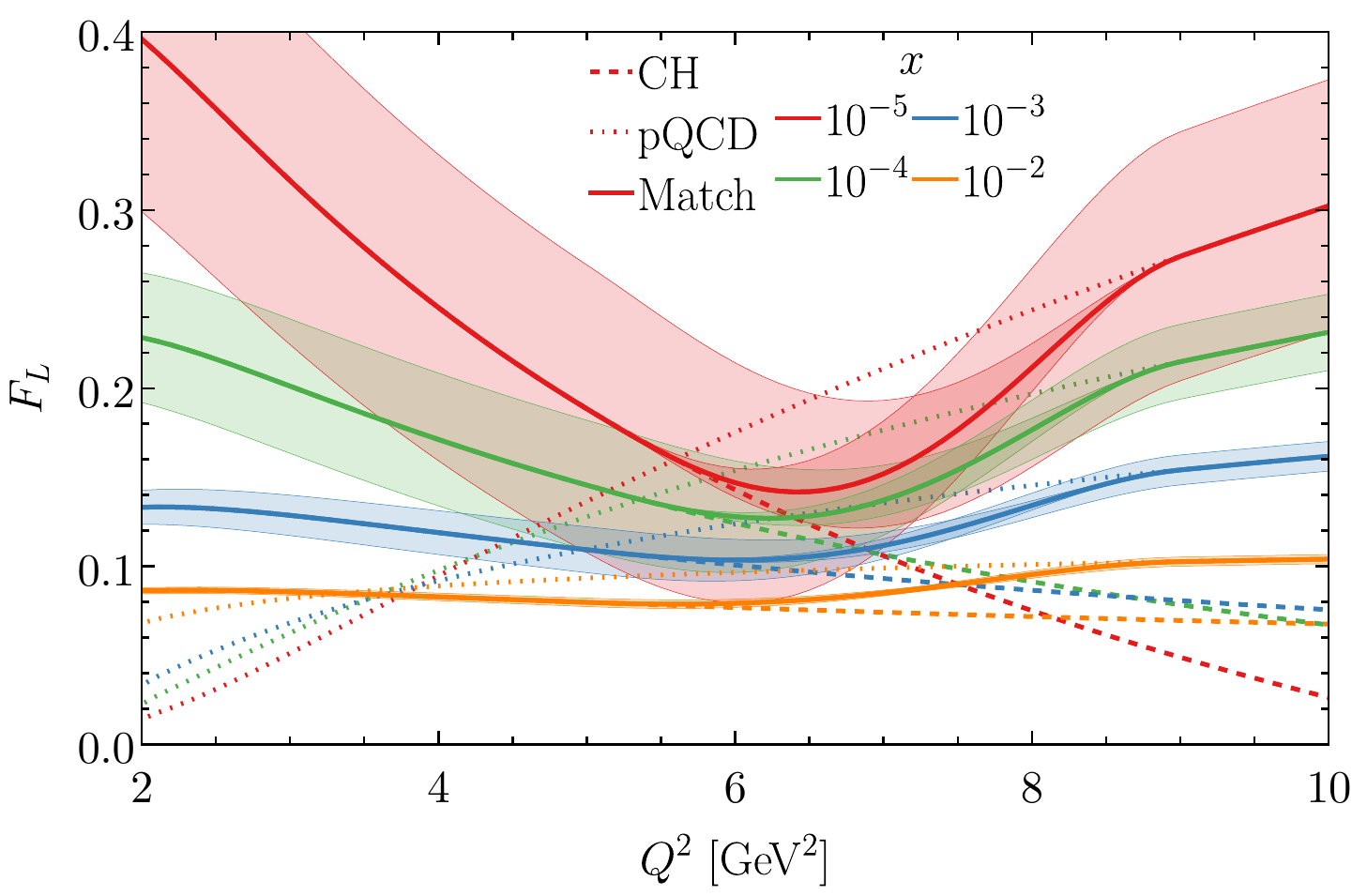}
		\includegraphics[width=0.515\textwidth]{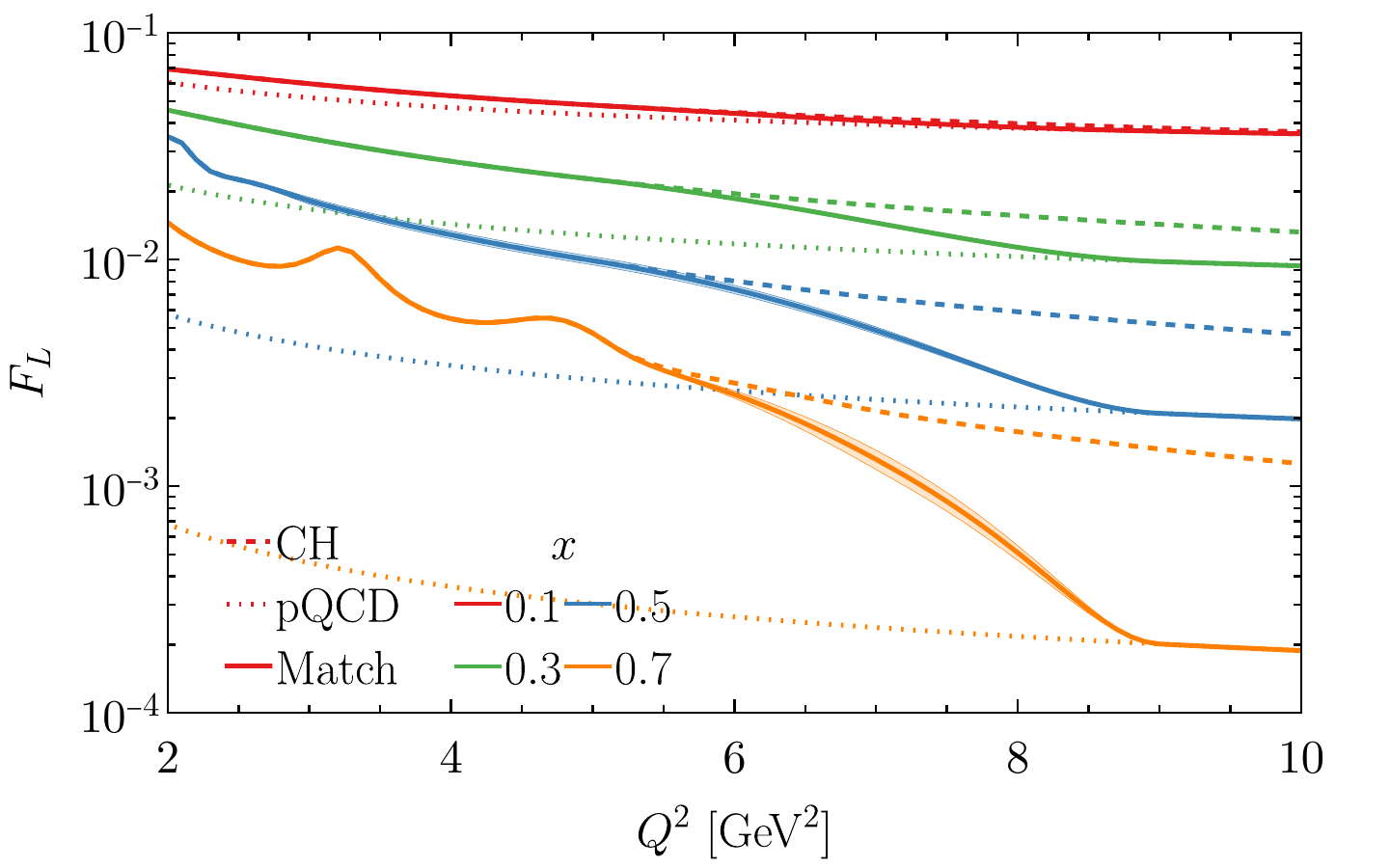}
		\caption{The neutron's structure functions ($F_2,~F_L$) in the smooth transition region $Q^2_{\rm lo}<Q^2<Q^2_{\rm PDF}$ from the low-$Q^2$ HERMES to high-$Q^2$ pQCD continuum. The error bands opening to the low-$Q^2$ direction are taken from the HERMES data with Eq.~(\ref{eq:neutronunc}), while the high-$Q^2$ bands indicate the pQCD PDF uncertainty.}
		\label{fig:match}
	\end{figure}

	\textbf{$Q^2$ transition}. Similar to the low-$W^2$ transition region, we also smoothly transit the SFs from the low-$Q^2$ HERMES to high-$Q^2$ pQCD continuum region when $Q^2_{\rm lo}<Q^2<Q^2_{\rm PDF}$, with $Q^2_{\rm lo}=5~\GeV^2$.
	For $Q^2_{\rm PDF}=5~\GeV^2$ case, we choose $Q^2_{\rm lo}=4~\GeV^2$.
	The smooth structure functions are constructed similarly as
	\begin{equation}
		F_{a}(x,Q^2)=(1-\rho)F_{a}^{\rm HERMES}+\rho F_a^{\rm pQCD},
	\end{equation}
	where
	\begin{equation}
		\rho=2\omega^2-\omega^4, ~
		\omega=\frac{Q^2-Q^2_{\rm lo}}{Q^2_{\rm PDF}-Q^2_{\rm lo}}.
	\end{equation}
	The structure functions $F_2,~F_L$ in this transition region with a few representative Bjorken-$x$'s are shown in Fig.~\ref{fig:match}.
	Different from the low-$W^2$ transition as the ``CH" shows in Fig.~\ref{fig:SFres}, we see sizable mismatches and error bands from the low-$Q^2$ HERMES to the high-$Q^2$ pQCD continuum region, especially at a small $x$. It mainly comes from the low-$x$ extrapolation of the HERMES fit, reflecting the lack of data in this region. 
	
	Here in Fig.~\ref{fig:match}, we also include the SF uncertainty at high $Q^2$ induced by the pQCD calculation. We see that in the moderate-$x$ region, such as $10^{-3}\lesssim x\lesssim0.5$, the HERMES results agree with the pQCD ones well in the matching region. However, with $x$ decreasing into the extrapolation region, $x\lesssim10^{-4}$, we see a clear deviation, as the HERMES GD11-D fit only parameterize the deuteron SFs $F_2^d$ in the region $0.89\times10^{-3}<x<0.9$~\cite{HERMES:2011yno}.
	It is questionable which prediction is more reliable in this small-$x$ and small-$Q^2$ region (denoted as ``sxQ2" later) between the HERMES extrapolation and the pQCD calculation. In this work, we take the HERMES fits as default to incorporate low-$Q^2$ SF data. Meanwhile, we also take the pQCD SFs as an alternative choice in this region ($x\lesssim10^{-4}$), with the difference as an additional quantification of the corresponding uncertainty. 
	
	\begin{figure}[h]
		\centering
		\includegraphics[width=0.515\textwidth]{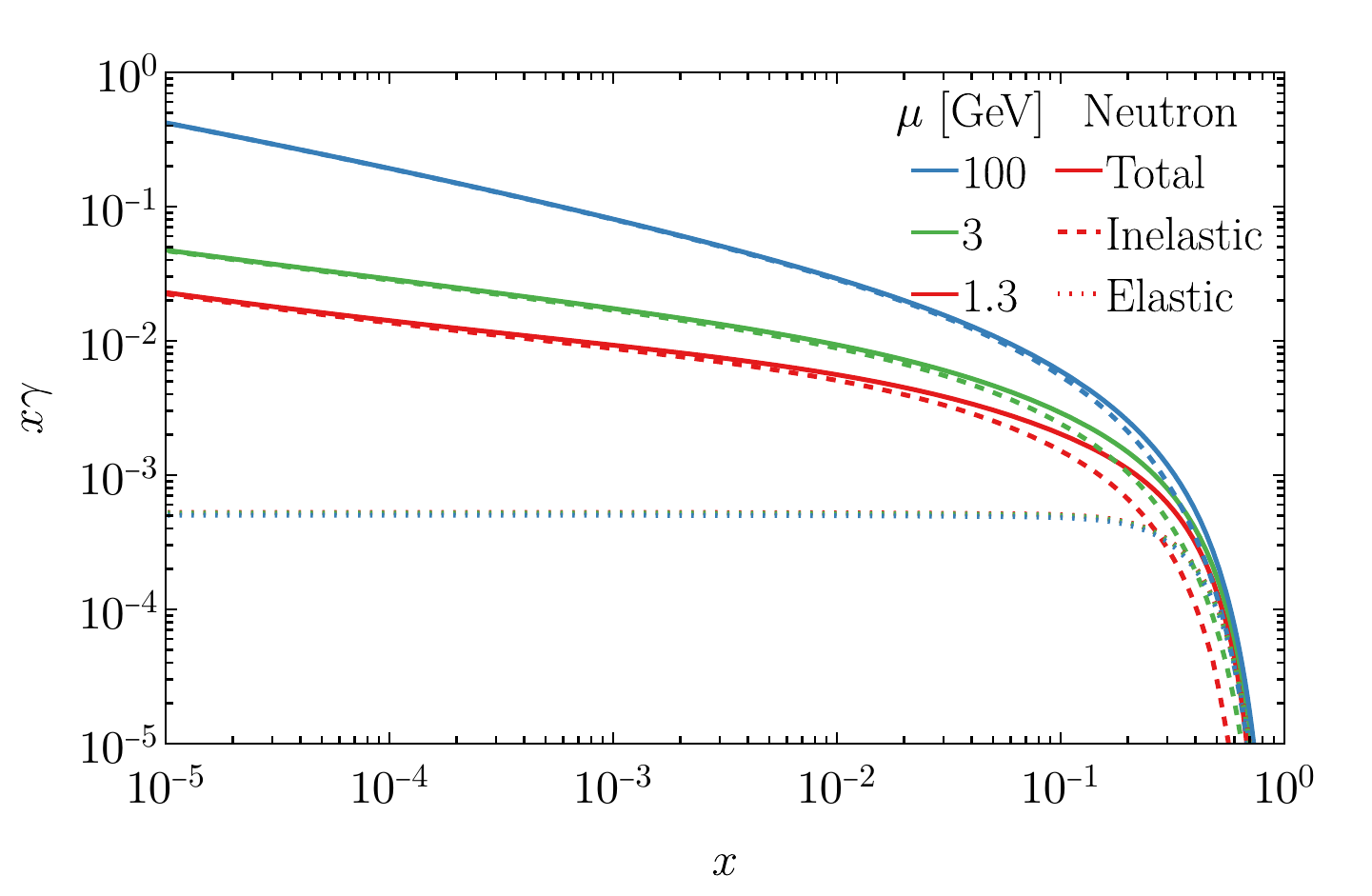}    
		\includegraphics[width=0.475\textwidth]{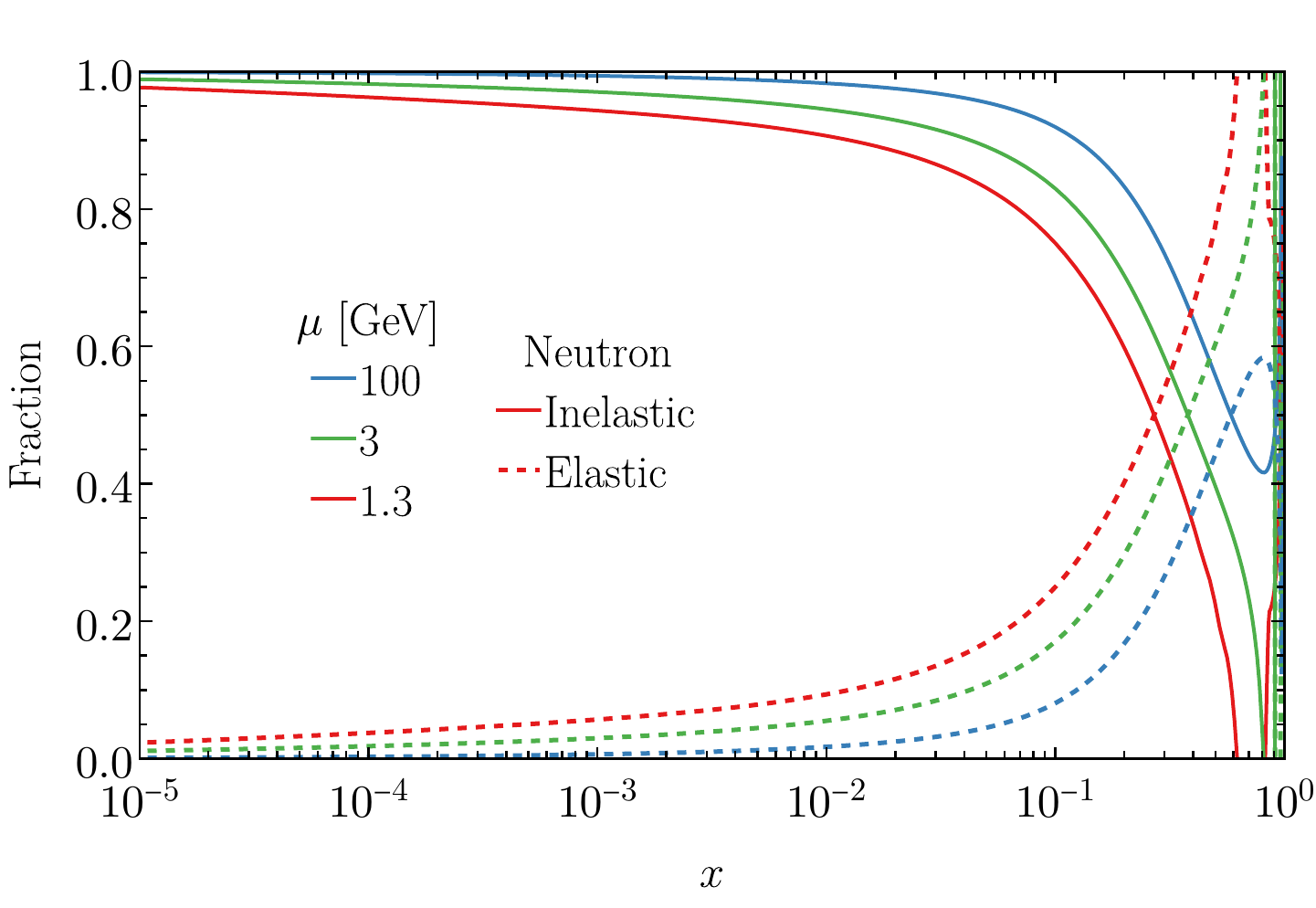}
		\caption{Left: neutron's photon PDF at $\mu=1.3,3,100~\GeV$.
			Right: the fraction of elastic and inelastic components of the neutron's photon PDF. }
		\label{fig:el2inel}
	\end{figure}
	At this stage, the inelastic photon PDF can be directly calculated with the LUXqed formalism, \emph{i.e.}, the CT18lux approach.
	In Fig.~\ref{fig:el2inel}, we show the inelastic as well as the total photon PDFs in comparison with the elastic components at a few representative scales, $\mu = 1.3, 3$, and 100 GeV.
	Similar to the proton case, we see the inelastic photon grows with scale very fast, especially in the small-$x$ region. In the right panel of Fig.~\ref{fig:el2inel}, we display the fractions of elastic and inelastic components. We see that due to the neutral electric charge, the elastic photon only contributes a small fraction to the total photon PDF within $x\lesssim\calO(\textrm{a few}\cdot10^{-1})$, which differs from the proton case where the elastic photon takes over at a low scale when $\mu\lesssim10~\GeV$.\footnote{See Fig. 27 of Ref.~\cite{Xie:2021equ} for details.} (For this reason, the CT14qed took a zero elastic component for neutron's photon PDF~\cite{Schmidt:2015zda}.)
	Only at a very large $x$, the neutron's elastic photon can play an important role, mostly induced by the neutron's magnetic form factor.
	
	We also notice a numerical subtlety here. As shown in Fig.~\ref{fig:el2inel} (right), we see that the inelastic component of the neutron's photon at $\mu=1.3~\GeV$ becomes zero around $x\sim0.6$, and becomes positive again when $x\gtrsim0.8$. It is resulted from the subtracted structure function $F_2(x/z,\mu^2)$ in Eq.~(\ref{eq:LUXqed})  falls into the non-perturbative resonance region, which exceeds the $Q^2$ integration. However, considering the numerical absolute smallness ($\lesssim10^{-6}$) and phenomenological irrelevance of the photon PDF at such a large $x$ value, we just enforce it as zero in this region. The specific $x$ range depends on the scale $\mu$ as well as other low-$Q^2$ SF treatments as well.
	
	In the high-$Q^2$ continuum region as shown in Fig.~\ref{fig:breakup}, the theoretical uncertainties for the pQCD structure functions propagate into the inelastic photon, including the higher-twist (HT) as well as target-mass (TM) corrections.
	
	The HT correction is explained as follows.
	In terms of the spirit of operator product expansion~\cite{Wilson:1969zs}, the DIS hadronic tensor can be expanded with coefficient functions together with local operator matrix elements,
	\begin{eqnarray}
		W_{\mu\nu}(q^2,\nu)
		&=\frac{1}{\pi}\Im T_{\mu\nu}(q^2,\nu)
		=\frac{1}{\pi}\Im\int\dd^4z e^{-iq\cdot z}\langle P|iT[J_{\nu}^\dagger(z)J_\mu(0)]|P\rangle\\
		&=\frac{1}{\pi}\Im\sum_{i,\tau,n}
		C^{i,\mu_1\cdots\mu_n}_{\tau,\mu\nu}(q)
		\langle P|\calO^{\tau}_{i,\mu_1\cdots\mu_n}|P\rangle,
	\end{eqnarray}
	where $\nu=q\cdot p/m=E-E'$ is the lepton's energy loss and $q^2=-Q^2$ is the photon virtuality.
	The fermionic operators can be written as
	\begin{eqnarray}
		\calO_{i,\mu_1\cdots\mu_n}^{\tau}=\frac{1}{2}\frac{i^{n-1}}{n!}
		\left\{\bar{q}(x)\gamma_{\mu_1}D_{\mu_2}\cdots D_{\mu_n}q(x)
		+\textrm{permutations}\right\}.
	\end{eqnarray}
	while the scalar and vector operators can be defined similarly~\cite{Cheng:1984vwu}. 
	The operator twist $\tau=d-j$ is defined as the difference between the mass dimension $d$ and the corresponding spin $j$~\cite{Gross:1971wn}.
	In the large virtuality limit $Q^2\to\infty$, the forward amplitude $T_{\mu\nu}(q^2,\nu)$ can be approximated as
	\begin{equation}
		T_{\mu\nu}(q^2,\nu)
		\approx \sum_{i,j}x^{-j}\tilde{C}_{i,\mu\nu}^{(j)}(Q^2)\calO_{i}^{(j)},
	\end{equation}
	where $x=Q^2/(2m\nu)$ is the Bjorken scaling.
	The Wilson coefficients scale as
	\begin{eqnarray}
		\tilde{C}_{i,\mu\nu}^{(j)}(Q^2)\sim\left(1/Q^2\right)^{\gamma_j/2},
	\end{eqnarray}
	where $\gamma_j$ is the anomalous dimension of operator $\calO_{i}^{(j)}$, which increase monotonically with $j$.
	In perturbative QCD, the leading-twist operators correspond to $\tau=2$ with $\gamma_2=0$ which are defined as parton distribution functions, while higher twists starting from $\tau=4$ serve as power corrections $\calO(\Lambda^2/Q^2)$.
	
\begin{figure}[h]
		\centering
		\includegraphics[width=0.49\textwidth]{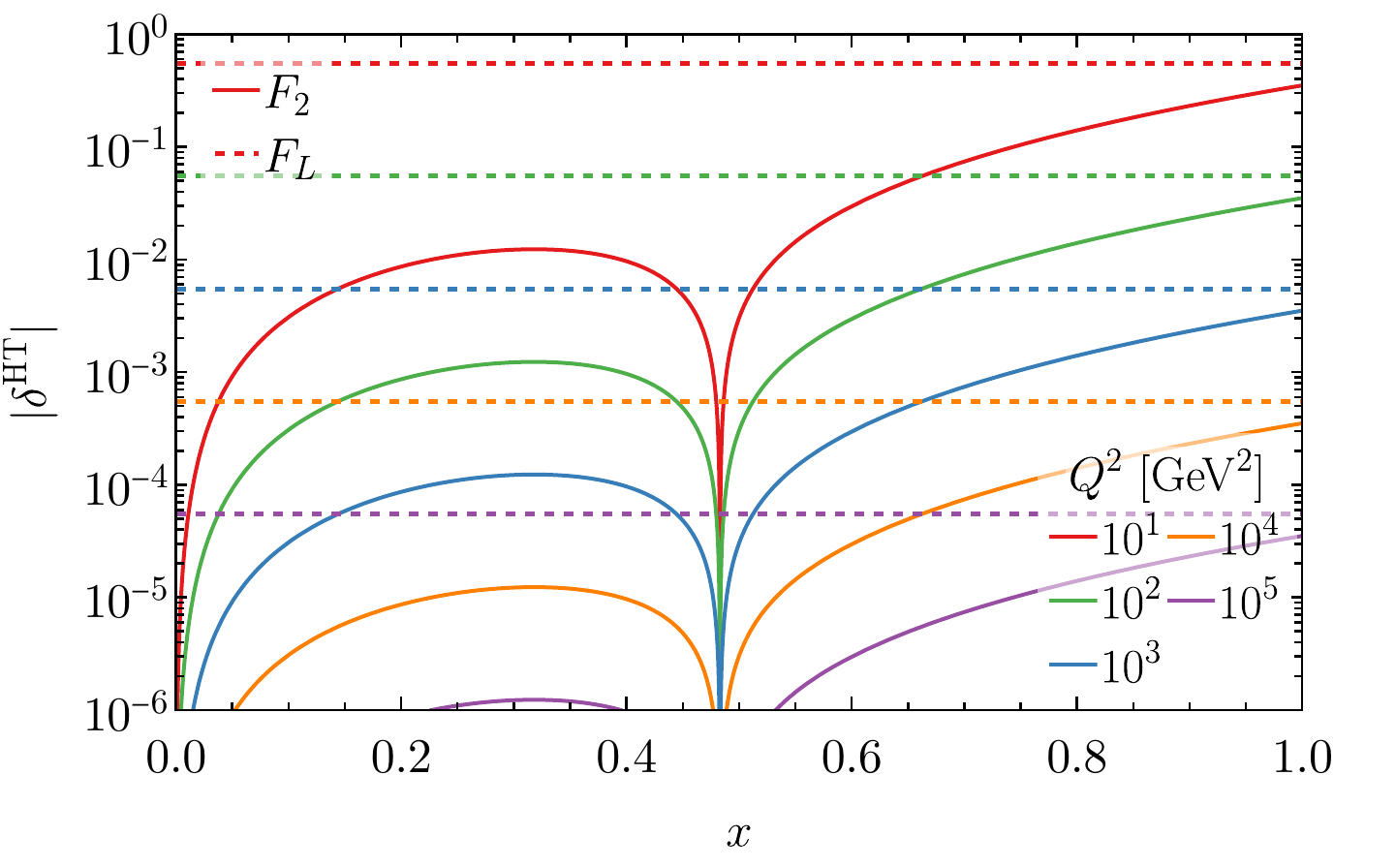}    
		\includegraphics[width=0.49\textwidth]{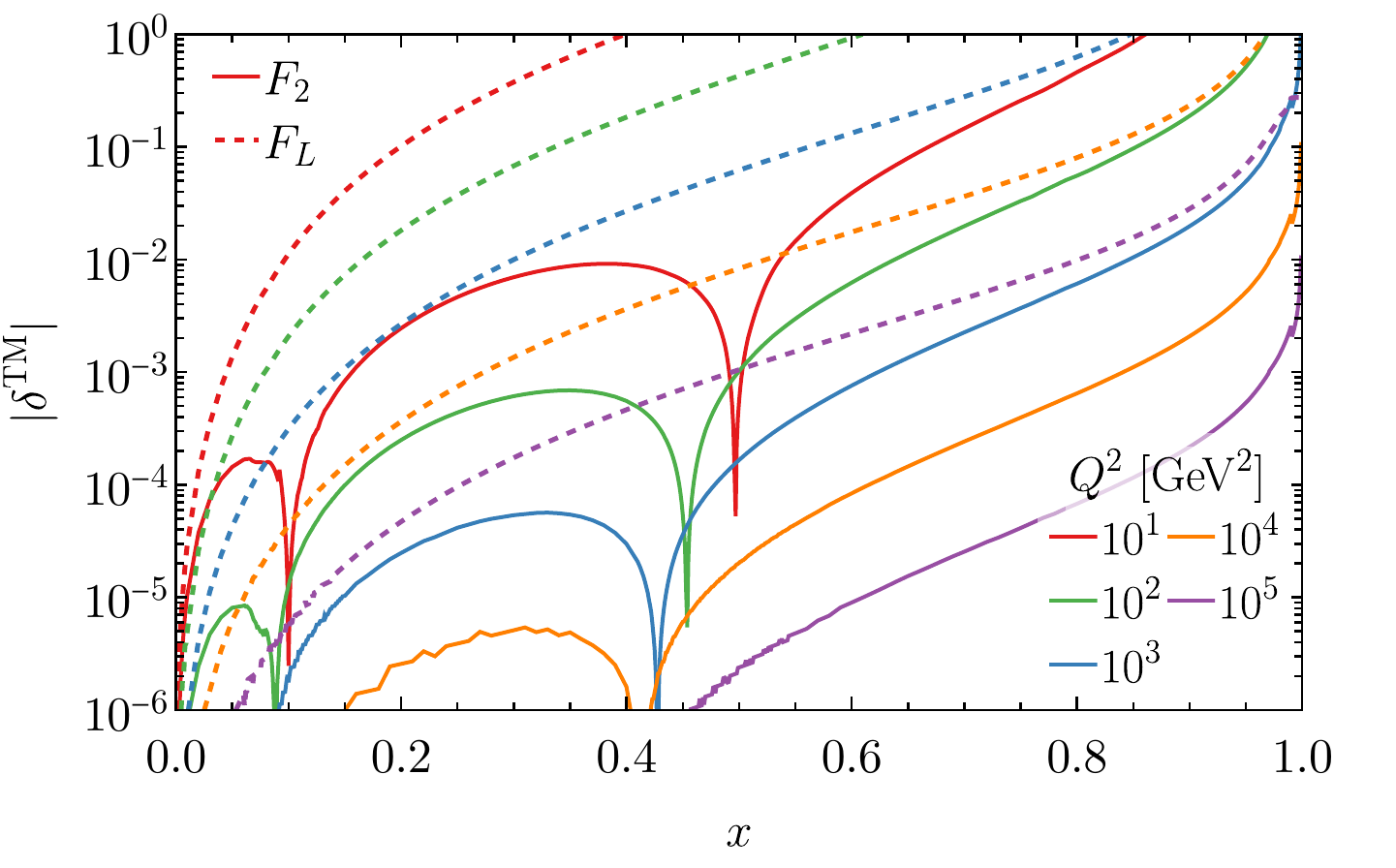}
		\caption{The higher-twist (left) and target-mass (right) corrections to the structure functions $F_{2,L}(x,Q^2)$ for a few representative $Q^2$ in the high-$Q^2$ continuum region.}
		\label{fig:HT}
	\end{figure}

	In this work, we take the same treatment as the CT18qed study~\cite{Xie:2021equ}. The HT corrections to the $F_2$ are taken from the CJ15 NLO fitting~\cite{Accardi:2016qay},
	\begin{equation}
		F_2^{\rm HT}(x,Q^2)=F_2^{\rm LT}(x,Q^2)\left(1+\frac{C_{\rm HT}(x)}{Q^2}\right),
	\end{equation}
	where $C_{\rm HT}=h_0x^{h_1}(1+h_2x)~\GeV^2$ with parameters $h_{0,1,2}$ fitted from data. 
	The higher twist contribution to the longitudinal SF is taken as
	\begin{equation}
		F_L^{\rm HT}(x,Q^2)=F_L^{\rm HT}(x,Q^2)\left(1+\frac{A_{\HT}}{Q^2}\right),
	\end{equation}
	where $A_{\rm HT}=5.5\pm0.6~\GeV^2$ from the HERA fit~\cite{Abt:2016vjh}. The corresponding HT corrections to SFs, defined as 
	\begin{equation}
		\delta_{2,L}^{\rm HT}(x,Q^2)=\frac{F_{2,L}^{\rm HT}(x,Q^2)}{F_{2,L}^{\rm LT}(x,Q^2)}-1,
	\end{equation}
	are shown in Fig.~\ref{fig:HT} (left).
	Also, we see the HT generally gives larger corrections to $F_L$ than $F_2$, mainly resulting from the smaller magnitude of $F_L$ with respect to that of $F_2$. 
	
	Besides the higher-twist corrections, the non-zero target mass (TM) of the scattered nucleus also introduces kinematic power corrections $\calO(x^2m^2/Q^2)$~\cite{Miramontes:1988fz}, which was first calculated by Georgi and Politzer~\cite{Georgi:1976ve}.
	With the Nachtmann variable~\cite{Nachtmann:1973mr}
	\begin{eqnarray}
		\xi=\frac{2x}{1+r}, ~
		\textrm{where}~r=\sqrt{1+4x^2m^2/Q^2},
	\end{eqnarray}
	we have the TM corrected SFs as~\cite{Schienbein:2007gr,Goharipour:2020gsw}
	\begin{eqnarray}
		&F_2^{\rm TM}(x,Q^2)=\frac{(1+r)^2}{4r^3}F_2^{(0)}(\xi,Q^2)+\frac{3x(r^2-1)}{2r^4}I_2(\xi,Q^2),\\
		&F_L^{\rm TM}(x,Q^2)=\frac{(1+r)^2}{4r}F_L^{(0)}(\xi,Q^2)+\frac{x(r^2-1)}{r^2}I_2(\xi,Q^2),
	\end{eqnarray}
	where
	\begin{eqnarray}
		I_2(\xi,Q^2)=\int_{\xi}^{1}\dd u\left[1+\frac{r^2-1}{2xr}(u-\xi)\right]\frac{F_2^{(0)}(u,Q^2)}{u^2},
	\end{eqnarray}
	and $F_{2,L}^{(0)}$ are the SFs in the $m^2/Q^2\to0$ limit.
	The TM corrected SFs normalized to the uncorrected ones as 
	\begin{eqnarray}
		\delta_{2,L}^{\rm TM}=\frac{F_{2,L}^{\rm TM}(x,Q^2)}{F_{2,L}^{(0)}(x,Q^2)},
	\end{eqnarray}
	are shown in Fig.~\ref{fig:HT} (right).
	Similar to the HT case, we see the TM corrections only become significant at large $x$ and small $Q^2$.
	
	\begin{figure}[h]
		\centering
		\includegraphics[width=0.49\textwidth]{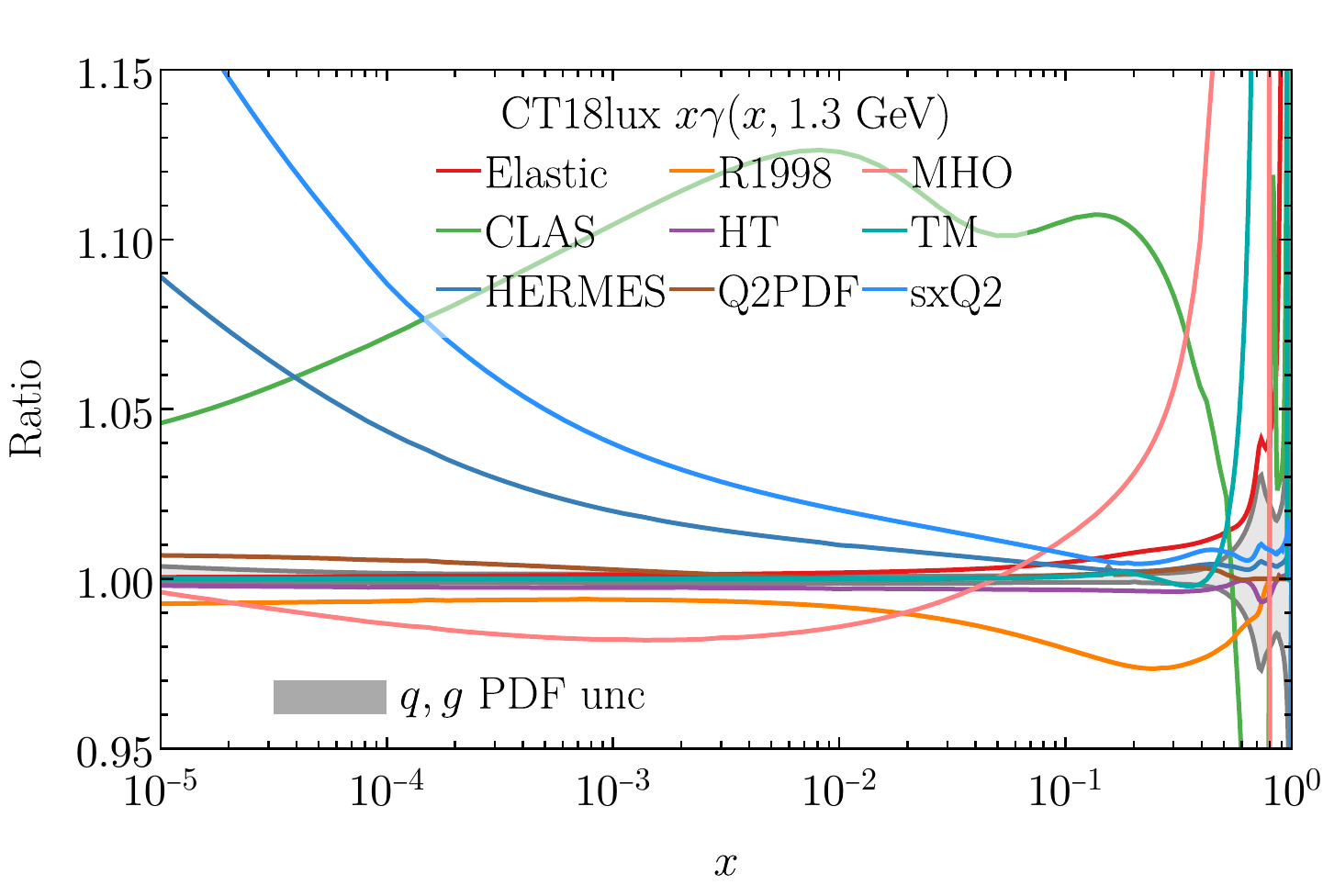}
		\includegraphics[width=0.49\textwidth]{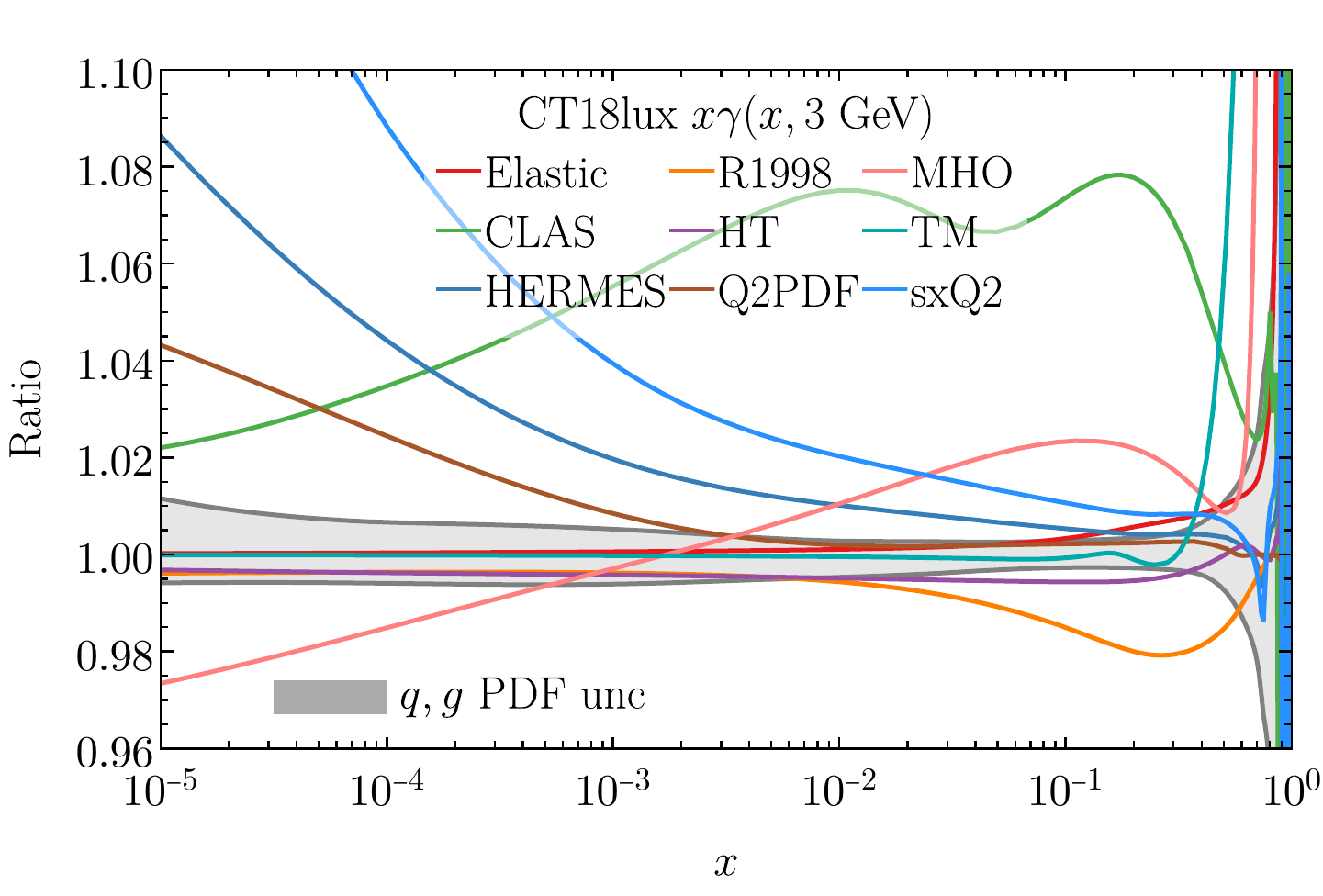}
		\includegraphics[width=0.49\textwidth]{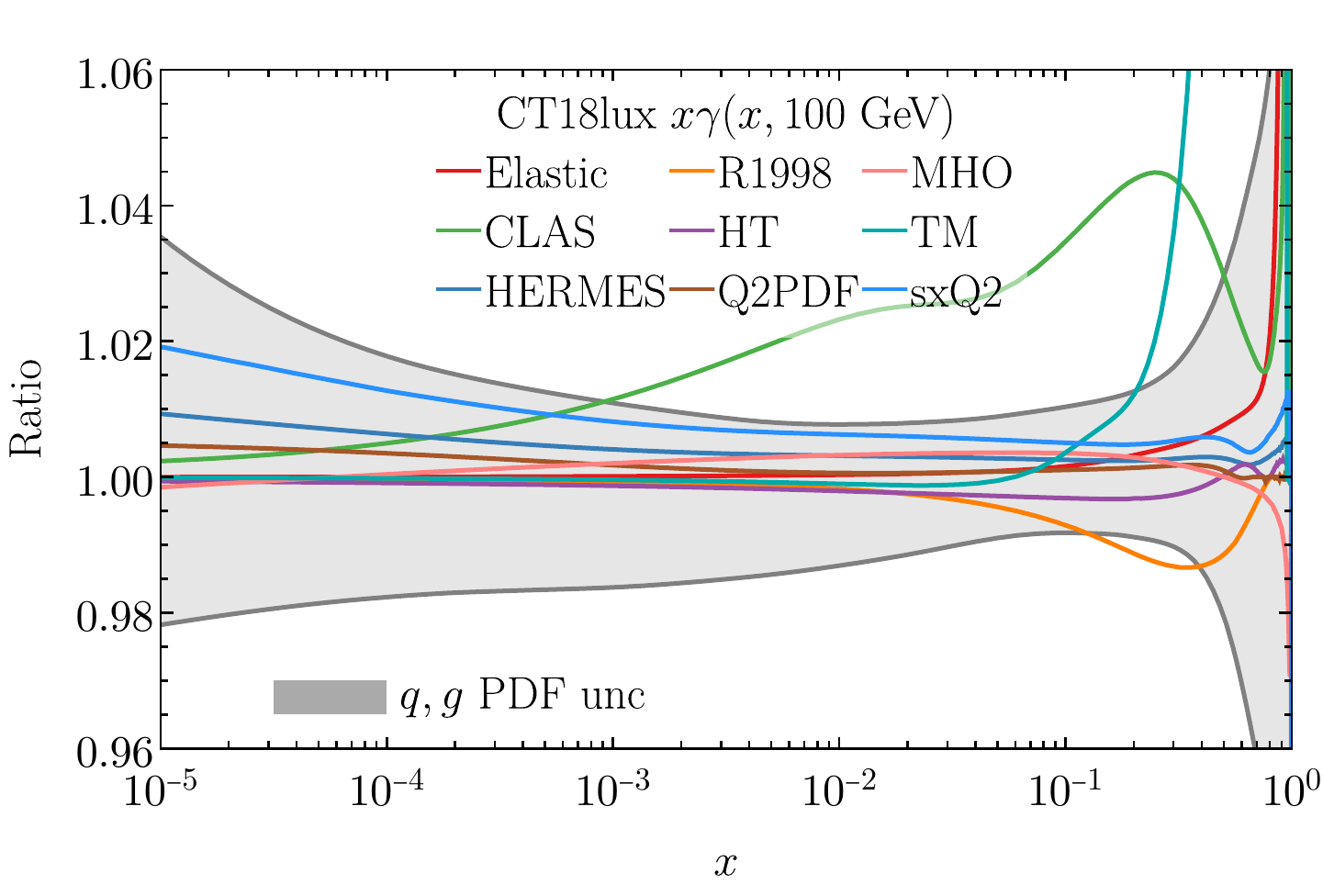}
		\caption{Various CT18lux uncertainties to the neutron's photon PDF at $\mu=1.3,3,100~\GeV$.}
		\label{fig:luxunc}
	\end{figure}
	
	In Fig.~\ref{fig:luxunc}, we show various resources contributing to the uncertainties of neutron's photon PDF (summing over elastic and inelastic components), based on the CT18lux methodology. Most of them are in parallel with the proton case studied in Ref.~\cite{Xie:2021equ}. We discuss them in sequence as follows.
	\begin{itemize}
		\item \textbf{Elastic.} The elastic photon variation is deduced from the uncertainty of electromagnetic form factors obtained in the Ye \emph{et al.}'s fit~\cite{Ye:2017gyb}.
		Due to a small contribution to the total photon, we see the elastic uncertainty is small in most of the $x$ range, which is only noticeable up to a percent level at a very large momentum fraction $x\gtrsim0.5$.
		
		\item \textbf{CB21.} The ``CB21" denotes the variation from our default CLAS resonance structure functions~\cite{CLAS:2003iiq} to the CB21~\cite{Christy:2021abc} fit as shown in Fig.~\ref{fig:SFres}. Different from the proton case, we see that the CB21 resonance dominates the neutron's photon uncertainty, suggesting the imprecise determination of the neutron structure functions in the resonance region. 
		We have made sure that the CB07~\cite{Christy:2007ve,Bosted:2007xd} fit gives a similar size of variation, as shown in Fig.~\ref{fig:CB2CLAS}.
		Indeed, the neutron's SFs are not directly measured from experiments, which are instead extracted from the scatterings of Deuterium/Helium and proton targets in this region, based on some assumptions of the nuclear corrections. See Ref.~\cite{Bosted:2007xd} for an example. For this reason, the neutron's SFs are not as well determined as the proton's, which introduces larger uncertainty on the inelastic photon PDF. In addition, we see that with an increment of the scale $\mu$, the CB21/07 resonance uncertainty decreases, because of the increasing contribution from the pQCD continuum region.
		\begin{figure}
			\centering
			\includegraphics[width=0.49\textwidth]{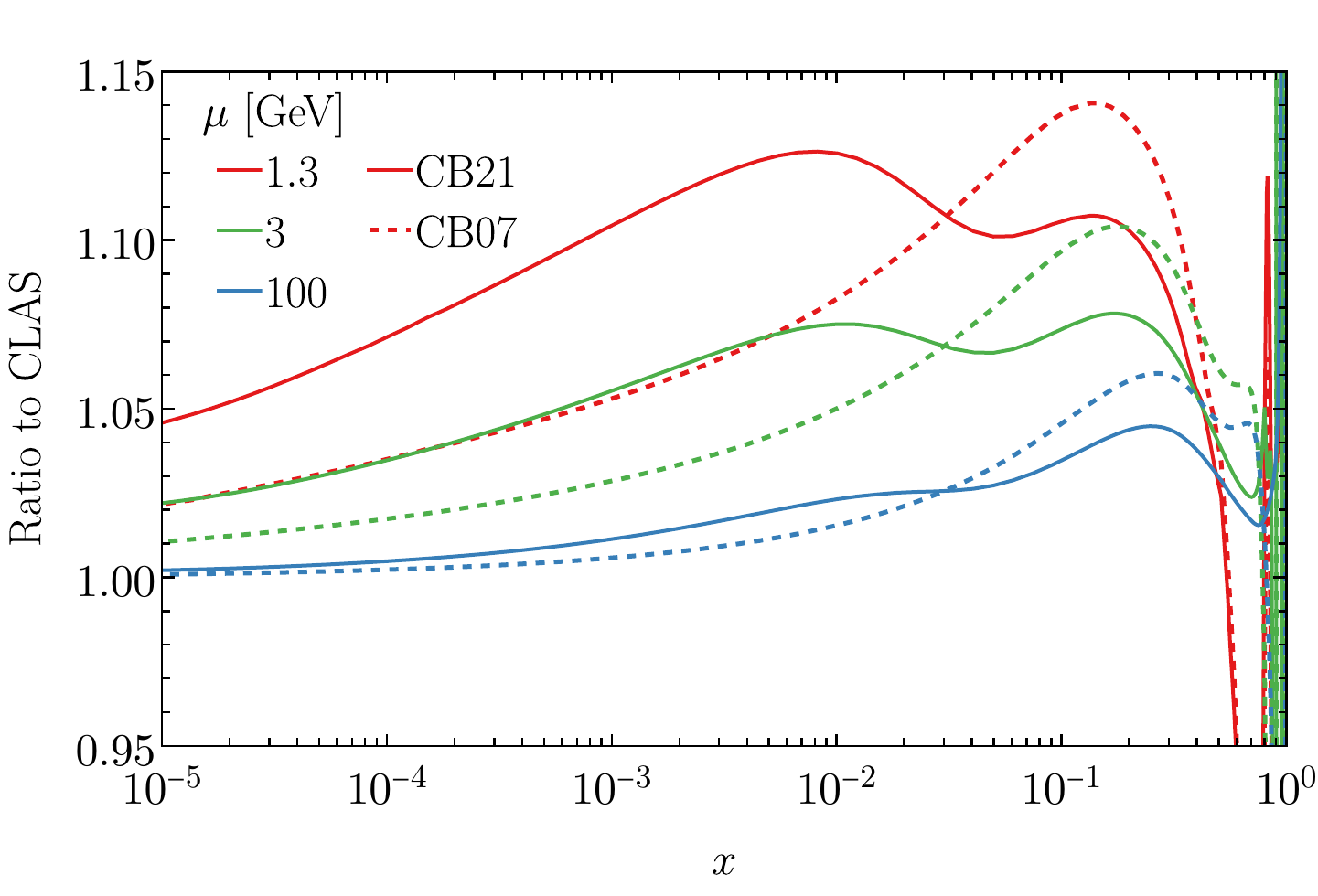}
			\caption{The variation of the CT18lux neutron's photon PDF at $\mu=1.3,3,100~\GeV$ with the resonance structure functions of CB07~\cite{Christy:2007ve,Bosted:2007xd}, CB21~\cite{Christy:2021abc} and CLAS~\cite{CLAS:2003iiq} fits.}
			\label{fig:CB2CLAS}
		\end{figure}
		\item \textbf{HERMES.} The neutron's SFs in the low-$Q^2$ HERMES region are constructed in terms of Eq.~(\ref{eq:neutron}). In this work, the nuclear corrections are neglected in the central relation, while the uncertainty is conservatively propagated with the squared sum according to Eq.~(\ref{eq:neutronunc}).
		We see that the HERMES uncertainty can induce up to an 8\% variation for the neutron's photon in the small-$x$ and small-$\mu$ region. The dying out HERMES' impact in the large $x$ direction is due to the better matching of SFs in the transition region as shown in Fig.~\ref{fig:match}. 
		With increasing scale $\mu$, the HERMES-induced uncertainty gets the most pronounced around $\mu=3~\GeV$.
		When $\mu>3~\GeV$, this impact decreases due to a smaller contribution from the HERMES region, similar to the CB21 resonance impact.
		
		\item \textbf{R1998.} As described above, we have assigned a large variation to the longitudinal-to-transverse cross-section ratio as $R_{L/T}=R_{1998}(1\pm50\%)$, which introduces the corresponding inelastic photon PDF uncertainty. This variation is as small as about $1\%\sim2\%$ relevant in the large $x$ region when $x>0.1$, due to the small contribution from the longitudinal SF $F_L$ in the HERMES region.
		
		\item \textbf{sxQ2.} As shown in Fig.~\ref{fig:match}, the low-$Q^2$ SFs of the HERMES fit deviates from the pQCD ones in the extrapolated small-$x$ region, $x\lesssim10^{-4}$. In this work, we take the HERMES fit as our default choice, while the pQCD SFs as an alternative choice gives an additional error set to quantify the SF variation in this small-$x$ and small-$Q^2$ region (shortened as ``sxQ2"). 
		In terms of Fig.~\ref{fig:luxunc}, the ``sxQ2" SF variation gives a more than 10\% variation for the photon PDF at a low $x$ ($x\lesssim10^{-4}$) and a low scale ($\mu=1\sim3~\GeV$), which is the largest error source in this region. However, with the increasing of the scale, \emph{e.g.}, at $\mu=100~\GeV$, the uncertainty from the sxQ2 SF variation falls quickly into the error band from other resources, as a result of the increasing contribution from the high-$Q^2$ pQCD region. 
		
		\item \textbf{HT.} The higher-twist corrections are described above for $F_{2,L}^{\rm HT}$. The corresponding uncertainty is negligible in most scenarios, as a result of the $1/Q^2$ suppression in the pQCD region.
		
		\item \textbf{Q2PDF.} We take $Q^{2}_{\rm PDF}$ (flatten as ``Q2PDF") to denote the matching scale between the low-$Q^2$ continuum HERMES and high-$Q^2$ pQCD regions. The matching uncertainty is quantified by varying the default $Q^2_{\rm PDF}=9~\GeV^2$ to $5~\GeV^2$. We see in both $\mu=1.3~\GeV$ and 100 GeV cases, the matching uncertainty is under control, while it can introduce a large variation up to 4\% at $x\sim10^{-5}$ when $\mu=3~\GeV$. At $\mu=1.3(100)~\GeV$, the contribution to the photon PDF from the pQCD (low-$Q^2$ HERMES) SFs is small. As a result, the variation of $Q^{2}_{\rm PDF}=9(5)~\GeV^2$ does not make a big difference in both cases. In contrast at $\mu=3~\GeV$, where the photon PDF has a large contribution from the matching region. For this reason, the mismatch between the pQCD and HERMES SFs will be reflected in the photon PDF variation. The larger photon PDF with $Q^2_{\rm PDF}=5~\GeV^2$ in the small $x$ region is due to the larger low-$Q^2$ SFs in the HERMES extrapolation as shown in Fig.~\ref{fig:match}.
		In comparison, the proton's photon uncertainty from $Q^{2}_{\rm PDF}$ is much smaller~\cite{Xie:2021equ},
		resulted from a much better matching for the proton SFs in these two regions.
		\item \textbf{MHO.} The missing higher order (MHO) uncertainty is quantified by varying the upper integration limit $\mu^2/(1-x)$ in Eq.~(\ref{eq:LUXqed}) to be $\mu^2$, with the corresponding $\msbar$ matching term~\cite{Manohar:2017eqh,Xie:2021equ}. We see that the MHO uncertainty only becomes significant in the $x\to1$ limit, as a result of the asymptotic divergence of $\mu^2/(1-x)$. 
		Nevertheless, the MHO only gives $1\%\sim2\%$ uncertainty in the normal $x$ region.
		Compared with $\mu=1.3$ and 100 GeV cases, the MHO variation of $\gamma(x,3~\GeV)$ is slightly larger, due to the amplification of the transition impact in the $Q^2_{\rm lo}<Q^2<Q^2_{\rm PDF}$ region.
		\item \textbf{TM.} In the small-$x$ region, the target-mass corrections give a negligible effect, which only significantly deviates when $x\to1$. In addition, we see the TM effect increases with scale $\mu$, which can even dominate when $x>0.2$ at $\mu=100~\GeV$. A similar behavior occurs in the proton case already but at a much larger momentum fraction around $x>0.6$~\cite{Xie:2021equ}.
		\item \textbf{$q,g$ PDF unc.} In Fig.~\ref{fig:luxunc}, the gray error bands denote the uncertainty induced by the quark/gluon PDF error sets. At $\mu=1.3~\GeV$, we see the $q,g$ induced uncertainty is negligible, suggesting a small contribution from pQCD SFs. In contrast, the $q,g$ induced uncertainty increases with scale, as a result of the increment of pQCD significance. At $\mu=100~\GeV$, the $q,g$ induced uncertainty is about $1\%\sim3\%$, the same level as the proton one~\cite{Xie:2021equ}.
	\end{itemize}
	
	To summarize this subsection, based on the CT18lux approach we present the neutron's photon PDFs (including the elastic and inelastic components) in Fig.~\ref{fig:el2inel}. Various photon PDF uncertainties are shown in Fig.~\ref{fig:luxunc}, with the details discussed above. Many variations share similarities with the proton case, while the impacts of the low-$Q^2$ CLAS/CB21 resonance and HERMES get enhanced, due to the larger uncertainties of neutron SFs.
	The neutron's elastic uncertainty is much smaller than the proton's, thanks to its smaller contribution.
	The HERMES and matching scale $Q_{\rm PDF}^2$ uncertainties get pronounced around $\mu=3~\GeV$, resulting from the mismatch between the low-$Q^2$ HERMES and high-$Q^2$ pQCD continuum regions.
	
	\subsection{The DGLAP evolution approach: CT18qed}
	\label{sec:qed}
	Similar to our previous study, the neutron's photon PDF can be determined through the DGLAP evolution, with the initialization at $\mu_0$ taken from the LUXqed photon, which we dub as the ``CT18qed" methodology~\cite{Xie:2021equ}.
	
	\begin{table}[h]
		\centering
		\begin{tabular}{c|c|c|c }
			\hline
			$\mu_{\min}~[\GeV]$ & 1.3 & 1.3 & 1 \\
			$\mu~[\GeV]$ &  CT18lux & CT18qed & MSHT20qed \\
			\hline
			1 &  -- & --  & $0.042\pm0.009$\\
			1.3 & $0.067\pm0.007$ & $0.067\pm0.007$ & $0.053\pm0.009$\\
			3 &  $0.098\pm0.007$ & $0.097\pm0.007$ & $0.086\pm0.009$\\
			100 & $0.230\pm0.008$ & $0.226\pm0.007$ & $0.215\pm0.009$ \\
			\hline
		\end{tabular}
		\caption{The momentum fraction of neutron's photon $\langle x\gamma\rangle(\mu^2)~[\%]$.}
		\label{tab:mom}
	\end{table}
	
	\textbf{Momentum sum rule.}
	As emphasized in Ref.~\cite{Xie:2021equ}, the additional photon in the CT18lux approach introduces a violation of the momentum sum rule, as
	\begin{eqnarray}
		\langle x(\Sigma+g+\gamma)\rangle(\mu_0^2)>\langle x(\Sigma+g)\rangle(\mu_0^2)=1,
	\end{eqnarray}
	where $\Sigma=\sum_{i}(q_i+\bar{q}_i)$ is the flavor singlet.
	At the initialization scale $\mu_0=1.3~\GeV$, the CT18lux neutron's photon momentum fraction for the elastic and inelastic components are
	\begin{equation}\label{eq:xph0}
		\langle x\gamma^{\rm el}\rangle(\mu_0^2)=(0.0191\pm0.0004)\%,~
		\langle x\gamma^{\rm inel}\rangle(\mu_0^2)=(0.0477\pm0.0071)\%,
	\end{equation}
	with the total listed in Table~\ref{tab:mom}.
	In comparison with the proton case, Eq.~(28) of Ref.~\cite{Xie:2021equ}, the neutron elastic photon is significantly smaller due to its neutral electric charge. The inelastic component is comparable to the proton one, with a slightly smaller size due to its smaller charge-weighted singlet $\Sigma_{e}=\sum_{i}e_i^2(q_i+\bar{q}_i)$,
	where $e_i$ is the charge of quark $q_i$.
	We take out a momentum fraction from the gluon PDF to enforce the momentum sum rule,
	\begin{equation}\label{eq:sum}
		\langle x(\Sigma+g'+\gamma^{\rm inel})\rangle(\mu_0^2)=1.
	\end{equation}
	In such a way, the initial gluon PDF is re-scaled by a factor of
	\begin{equation}
		g'(x,\mu_0^2)=\frac{\langle x(g-\gamma^{\rm inel})\rangle(\mu_0^2)}{\langle xg\rangle(\mu_0^2)}g(x,\mu_0^2).
	\end{equation}
	Note that there exists an ambiguity about whether the momentum sum rule should include the elastic photon. Different from our previous treatment in the proton's QED PDFs~\cite{Xie:2021equ}, we only include the inelastic component in Eq.~(\ref{eq:sum}) based on two arguments.
	First, in the elastic scattering, the nucleon remains intact and therefore the elastic photon does not become an internal parton. Moreover, the DGLAP equations conserve the momentum sum rule in Eq.~(\ref{eq:sum}), which only involves the inelastic photon in evolution.
	
	\textbf{Isospin symmetry violation (ISV).}
	In the DGLAP QCD+QED co-evolution, the quark and gluon PDFs will receive modification from the photon splittings with respect to the pure QCD evolution.
	In the CT18lux calculation of the neutron's photon PDF, we have to rely on the assumption of the isospin symmetry, Eq.~(\ref{eq:IS}), to obtain the quark and gluon PDFs at all scales. However, the isospin symmetry will be violated starting from the $\calO(\alpha)$ order, due to different electric charges of $u/d$ quarks. 
	
	In the NNPDF2.3qed, the initial neutron's quark/gluon PDFs are related to the proton's with the isospin symmetry, while the violation only occurs at a higher scale as a result of the QED evolution~\cite{Ball:2013hta}.
	However, the ISV can emerge at the starting scale $\mu_0$ as well.
	In MRST2004qed, the initial ISV is modeled with the leading-order splitting of the valence quarks since current/constituent masses~\cite{Martin:2004dh}.
	This idea is inherited in the CT14qed PDFs, while a universal low cutoff scale $Q_{\rm cut}=71~\MeV$ is adopted~\cite{Schmidt:2015zda}.
	In this work, we follow the MMHT2015qed/MSHT20qed treatment~\cite{Harland-Lang:2019pla,Cridge:2021pxm} to parameterize the initial ISV as
	\begin{eqnarray}\label{eq:ISV}
		&\Delta d_{V,n}(x,\mu_0^2)=d_{V,n}(x,\mu_0^2)-u_{V,p}(x,\mu_0^2)
		=\epsilon\left(1-\frac{e_d^2}{e_u^2}\right)u_{V,p}^{\rm(QED)}(x,\mu_0^2),\\
		&\Delta u_{V,n}(x,\mu_0^2)=u_{V,n}(x,\mu_0^2)-d_{V,p}(x,\mu_0^2)
		=\epsilon\left(1-\frac{e_u^2}{e_d^2}\right)d_{V,p}^{\rm(QED)}(x,\mu_0^2),
	\end{eqnarray}
	where $q_{V,p}^{\rm(QED)}\propto P_{qq}^{\rm(QED)}\otimes q_{V,p}$. Note that the quark charge $e_q$ is implicitly contained in $P_{qq}^{\rm(QED)}$. The $\epsilon$ can be fixed through the momentum sum rule of proton and neutron PDFs, as
	\begin{eqnarray}\label{eq:epsilon}
		\epsilon=\frac{\int\dd x x(\gamma_p^{\rm inel}(x,\mu_0^2)-\gamma_n^{\rm inel}(x,\mu_0^2))}
		{\int\dd x x\left(\frac{3}{4}u_{V,p}^{\rm(QED)}(x,\mu_0^2)-3d_{V,p}^{\rm(QED)}(x,\mu_0^2)\right)}.
	\end{eqnarray}
	Here the ISV only involves inelastic component $\gamma_{p,n}^{\rm inel}$, consistent as Eq.~(\ref{eq:sum}). 
	
	\begin{figure}[h]
		\centering
		\includegraphics[width=0.49\textwidth]{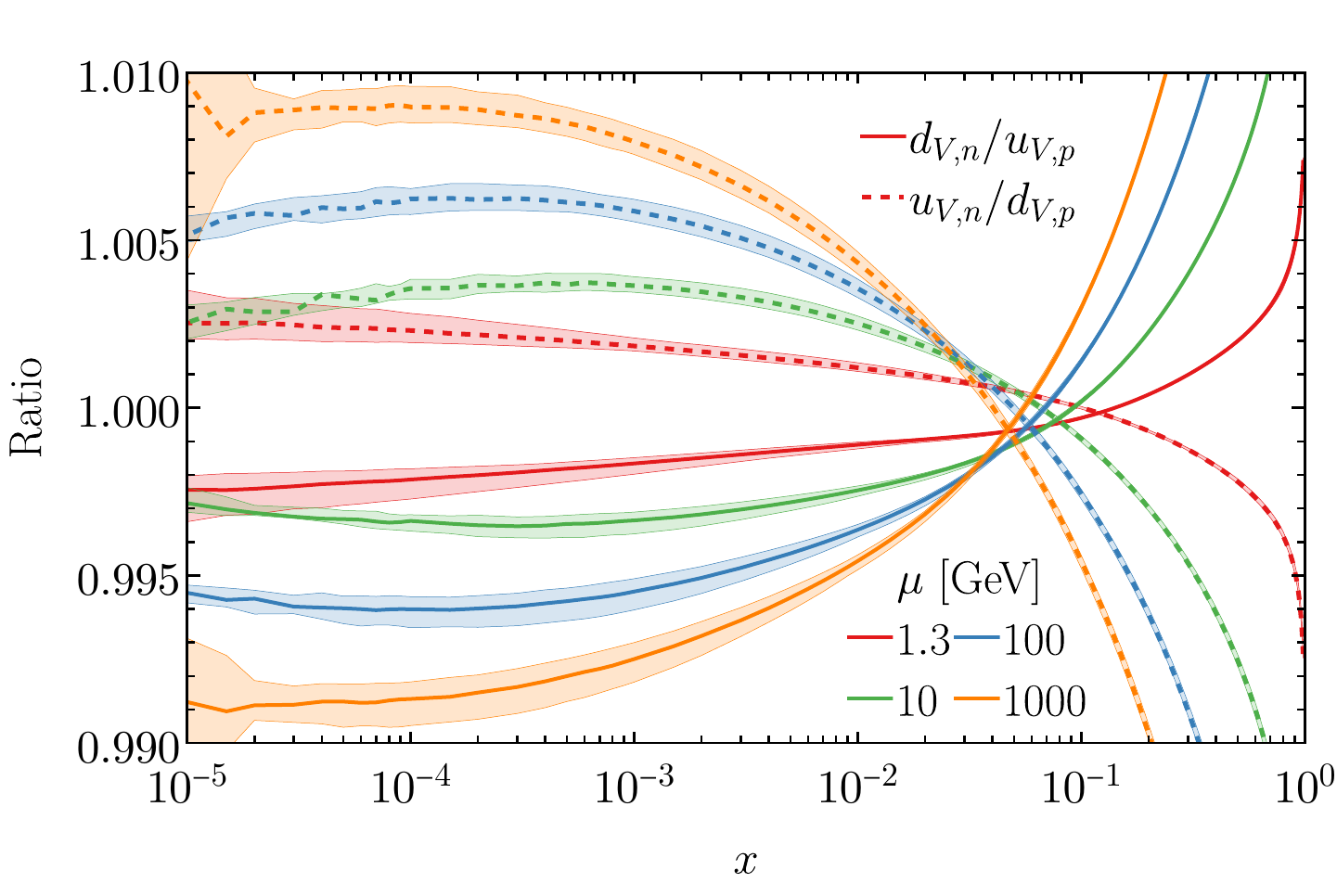}
		\includegraphics[width=0.49\textwidth]{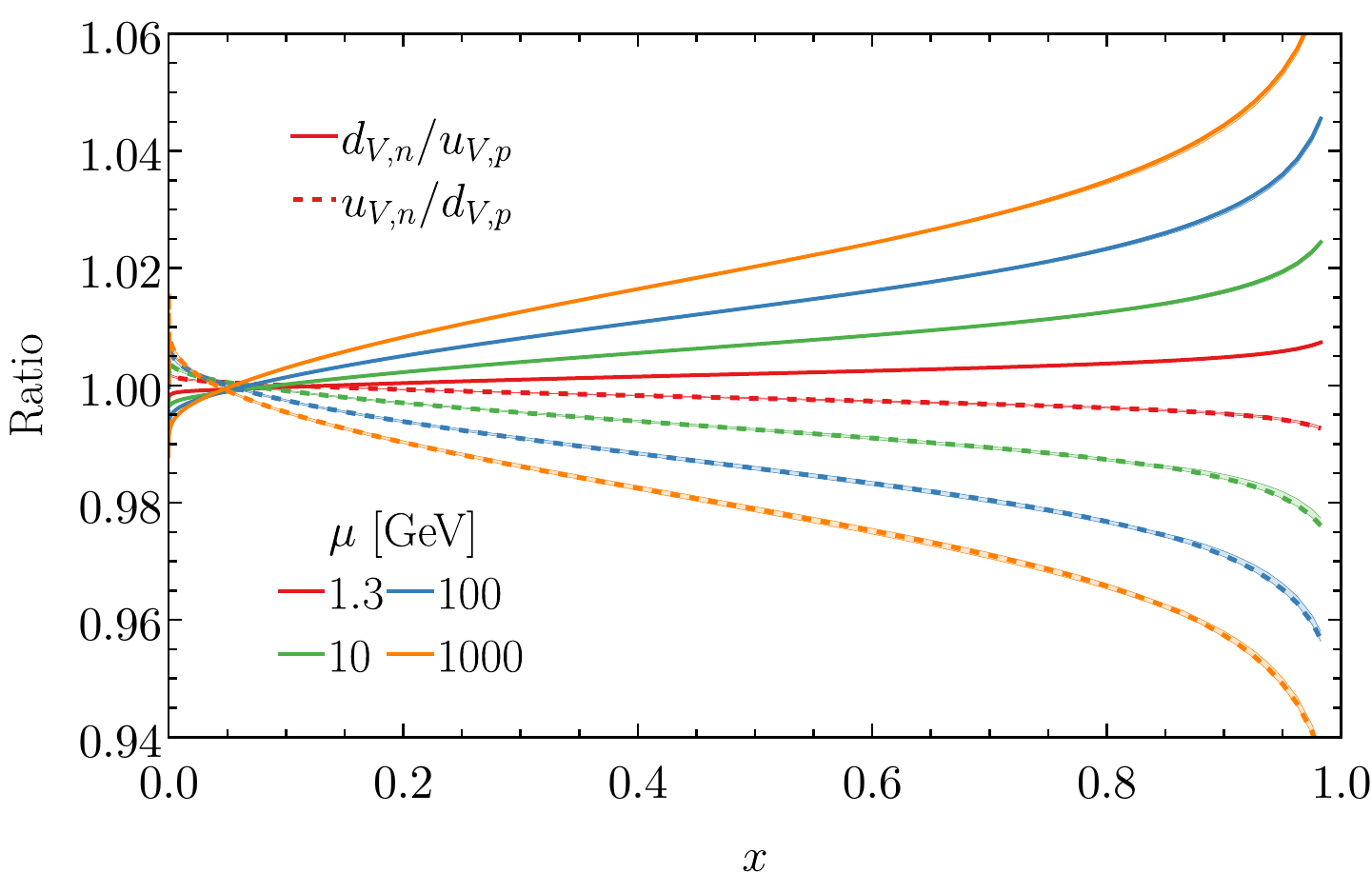}    
		\caption{Left: ratios of neutron valence quarks $d_{V,n},u_{V,n}$ to the proton isospin partners at $\mu=1.3,100,1000~\GeV$. Right: the same plot with a linear scale to emphasize the large $x$ behavior. The error bands indicate the envelope of CT18 eigenvector sets~\cite{Hou:2019efy}.}
		\label{fig:ISV}
	\end{figure}
	In Fig.~\ref{fig:ISV}, we compare the ratios of neutron's valence quarks to their proton isospin partners with the scale from $\mu=1.3~\GeV$ up to 1 TeV, which quantifies the size of isospin symmetry violation. The initial ISV effect is about 0.3\% (0.7\%) at small (large) $x$, which is less than the one found in the MMHT2015qed PDFs~\cite{Harland-Lang:2019pla}, as the elastic component is not included in Eq.~(\ref{eq:epsilon}) here. 
	The ISV effect increases with energy scale, which reaches up 1\% (6\%) at small (large) $x$ when $\mu=1~\TeV$,
	as a result of the QED DGLAP evolution, consistent with the CT14qed result~\cite{Schmidt:2015zda}.
	In such a way, the ISV corrected neutron's valence quarks are obtained through Eq.~(\ref{eq:ISV}), while the sea quarks are still assumed to obey the isospin symmetric assumption, \emph{i.e.}, Eq.~(\ref{eq:IS}).
	
	
	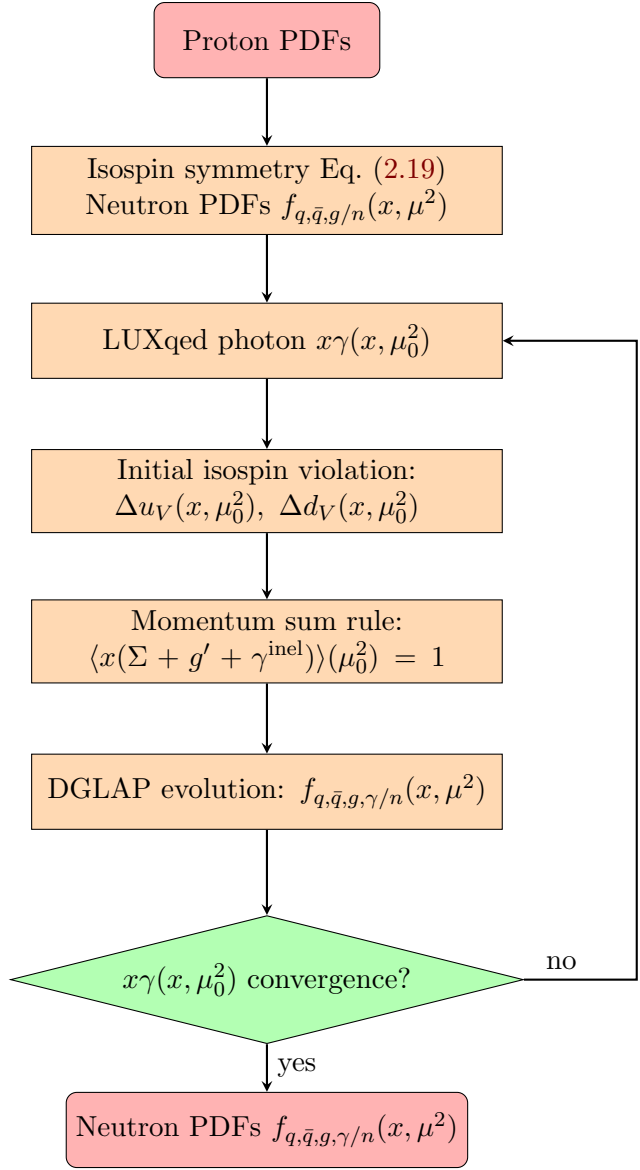
\begin{figure}[h]
		\centering
		\begin{tikzpicture}[node distance=2cm]
			\node (start) [startstop] {Proton PDFs};
			\node (pro1) [process, below of=start] {Isospin symmetry Eq.~(\ref{eq:IS})\\
				Neutron PDFs $f_{q,\bar{q},g/n}(x,\mu^2)$};
			\node (pro2) [process, below of=pro1] {LUXqed photon $x\gamma(x,\mu_0^2)$};
			\node (pro3) [process, below of=pro2] {Initial isospin violation:\\
				$\Delta u_V(x,\mu_0^2),~\Delta d_{V}(x,\mu_0^2)$};
			\node (pro4) [process, below of=pro3] {Momentum sum rule: \\
				$\langle x(\Sigma+g'+\gamma^{\rm inel})\rangle(\mu_0^2)=1$};
			\node (pro5) [process, below of=pro4] {DGLAP evolution: $f_{q,\bar{q},g,\gamma/n}(x,\mu^2)$};
			\node (dec1) [decision, below of=pro5, yshift=-0.5cm] {$x\gamma(x,\mu_0^2)$ convergence?};
			\draw [arrow] (dec1.east) -- ++(0.5,0) node[above] {no}
			-- ++(,0) |- (pro2.east);
			\node (stop) [startstop, below of=dec1] {Neutron PDFs $f_{q,\bar{q},g,\gamma/n}(x,\mu^2)$};
			
			\draw [arrow] (start) -- (pro1);
			\draw [arrow] (pro1) -- (pro2);
			\draw [arrow] (pro2) -- (pro3);
			\draw [arrow] (pro3) -- (pro4);
			\draw [arrow] (pro4) -- (pro5);
			\draw [arrow] (pro5) -- (dec1);
			\draw [arrow] (dec1) -- node[anchor=west] {yes} (stop);
			
		\end{tikzpicture}
		\caption{The flowchart to determine the neutron's inelastic photon with the CT18qed (DGLAP evolution) approach.}
		\label{fig:flowchart}
	\end{figure}
	
	\textbf{The DGLAP evolution.} As explicitly demonstrated in Ref.~\cite{Xie:2021equ}, the QED impact on the fitted $\chi^2$ and PDFs in the global analysis is minimal in the CTEQ-TEA framework. Based on this observation, the photon PDF can be obtained with a few iterations, which is enough to give a consistent co-evolved PDF set. 
	
	We sketch our procedure as a flowchart shown in Fig.~\ref{fig:flowchart}. As discussed in the previous subsection, we start with the CT18 proton PDFs and obtain the neutron PDFs with the isospin symmetry approximation.
	Then, the neutron's photon content at the starting scale $x\gamma(x,\mu_0^2)$ can be constructed using the LUXqed formalism.
	Afterward, the isospin violation induced by the QED effect can be determined correspondingly through Eq.~(\ref{eq:ISV}). The momentum sum rule is enforced by taking out the corresponding photon momentum fraction from the gluon component afterward. Then, the PDFs at a high scale can be self-consistently determined through the DGLAP evolution. With respect to the isospin symmetry approximation (ISA), the DGLAP evolved PDFs can be treated as a small perturbation, which will converge very fast within a few iterations, as depicted in Fig.~\ref{fig:flowchart}. We take the variation of photon and ISV-corrected valence quark PDFs within the permille level as the criterion to determine the convergence.
	
	\begin{figure}[h]
		\centering
		\includegraphics[width=0.49\textwidth]{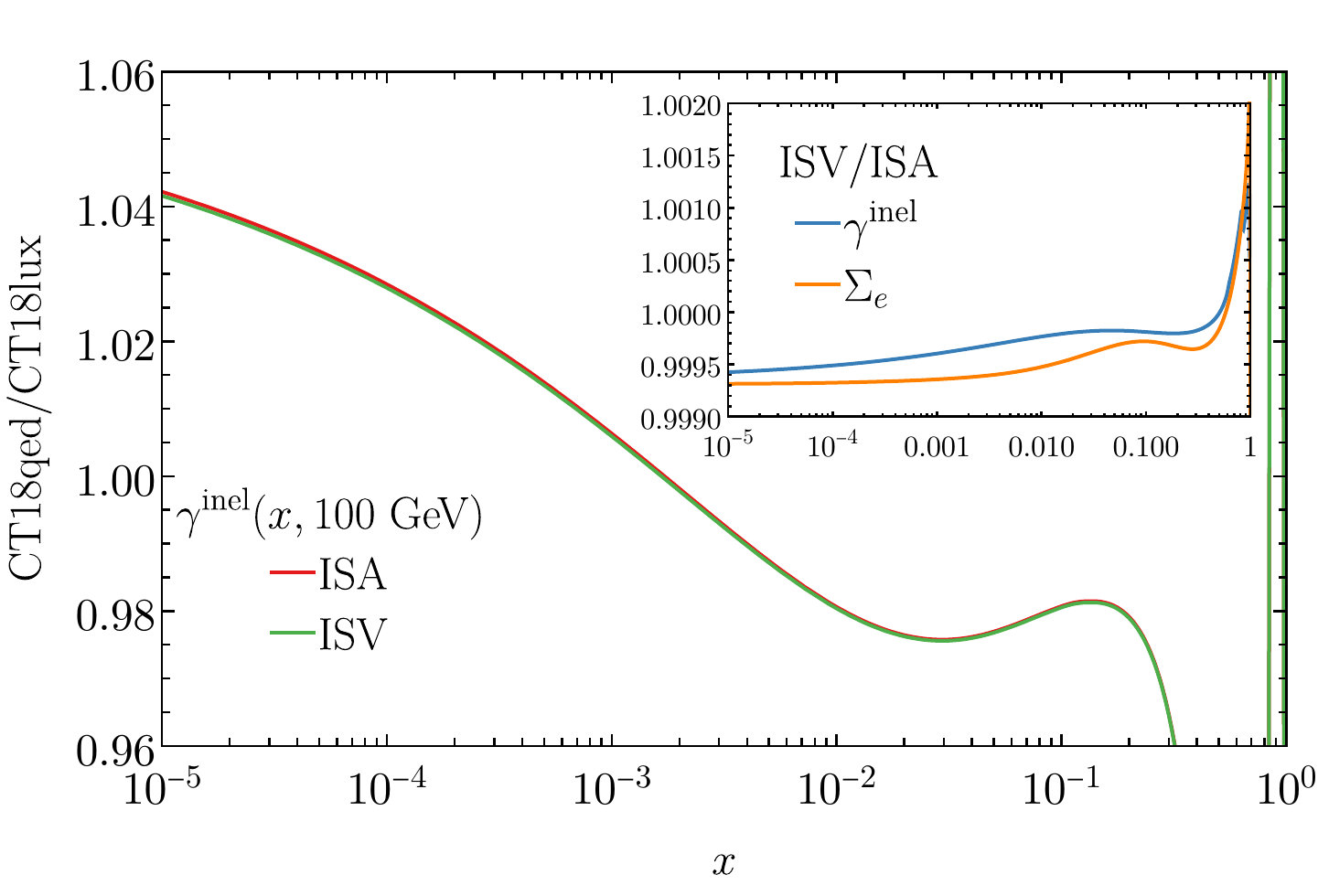}
		\includegraphics[width=0.49\textwidth]{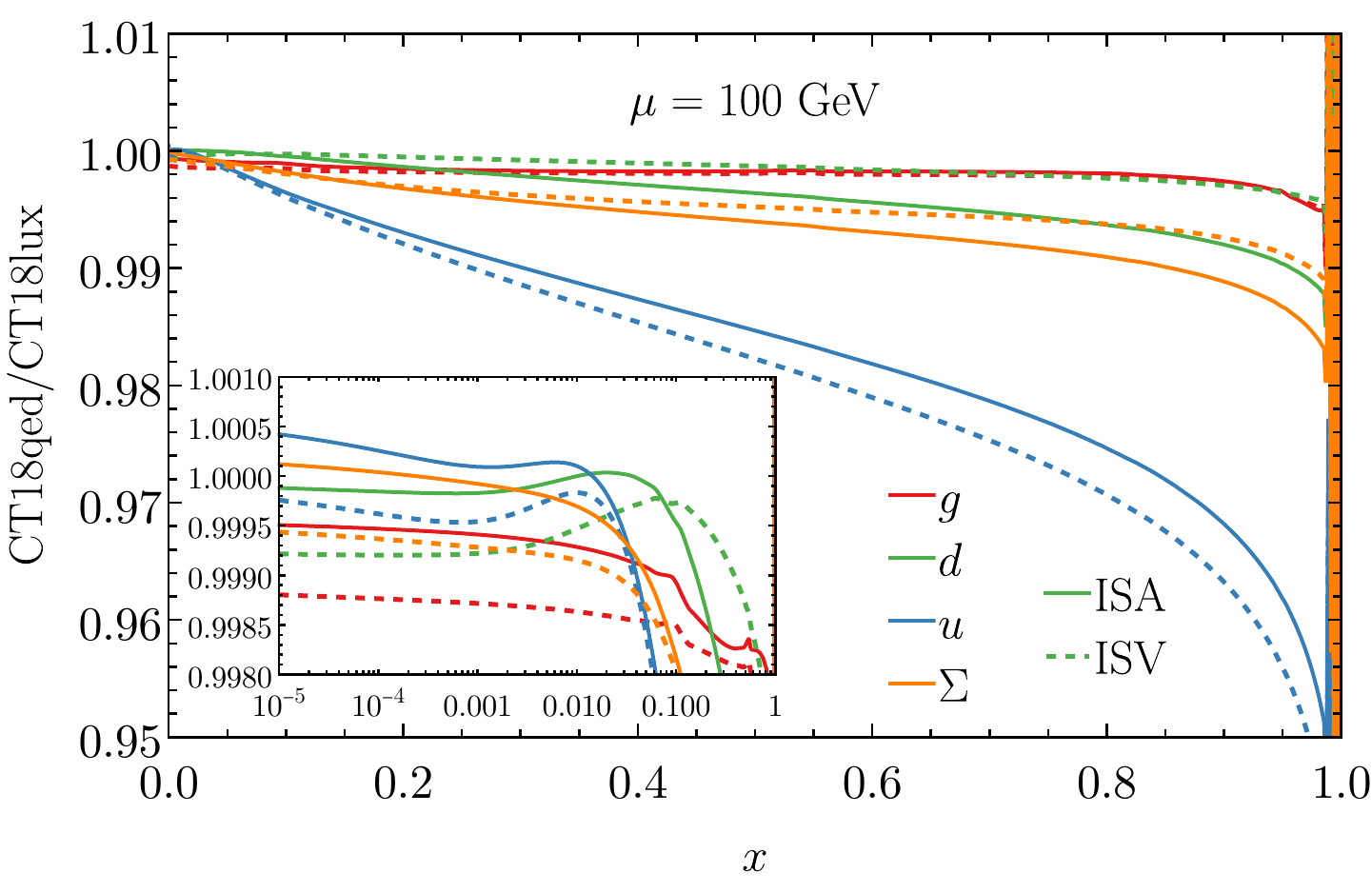}
		\caption{The PDF ratios of neutron's $\gamma^{\rm inel},\Sigma_{e}$ (left) and $g,u,d,\Sigma$ (right) at $\mu=100~\GeV$ in the CT18qed framework with respect to the CT18lux ones, with isospin symmetry approximation (ISA) or violation (ISV) at starting scale. The ``ISV" includes the enforcement of the momentum sum rule as well. }
		\label{fig:qed2lux}
	\end{figure}
	
\begin{figure}[h]
		\centering
		\includegraphics[width=0.49\textwidth]{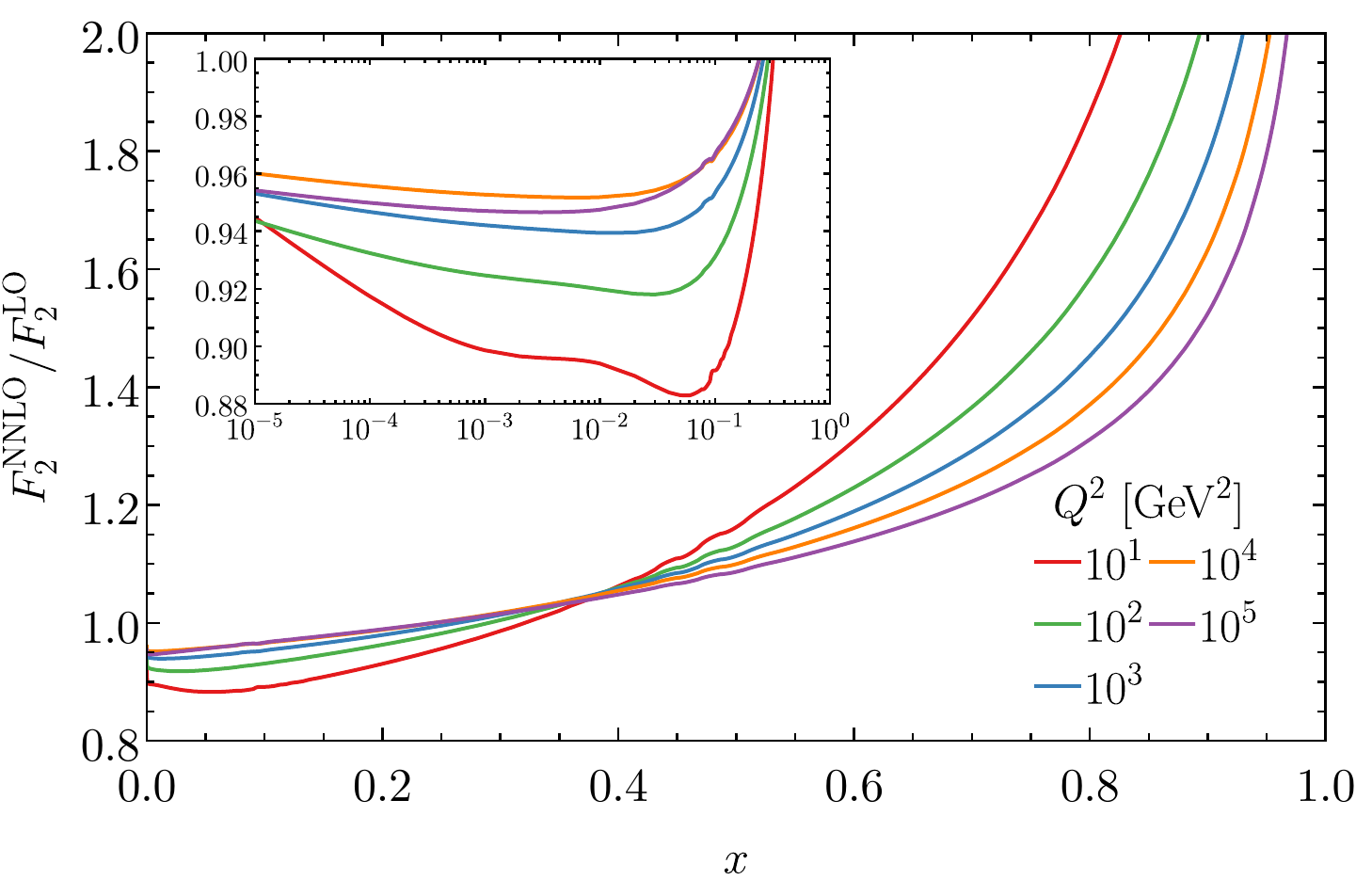}
		\includegraphics[width=0.49\textwidth]{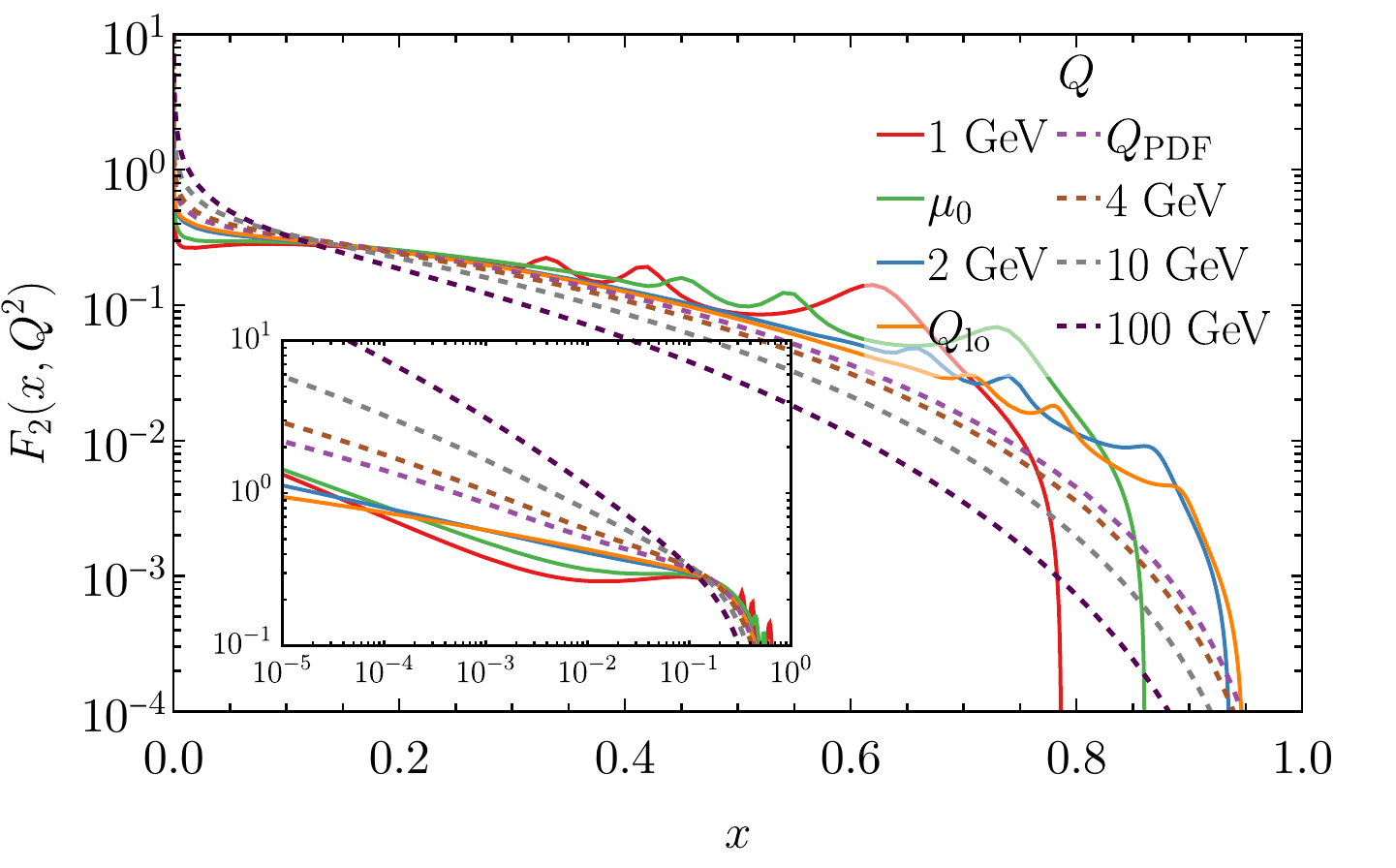}
		\caption{Left: the ratios of the neutron's structure function $F_2$ at NNLO to the LO one in the high-$Q^2$ pQCD region. Right: the comparison with low-$Q^2$ non-perturbative $F_2$ (solid curves) with the high-$Q^2$ pQCD one (dashed curves).}
		\label{fig:F2}
	\end{figure}

	In Fig.~\ref{fig:qed2lux}, we present the neutron's inelastic photon, charge-weighted singlet $\Sigma_e$, as well as $g,d,u,\Sigma$ PDFs in the CT18qed framework in comparison with the CT18lux ones. Similar to the proton case~\cite{Xie:2021equ}, we see the CT18qed photon PDF gets a few percent enhancement at small $x$, while agrees quite well in the moderate-$x$ region.
	The low-$x$ enhancement of CT18qed is induced by the DGLAP evolution, which equivalently integrated out the LO pQCD SFs, larger than the higher-order (NNLO) one employed in the LUXqed formalism, as shown in Fig.~\ref{fig:F2} left panel.
	In the extremely large-$x$ region, the CT18lux photon becomes significantly larger than the CT18qed one. It is mainly driven by the larger $\msbar$ conversion term, induced by a smaller high-$Q^2$ pQCD SF $F_2$ than the low-$Q^2$ no-perturbative one\footnote{Pay attention to the minus sign in the $\msbar$ conversion term in Eq.~(\ref{eq:LUXqed}).}, as shown in Fig.~\ref{fig:F2} right panel. Similar behaviors were found in the proton case already~\cite{Xie:2021equ}.
	
	In addition, we also compare the evolved photon with and without isospin symmetry approximation (ISA) at the starting scale. 
	The ISV result includes the enforcement of the momentum sum rule as well.
	We see that the corresponding difference only shows up at a sub-percent level. At a small $x$, the ISV photon gets slightly smaller with respect to ISA one, mainly resulting from the momentum sum rule. In comparison, the large-$x$ ISV photon gets a small enhancement induced by the larger charge-weighted singlet.
	In the right panel of Fig.~\ref{fig:qed2lux}, we present the CT18qed evolved PDFs of $g,u,d$ and singlet $\Sigma$, normalized to the CT18lux ones. We see the ISV reduces (enhances) $u$- and $d$-quark PDFs at large $x$, as implied in Fig.~\ref{fig:ISV} already.
	At small $x$, the singlet and gluon PDFs get a slight reduction, mainly driven by the enforcement of the momentum sum rule, which turns out to be a small effect, consistent with Eq.~(\ref{eq:xph0}). 
	
	\begin{figure}[h]
		\centering
		\includegraphics[width=0.49\textwidth]{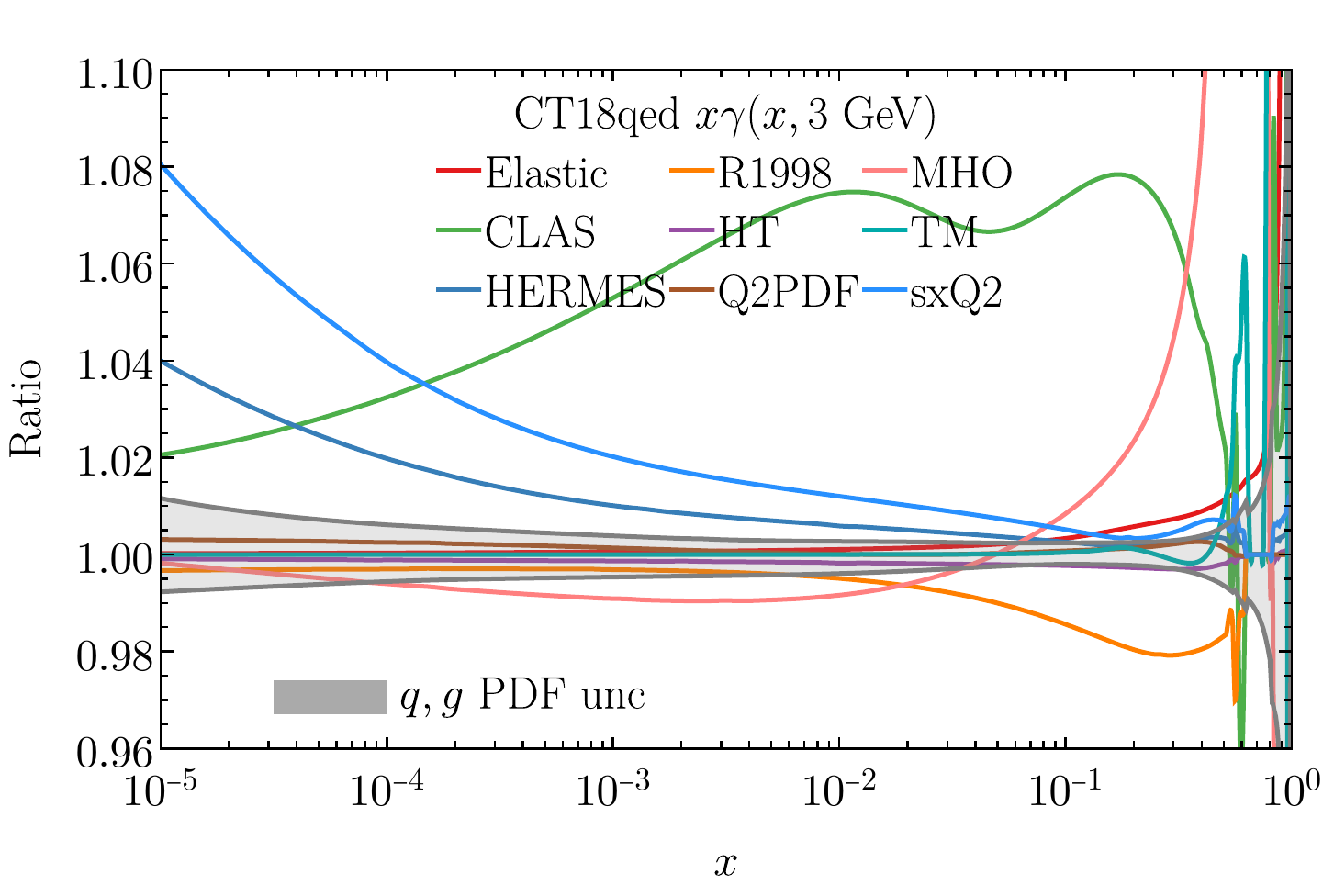}    
		\includegraphics[width=0.49\textwidth]{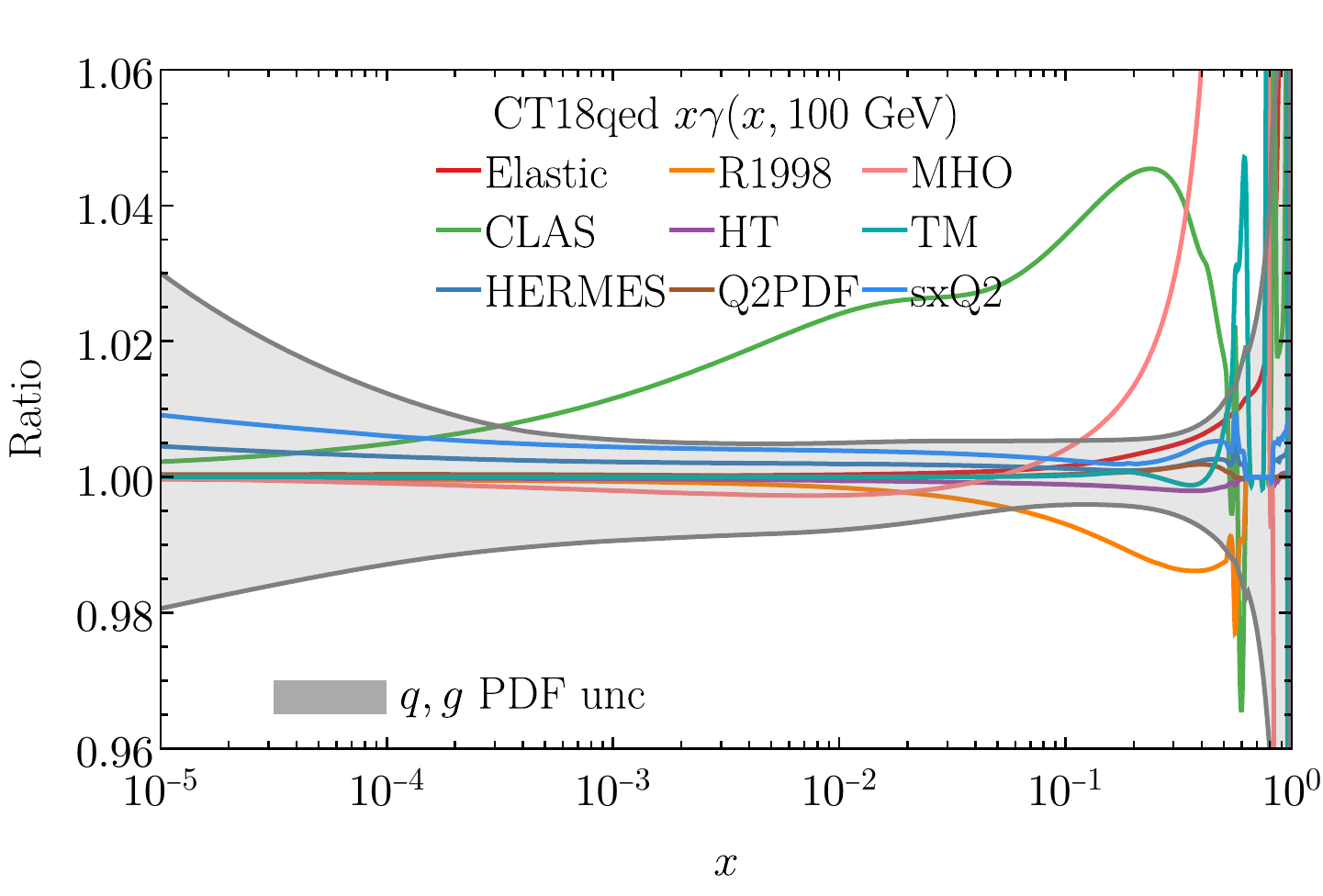}
		\caption{Similar to Fig.~\ref{fig:luxunc}, but neutron's photon PDF at $\mu=3,100~\GeV$ in the CT18qed framework.}
		\label{fig:CT18qed}
	\end{figure}

	Similar to the CT18lux variation, we present various low-$Q^2$ resources contributing to the neutron's photon PDF uncertainties in the CT18qed framework in Fig.~\ref{fig:CT18qed}. Many features are similar to the CT18lux results in Fig.~\ref{fig:luxunc}, with some minor differences. For example, the variations induced by Ye et al.'s elastic form factors and R1998 ratios remain at the same level.
	The dominant uncertainty comes from the resonance structure functions, by switching the CLAS result to the CB21 fit. In comparison, the relative contribution from the HERMES low-$Q^2$ continuum region (HERMES and sxQ2) is under control in comparison with the CT18lux case, mainly due to the increase of the perturbative inelastic component through the DGLAP evolution. For this reason, we see the $q,g$ PDF induced uncertainty get enhanced, slightly. In the $x\to1$ limit, the dominant uncertainty resources are the MHO and TM corrections, propagating from the uncertainties at $\mu=1.3~\GeV$.
	The HT uncertainty for $\mu=100~\GeV$ gets reduced with respect to the CT18lux ones, mainly driven by the small contribution to the initial scale. More specifically, when comparing with the CT18lux photon PDF at $\mu=3~\GeV$ as in Fig.~\ref{fig:luxunc}, we see the CT18qed uncertainties from HERMES, sxQ2, Q2PDF, and MHO get smaller, as a result of its under control at $\mu_0=1.3~\GeV$. For this reason, we will stay with the $\mu_0=1.3~\GeV$ as our CT18qed default choice for neutron's photon PDF, different from the proton treatment~\cite{Xie:2021equ}.
	
	\subsection{A comparison with other neutron's photon PDF sets}
	\label{sec:compare}
	Similar to the CT18qed proposal~\cite{Xie:2021equ}, we treat these nine low-$Q^2$ non-perturbative resources as orthogonal eigenvectors for the neutron's Hessian error sets. 
	For Ye et al.'s Elastic form factors, HERMES low-$Q^2$ continuum SFs, and R1998 ratio, we take the two-directional $(\pm)$ variations as the corresponding eigenvector sets.
	For the CB21 resonance SFs, sxQ2 SF variation, Q2PDF matching, MHO, HT, and TM corrections, we symmetrize the difference with respect to the central set to obtain the opposite-directional eigenvector sets. The final combined photon PDF uncertainties can be constructed through 
	\begin{eqnarray}\label{eq:unc}
		&\delta X=\sqrt{\sum_{i=1}^{N_{\rm PDF}+n_{\textrm{low-}Q^2}}
			\left(\frac{X_i^+-X_i^-}{2}\right)^2},\\
		&\delta X^+=\sqrt{\sum_{i=1}^{N_{\rm PDF}+n_{\textrm{low-}Q^2}}
			\left[\max(X_i^+-X_0,X_i^--X_0,0)\right]^2},\\
		&\delta X^-=\sqrt{\sum_{i=1}^{N_{\rm PDF}+n_{\textrm{low-}Q^2}}
			\left[\max(X_0-X_i^+,X_0-X_i^-,0\right]^2},
	\end{eqnarray}
	where $N_{\rm PDF}=28$ refers to the CT18 Hessian eigenvector sets, while $n_{\textrm{low-}Q^2}=9$ corresponds the eight low-$Q^2$ non-perturbative resources.
	
	\begin{figure}[h]
		\centering
		\includegraphics[width=0.49\textwidth]{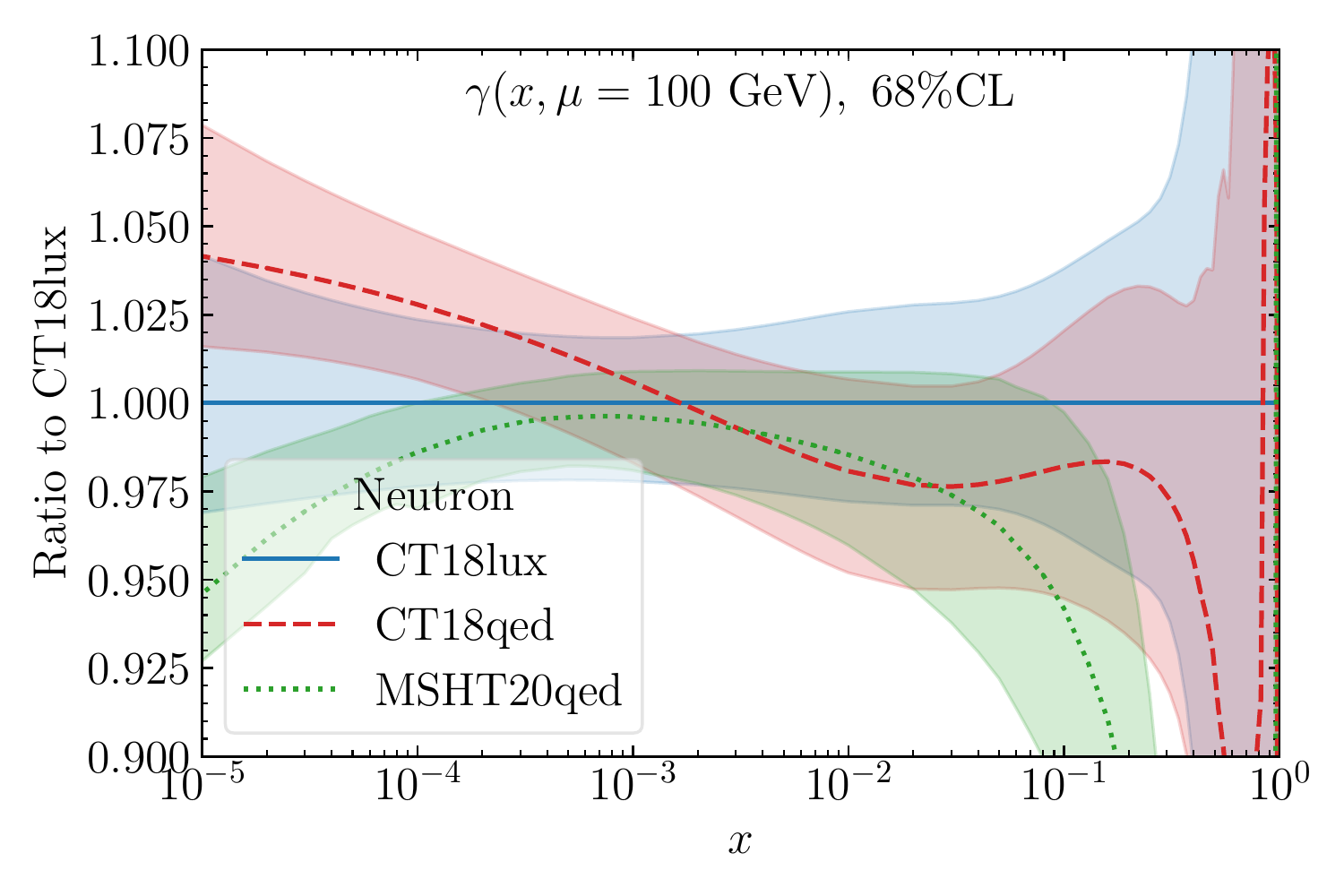}
		\includegraphics[width=0.49\textwidth]{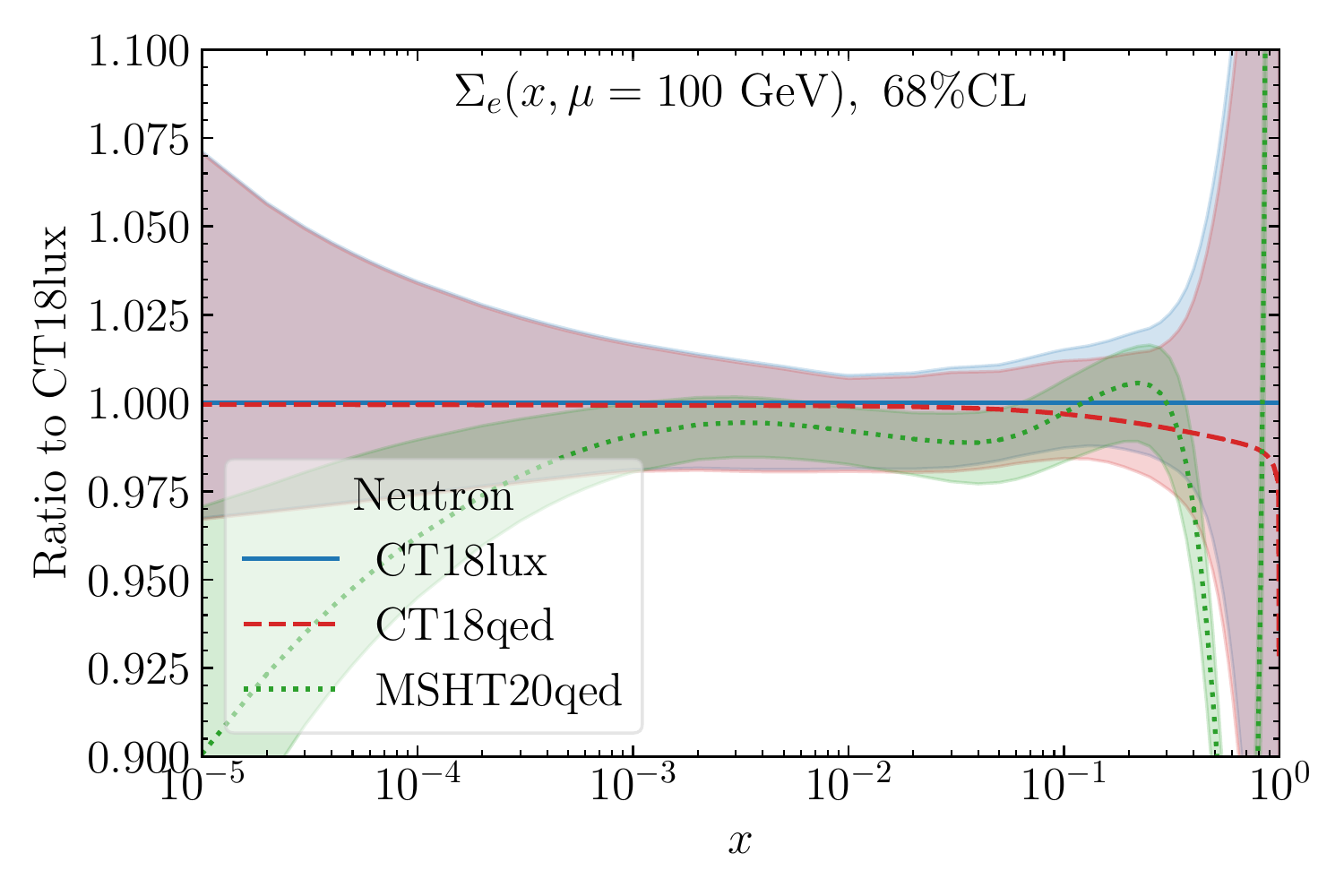}
		\includegraphics[width=0.49\textwidth]{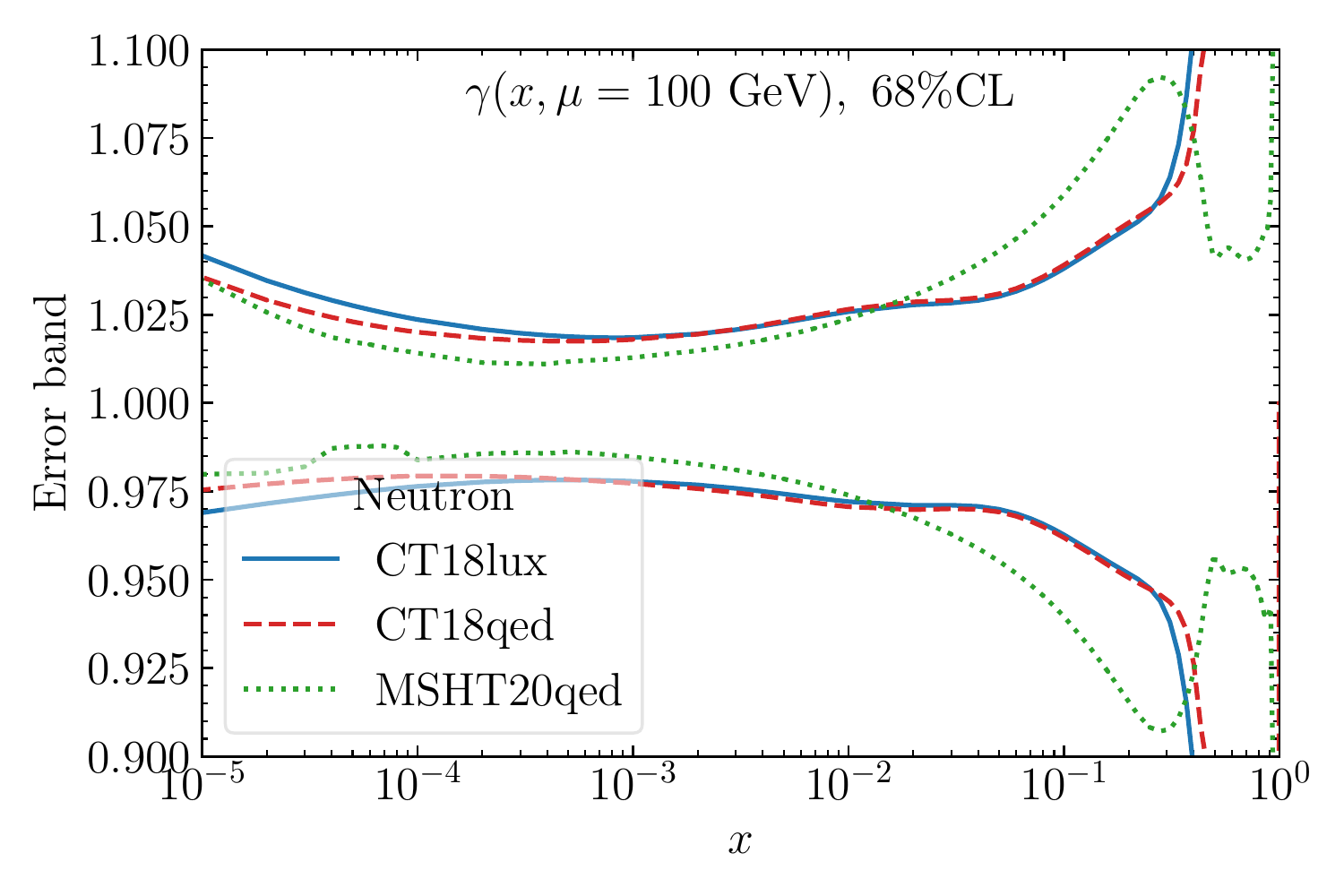}
		\includegraphics[width=0.49\textwidth]{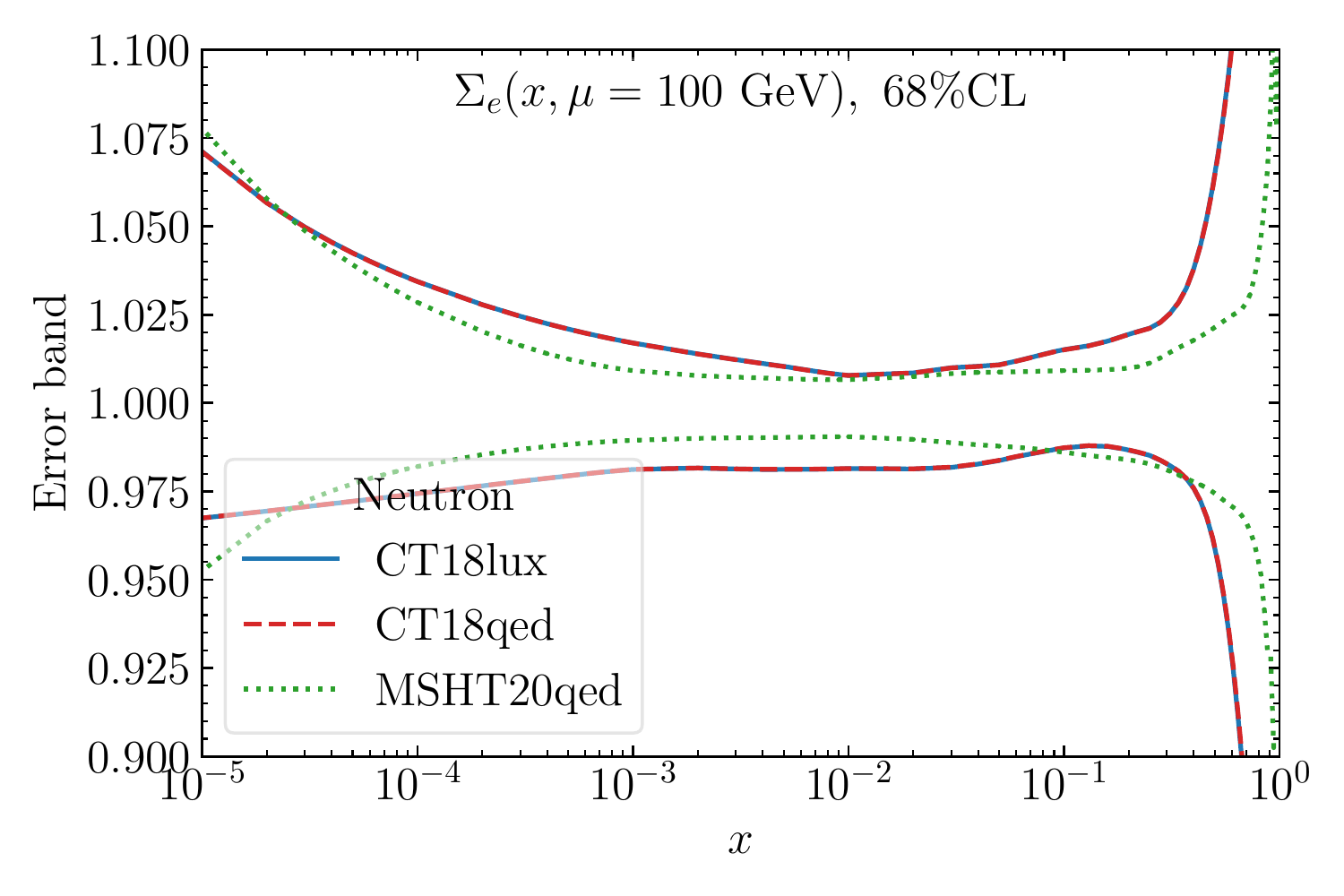}    
		\caption{The PDF comparison of the central values (upper) and the error bands (lower) for the neutron's photon (left) and charge-weighted singlet (right) at $\mu=100~\GeV$ among CT18lux, CT18qed, and MSHT20qed.}
		\label{fig:MSHT20}   
	\end{figure}
	
	In Fig.~\ref{fig:MSHT20}, we compare the central values and the error bands for neutron's photon and charge-weighted singlet $\Sigma_e=\sum_i e_i^2(q_i^2+\bar{q}_i)$ in the CT18lux, CT18qed, as well as the MSHT20qed PDFs. 
	In comparison with the CT18lux neutron's $\Sigma_e$ PDF, the CT18qed one gets a 1\% reduction around $x\sim0.6$, as a result of the photon's takeaway in the $q_i\to q_i\gamma$ QED splitting. The relative reduction size is smaller than the proton's case~\cite{Xie:2021equ} because the dominant valence quark $d_{V,n}$ has a smaller electric charge than the corresponding isospin partners $u_{V,p}$ inside of proton. In comparison with the uncertainty of CT18lux photon at $\mu=100~\GeV$, we see the CT18qed one is slightly smaller, both at small $x$ and large $x$, due to a small reduction of the non-perturbative impact in the propagation during the DGLAP evolution. For this reason, we advocate the CT18qed as the primary choice, which is used in our study of the phenomenological implications (Sec.~\ref{sec:imp}). But overall, we get quite a good agreement for the size of the photon PDF uncertainty between CT18lux and CT18qed.
	
	We also compare our photon and charge-weighted singlet PDFs with the MSHT20qed ones in Fig.~\ref{fig:MSHT20}. 
	We remind that similar to our CT18qed framework, the MSHT20qed initializes the photon PDF with the LUXqed formalism at a low scale, $\mu_0=1~\GeV$, and evolves the PDF to higher scales~\cite{Cridge:2021pxm}.
	Both photon and $\Sigma_e$ PDFs between CT18qed and MSHT20qed agree well within the moderate-$x$ region ($10^{-3}<x<10^{-1}$). In both small- and large-$x$ regions, the MSHT20qed photon PDF is smaller than the CT18qed ones, similar to the proton case as well~\cite{Xie:2021equ}.
	MSHT20qed small-$x$ ($x\lesssim10^{-3}$) photon gets softer, mainly driven by its smaller charge-weighted singlet PDF, as shown in Fig.~\ref{fig:MSHT20} (right). 
	
	In comparison, the MSHT20qed large-$x$ ($x\gtrsim10^{-1}$) photon gets smaller than CT18qed, resulted from various resources. First, MSHT20qed's $\Sigma_e$ at large $x$ is noticeably smaller than the CT18qed one, as shown in Fig.~\ref{fig:MSHT20} (right).
	It can be understood in terms of the more flexible parameterization of the MSHT20 fit, which allows a different high-$x$ power for the down quark from the up-quark one, while the CT18 parameterization fixes high-$x$ powers for the up and down quarks to be the same. As a result, the MSHT20qed's neutron-to-proton ratio of $\Sigma_e$ approaches to zero while CT18 one does not.
	Second, as noticed in the CT18qed proton PDF study~\cite{Xie:2021equ}, different treatments of the photon initialization with the LUXqed formalism have been adopted in CT18qed and MSHT20qed. Similar to the MMHT15qed~\cite{Harland-Lang:2019pla}, MSHT20qed takes an $\mu_0^2$ rather than $\mu_0^2/(1-z)$ with Eq.~(\ref{eq:LUXqed}), while the integration in $\mu^2\in[\mu_0^2,\mu_0^2/(1-z)]$ interval is done with a stationary approximation that 
	\begin{eqnarray}\label{eq:MSHT20}
		F_2(x,\mu^2)=F_2(x,\mu_0^2), ~
		F_L(x,\mu_0^2)=0,~
		\alpha(\mu^2)=\alpha(\mu_0^2),
	\end{eqnarray}
	which gives a smaller initial photon $x\gamma(x,\mu_0^2)$ than the CT18qed one in the $x\to1$ limit.
	
	In Fig.~\ref{fig:MSHT20}, we also compare the relative error bands of photon and charged-weighted singlet PDFs, which are normalized to the corresponding central sets. 
	We see that both PDF sets (CT18 and MSHT20) give a compatible estimation of the photon uncertainty.
	In the small $x$ region ($x<10^{-2}$), the MSHT20qed photon PDF gets a smaller error band, induced by its smaller charged-weighted singlet error band, as shown in Fig.~\ref{fig:MSHT20} (right).
	In the region $10^{-2}<x<0.3$, the MSHT20qed gives a slightly larger error, induced by its different error estimation~\cite{Harland-Lang:2019pla,Cridge:2021pxm}. Following MMHT2015qed~\cite{Harland-Lang:2019pla}, MSHT20qed~\cite{Cridge:2021pxm} includes variation of elastic form factors, R1998, CLAS/CB resonance SFs, HERMES fits (``Continuum" there), HT (``Renormalon" there) as well as $q,g$ PDFs. Additionally, different from our smooth transition between the HERMES and resonance region, MSHT20qed takes a threshold $W^2_{\rm cut}=3.5~\GeV^2$ and varies it within $3<W^2_{\rm cut}<4~\GeV^2$ to present a separate error set, which gives a larger uncertainty in this region.
	When $x\gtrsim0.65$, MSHT20qed freezes the photon PDF variation at the starting scale $\mu_0$, due to the numerical subtlety mentioned above. As a consequence, the photon PDF uncertainty, only driven by the DGLAP evolution, becomes smaller in this region. 
	In comparison, different from the MSHT stationary approximation in Eq.~(\ref{eq:MSHT20}), we take the complete $\mu^2$-dependent SFs in the region $\mu^2\in[\mu_0^2,\mu_0^2/(1-z)]$. It allows us to explore the missing higher order (MHO) uncertainty by varying the scale between $\mu^2_0$ and $\mu_0^2/(1-z)$ in the LUXqed formalism of Eq.~(\ref{eq:LUXqed}), which gives an additional error set and drives the increasing of CT18qed photon PDF uncertainty when $x\to1$.
	
	The neutron's photon momentum fractions at a few scales among CT18lux, CT18qed, and MSHT20qed are compared in Table~\ref{tab:mom}. We see the CT18lux and CT18qed results get a quite good agreement, except that the CT18lux result suffers a slightly larger uncertainty at high scale $\mu=100~\GeV$, due to the mentioned TM corrections. The MSHT20 result is smaller, as a result of its smaller elastic photon, as mentioned in Sec.~\ref{sec:el}.
	
	\begin{figure}[h]
		\centering
		\includegraphics[width=0.49\textwidth]{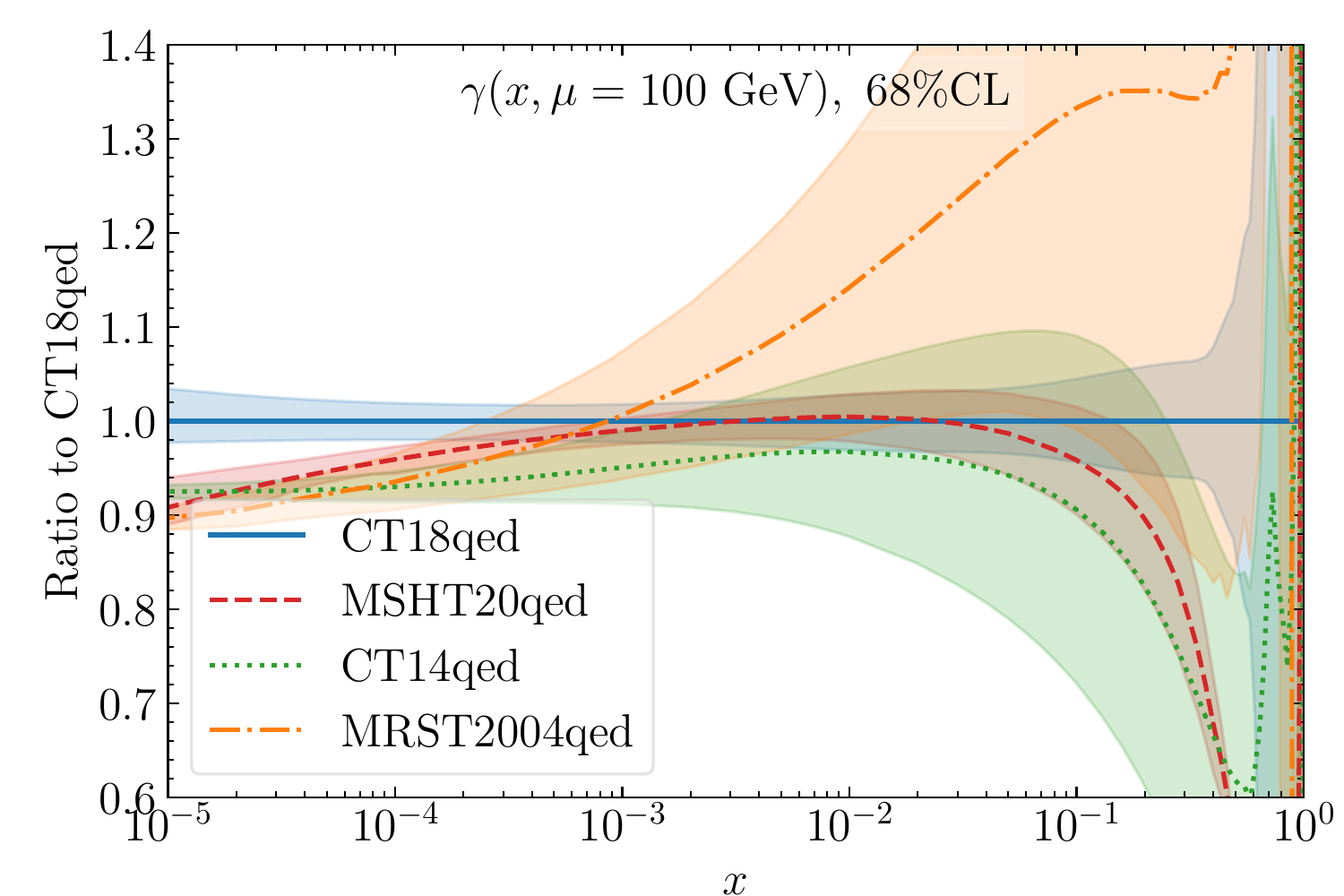}   
		\includegraphics[width=0.49\textwidth]{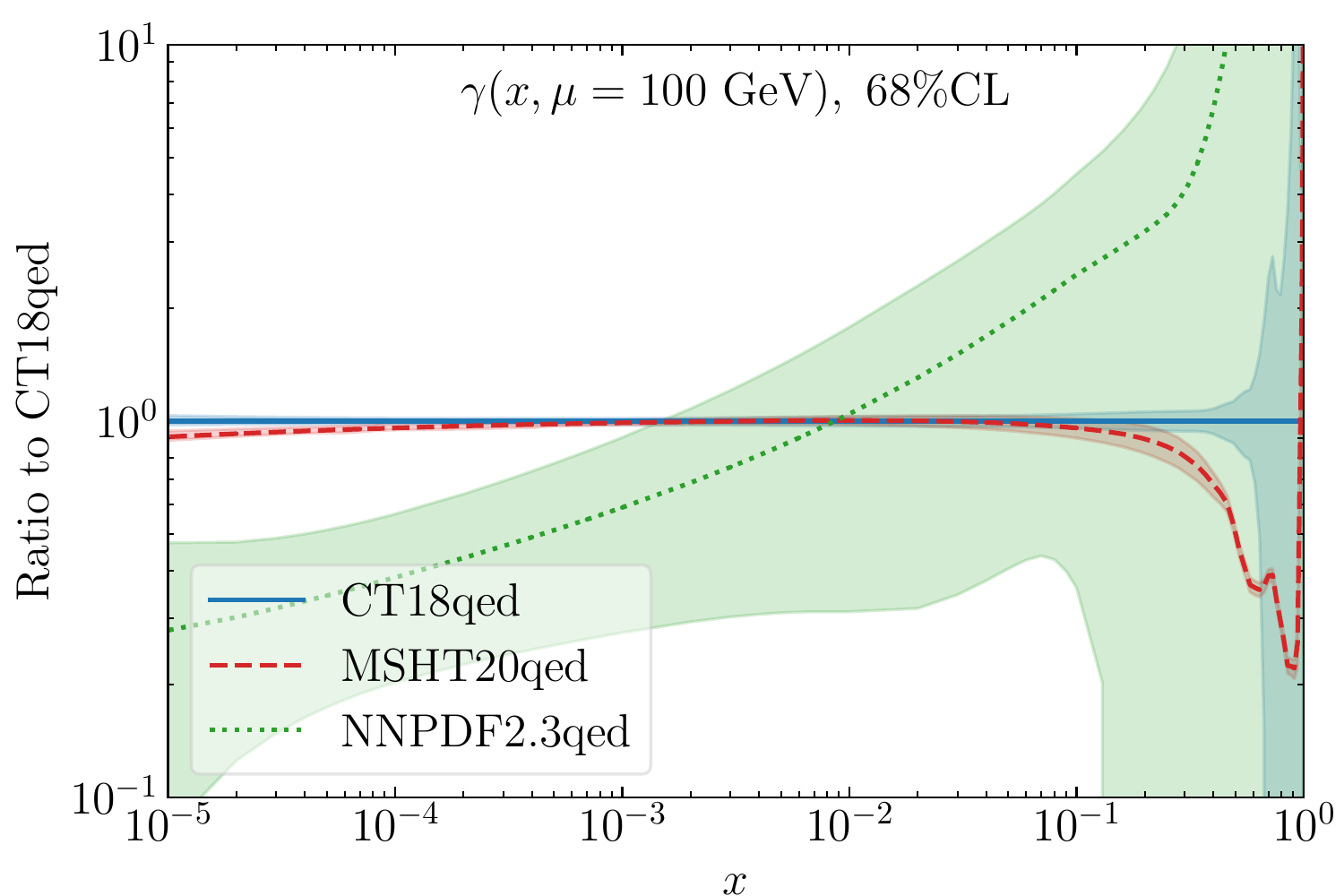} 
		\caption{The comparison between the first (CT14qed, MRST2004qed, and NNPDF2.3qed) and second (CT18qed and MSHT20qed) generations of neutron's QED PDFs.}
		\label{fig:1stgen}
	\end{figure}
	
	Meanwhile, we also include the comparison with the first generation of neutron's photon PDF sets, \emph{i.e.}, CT14qed~\cite{Schmidt:2015zda}, MRST2004qed~\cite{Martin:2004dh}, and NNPDF2.3qed~\cite{Ball:2013hta}, in the Fig.~\ref{fig:1stgen}.
	The central PDF of the MRST2004qed set is taken as an average between the two public sets, corresponding to the current- or constitute-quark mass parameterization, respectively, while the error band is their half difference~\cite{Martin:2004dh}. The CT14qed PDF error PDFs are taken as the 0-th and 11-th sets, which correspond to the initial proton's photon momentum fraction $\langle x\gamma_p\rangle(\mu_0^2)=0\%$ and $0.11\%$, respectively, in the 68\% confidence level~\cite{Schmidt:2015zda}.
	We see that at a low $x$, all these three first-generation sets give a softer photon, while the MRST2004qed and NNPDF2.3qed give a larger photon at large $x$, different from CT14qed.
	The NNPDF2.3qed gives a significantly larger error band than other PDF sets, due to its different methodology, which fits the available deep inelastic scattering and Drell-Yan data~\cite{Ball:2013hta}.
	In Fig.~\ref{fig:1stgen}, we also include the MSHT20qed in this comparison.
	
	\begin{figure}[h]
		\centering
		\includegraphics[width=0.49\textwidth]{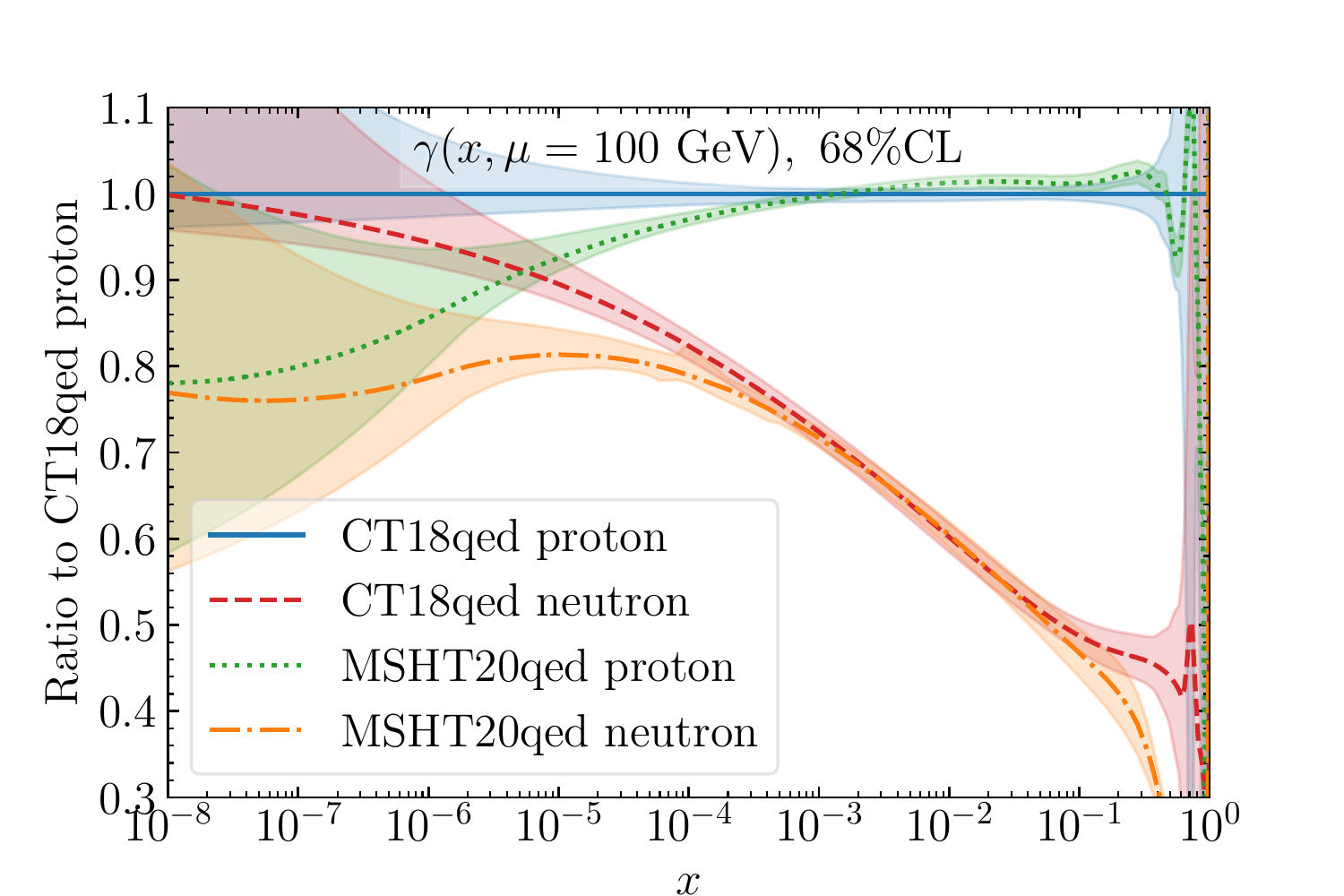}
		\includegraphics[width=0.49\textwidth]{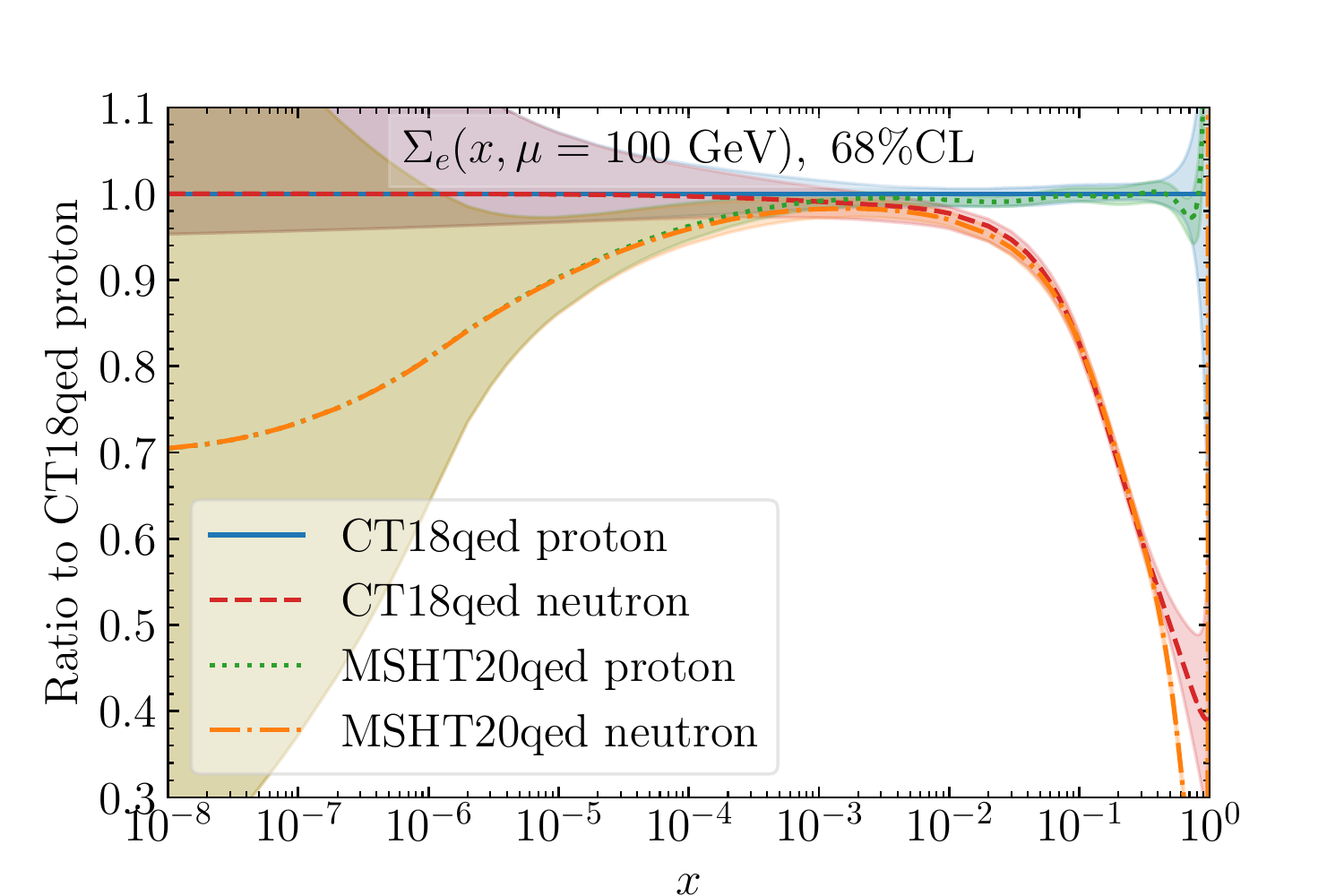}    
		\caption{The neutron-proton ratios for photon and charge-weighted singlet PDFs at $\mu=100~\GeV$.}
		\label{fig:n2p}
	\end{figure}
	
	Finally, we compare neutron's and proton's photon and charge-weighted singlet PDFs in both CT18qed and MSHT20qed in Fig.~\ref{fig:n2p}. We see that the neutron's photon is significantly smaller than the proton's, resulting from its smaller elastic component as well as the suppression of charge-weighted singlet PDFs. In the small-$x$ limit, the neutron's photon approaches the proton's, driven by the unity sea-quark ratio, which is shown explicitly in both the CT18qed and MSHT20qed scenarios.
	
	\section{Phenomenological implications}
	\label{sec:imp}
	The neutron's PDFs are essential for many phenomenological studies, especially for the processes involving nucleus scattering. In this section, we take the neutrino-nucleus $W$-boson production (WBP) and the GeV axion-like particle production in a high-energy muon beam dump experiment to explore the implications of our neutron's photon PDF.
	
	\subsection{$W$-boson production in the neutrino-nucleus scattering}
	
	The neutrino-nucleus $W$-boson production (WBP) is an important scattering process for TeV--PeV neutrino detections~\cite{Zhou:2019frk, Ackermann:2022rqc}. Previous work shows that the cross sections of WBP are up to $\simeq 7.5\%$ of the charged current deep-inelastic scattering, which is the dominant process for TeV--PeV neutrino detections by IceCube~\cite{IceCube:2015gsk}, KM3NeT~\cite{KM3Net:2016zxf}, Baikal-GVD~\cite{BAIKAL:2013jko}, etc. The rapidly increasing data due to more and more running and proposed experiments (e.g., IceCube-Gen2~\cite{Blaufuss:2015muc}, P-ONE~\cite{P-ONE:2020ljt}, TRIDENT~\cite{Ye:2022vbk}, FASER$\nu$~\cite{FASER:2023zcr}) necessitate a precision calculation of WBP. 
	
	\begin{figure}[h]
		\centering
		\includegraphics[width=0.33\textwidth]{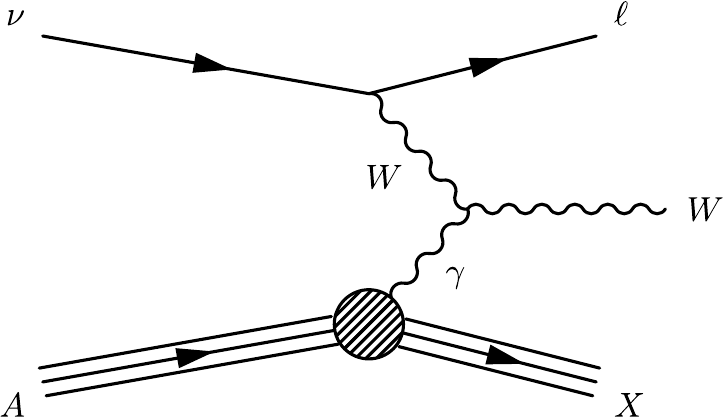}
		\includegraphics[width=0.33\textwidth]{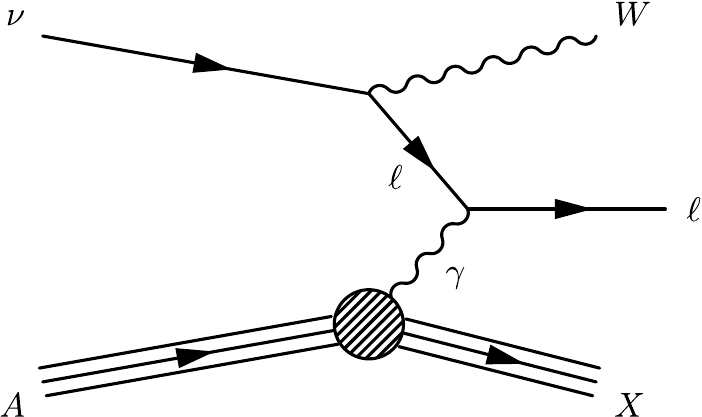}
		\caption{The $t,u$-channel diagrams for the $W$-boson production in the neutrino-nucleus scattering.}
		\label{fig:feynWBP}
	\end{figure}
	Fig.~\ref{fig:feynWBP} shows the Feynman diagrams of WBP, which is a \textit{photon-induced} process. Depending on the momentum of the photon, the scattering is in three different kinematic regimes: 1) coherent, in which the photon couples to the nucleus, which remains intact after the scattering, 2) diffractive, in which the photon couples to a nucleon, which remains intact after the scattering, and 3) inelastic, in which the photon couples to a nucleon or quark and the nucleon breaks after the scattering. The inelastic regime has the largest cross section and the largest uncertainty, mainly due to the uncertainties in the photon PDF~\cite{Zhou:2019vxt}.
	
	\begin{figure}[h]
		\centering
		\includegraphics[width=0.495\textwidth]{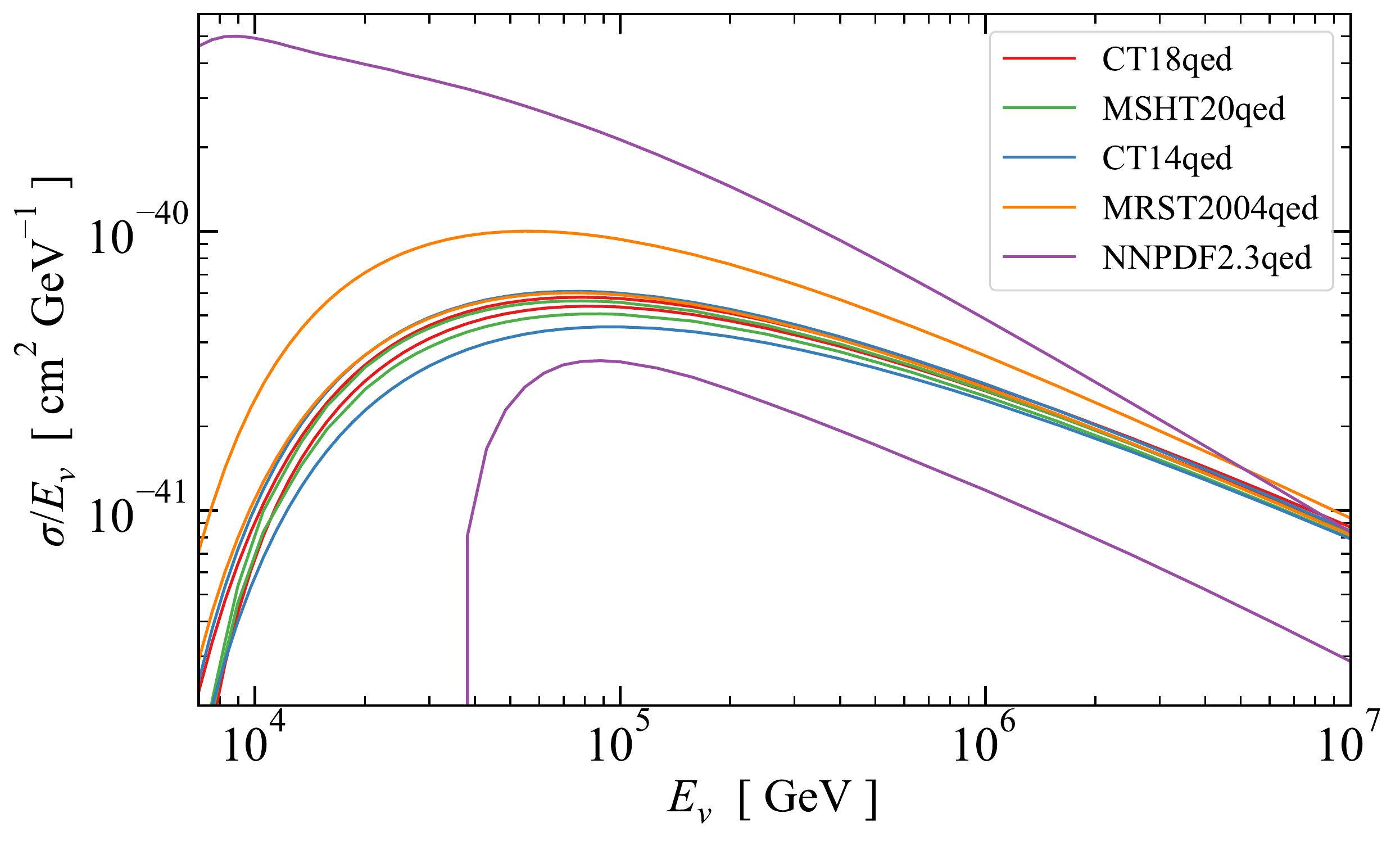}
		\includegraphics[width=0.47\textwidth]{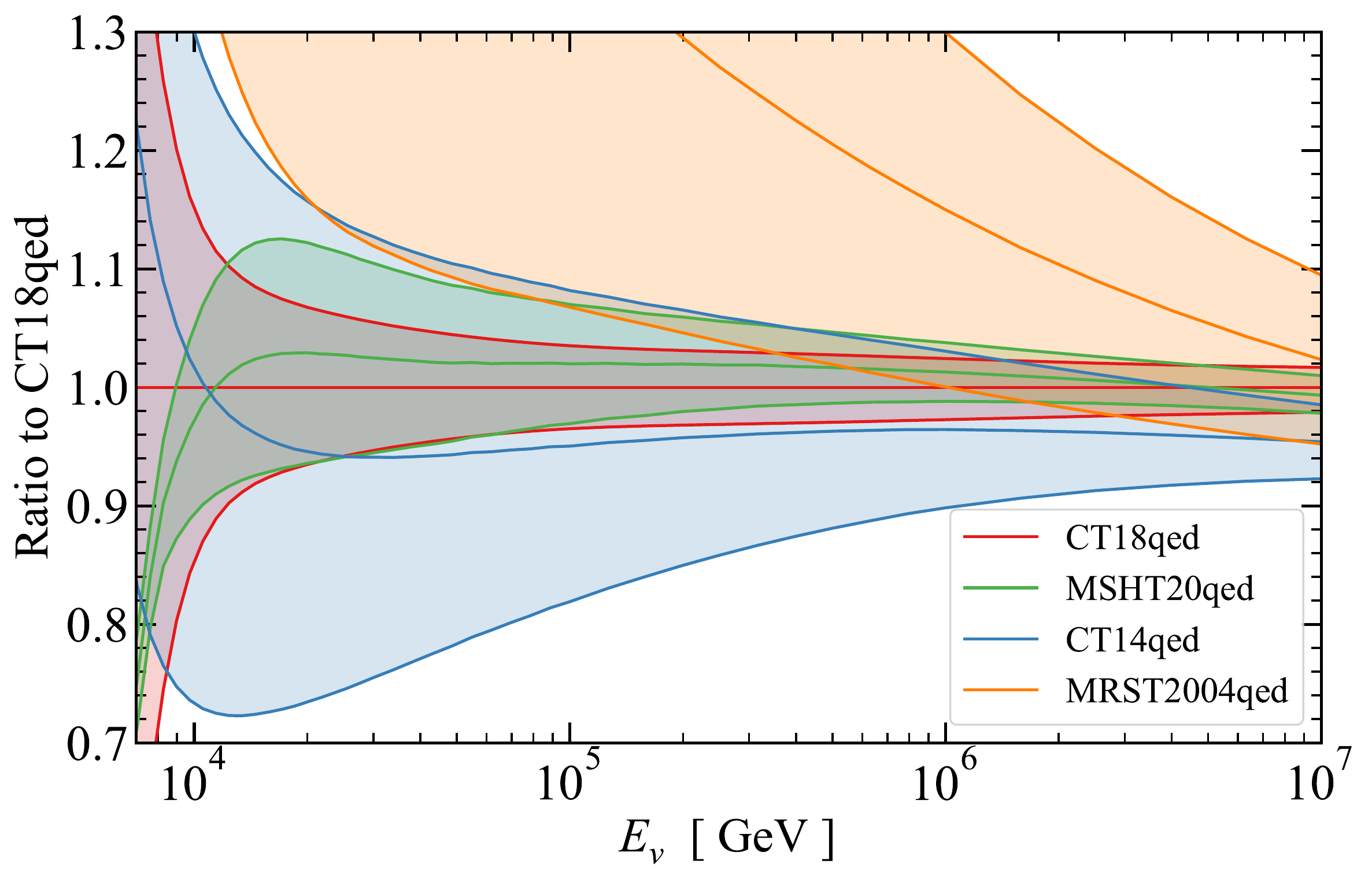}
		\includegraphics[width=0.47\textwidth]{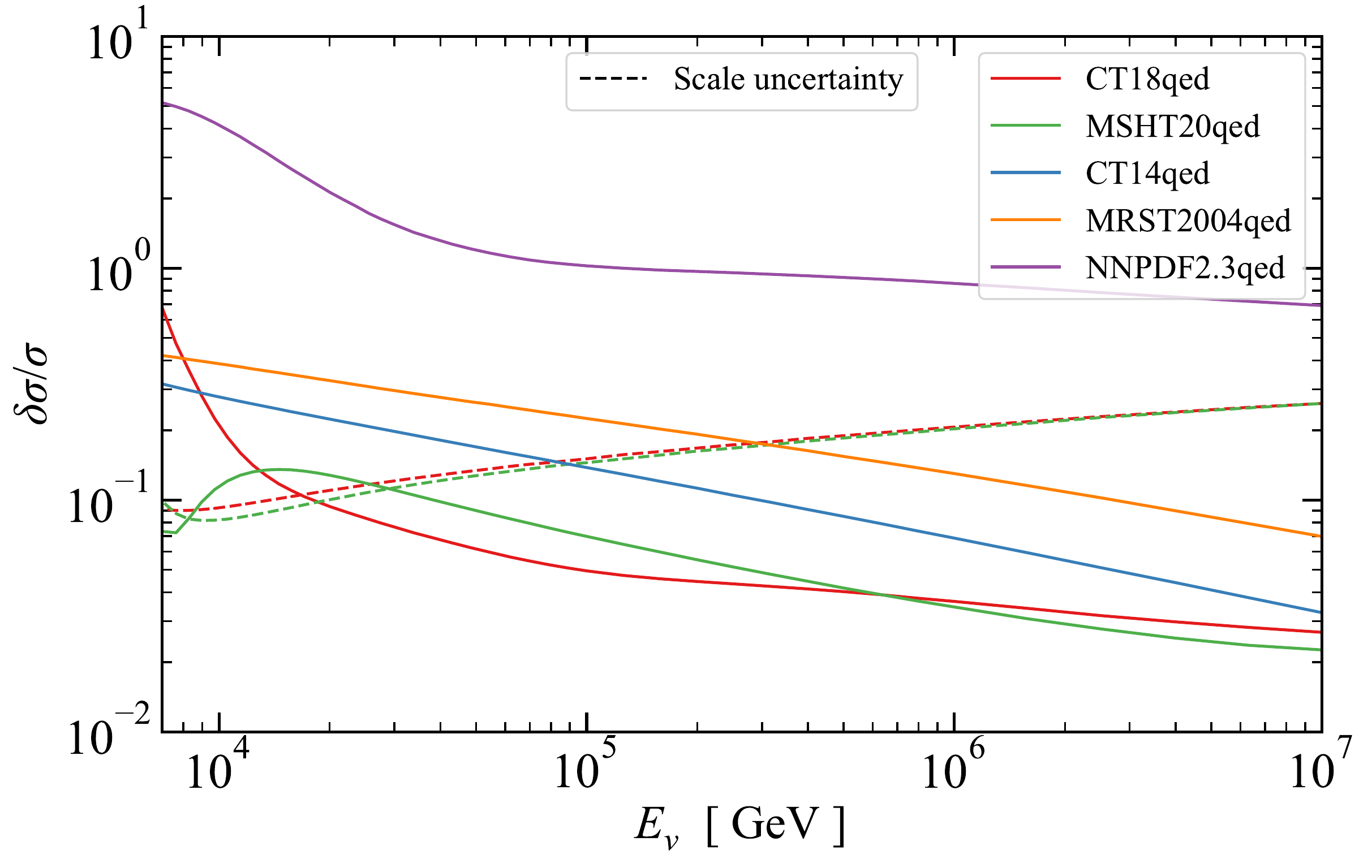}
		\caption{Cross sections, $\sigma$, of $\nu_e$-induced WBP in the inelastic regime on a neutron, $\overset{(-)}{\nu_e}n\to e^{\mp}W^\pm$. Upper left: $\sigma/E_\nu$ from different inelastic neutron photon PDFs. The two curves in the same color represent the uncertainty at a 68\% confidence level.
			Upper right: cross section ratios with uncertainties between CT18qed and others. 
			Lower: Relative uncertainties, $\delta \sigma / \sigma$. 
		}
		\label{fig:xsWBP}
	\end{figure}
	
	In Fig.~\ref{fig:xsWBP}, we show the improvement in the precision of the WBP inelastic cross section, thanks to the improvement in the precision of the photon PDF. We follow the computational procedure in Ref.~\cite{Zhou:2019vxt}. 
	The central factorization and renormalization scales are chosen to be the partonic collision energy $\sqrt{s_{\nu\gamma}}$, and the corresponding uncertainty is quantified with the 9-point variation by a factor of two.
	Here we show the WBP cross section of $\nu_e$ scattered with a neutron target, which is the main focus of this work.
	The same conclusion can be drawn for proton and for the other lepton flavors.
	
	The upper left panel of Fig.~\ref{fig:xsWBP} shows the cross sections (divided by neutrino energy to make the plots more visible).
	The upper right panel shows the cross section ratios with uncertainties and the lower panel shows the relative uncertainties.
	We see that among the first-generation photon PDF sets, the CT14qed gives the smallest error bands, which was used in Ref.~\cite{Zhou:2019vxt}.
	The second-generation photon PDFs (MSHT20qed and CT18qed) give much more precise predictions of WBP inelastic cross section ($\lesssim 10\%$ for $E_\nu \gtrsim 10^4$~GeV as in the right panel).  In the $E_\nu \lesssim 10^4$~GeV regime, the central of MSHT20qed prediction gets smaller than the CT18qed one, due to its smaller large-$x$ photon PDF, as shown in Figs.~\ref{fig:MSHT20}-\ref{fig:n2p}. With respect to the CT18qed PDF uncertainty, the MSHT20qed one is also smaller and decreases slightly in the small $E_\nu$ direction, due to its different treatment of photon PDF uncertainty, as discussed in Sec.~\ref{sec:compare}.
	
	The updated cross sections of WBP using the state-of-the-art CT18qed photon PDF will be uploaded to the GitHub repository under
	\href{https://github.com/beizhouphys/neutrino-W-boson-and-trident-production}{this link} \github{beizhouphys/neutrino-W-boson-and-trident-production}.
	

	\subsection{An Axion-like particle in a muon beam dump experiment}
	The axion, as a CP-odd scalar particle, was originally postulated by Peccei and Quinn~\cite{Peccei:1977hh} to resolve the strong charge-parity (CP) problem in quantum chromodynamics~\cite{Weinberg:1977ma,Wilczek:1977pj}. Afterward, many beyond-the-Standard-Model (BSM) theories predict axion or axion-like particles (ALPs) in a wide mass range~\cite{Kim:2008hd,DiLuzio:2020wdo,Marsh:2015xka}. See Refs.~\cite{ParticleDataGroup:2022pth,Choi:2020rgn} for the latest progress and review for axions and other similar particles.
	
	\begin{figure}[h]
		\centering
		\includegraphics[width=0.32\textwidth]{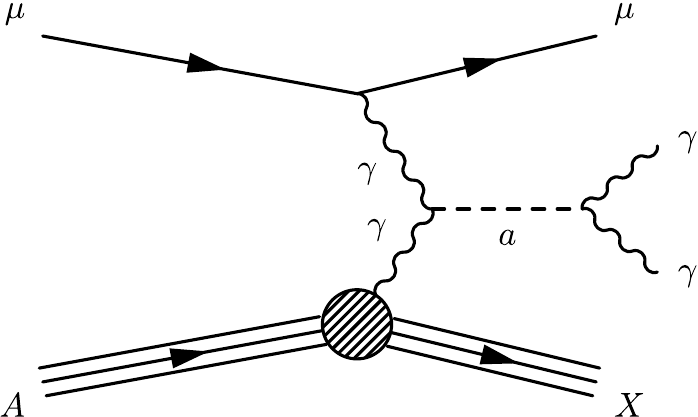}
		\includegraphics[width=0.33\textwidth]{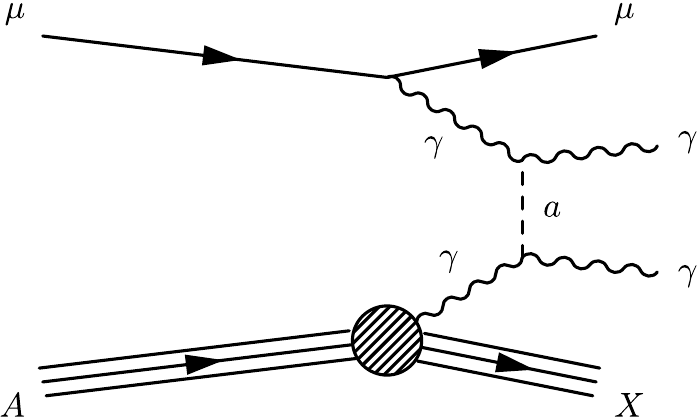}
		\includegraphics[width=0.33\textwidth]{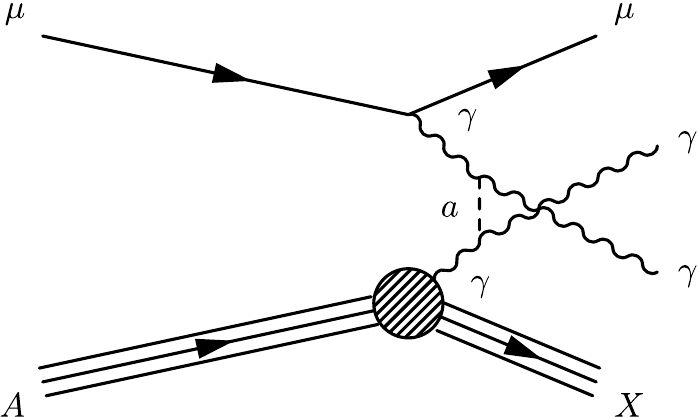}    
		\caption{The $s,t,u$-channel diagrams for the axion-like particle production at a muon beam dump experiment.}
		\label{fig:feynAxion}
	\end{figure}
	
	In this subsection, we focus on an ALP ($a$) with a two-photon interaction with an effective Lagrangian
	\begin{eqnarray}\label{eq:lag}
		\Delta\mathcal{L}=\frac{1}{2}\partial_\mu a\partial^\mu a-\frac{1}{2}m_a^2a^2-\frac{1}{f_a}aF_{\mu\nu}\tilde{F}^{\mu\nu},
	\end{eqnarray}
	where $F^{\mu\nu}$ is the electromagnetic field-strength tensor, with the dual $\tilde{F}^{\mu\nu}=\epsilon^{\mu\nu\lambda\rho}F_{\lambda\rho}/2$, and $m_a$ is the axion mass.
	As a consequence, this ALP can be produced through a photon fusion, as shown in Fig.~\ref{fig:feynAxion}, which happens in many high-energy collider environments. For simplicity, we consider a one-side neutron's photon involvement, which can be directly measured at a future high-energy muon beam dump experiment~\cite{Cesarotti:2022ttv} or the muon-ion collider~\cite{Acosta:2021qpx}.
	
	\begin{figure}[h]
		\centering
		\includegraphics[width=0.48\textwidth]{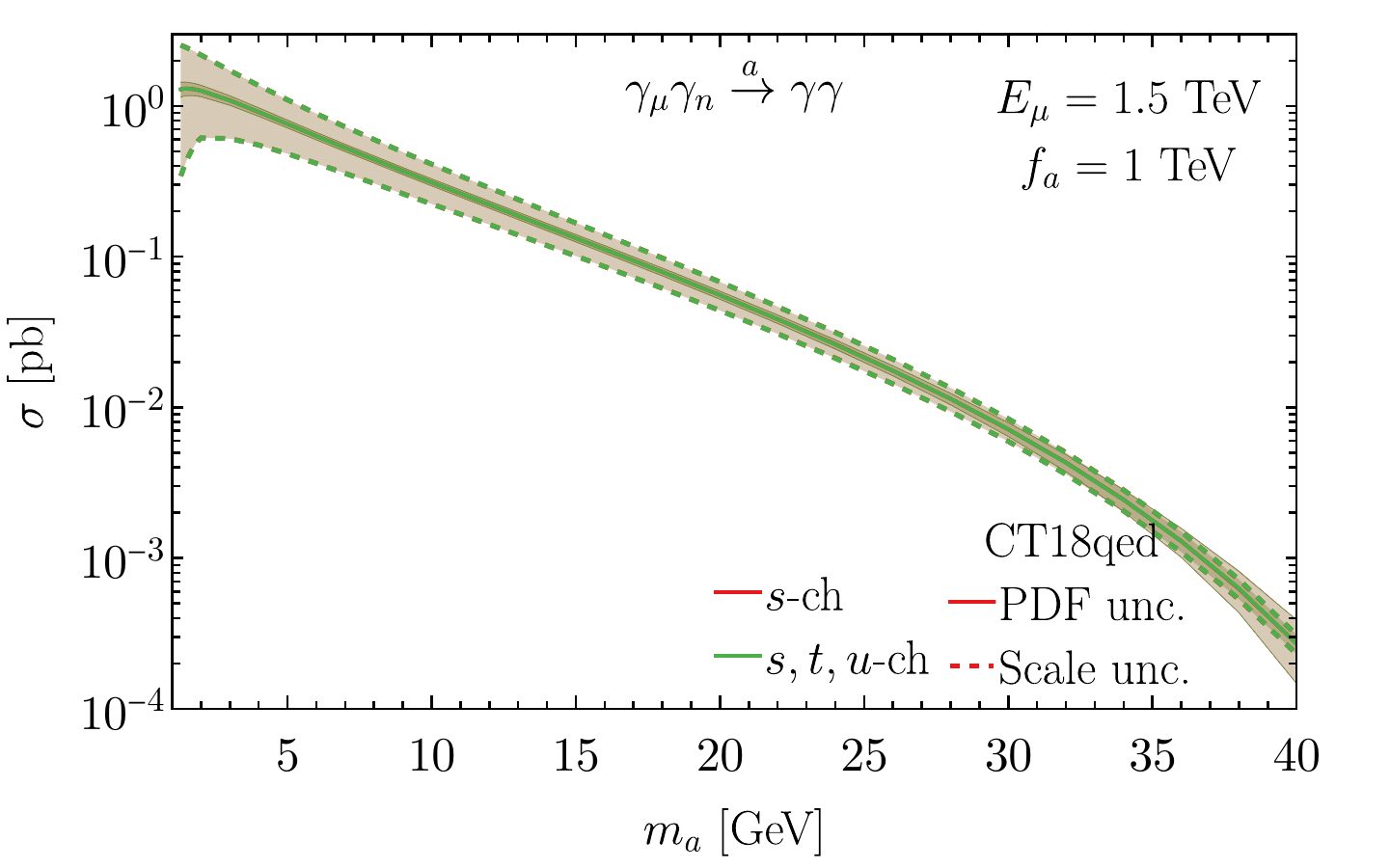}
		\includegraphics[width=0.51\textwidth]{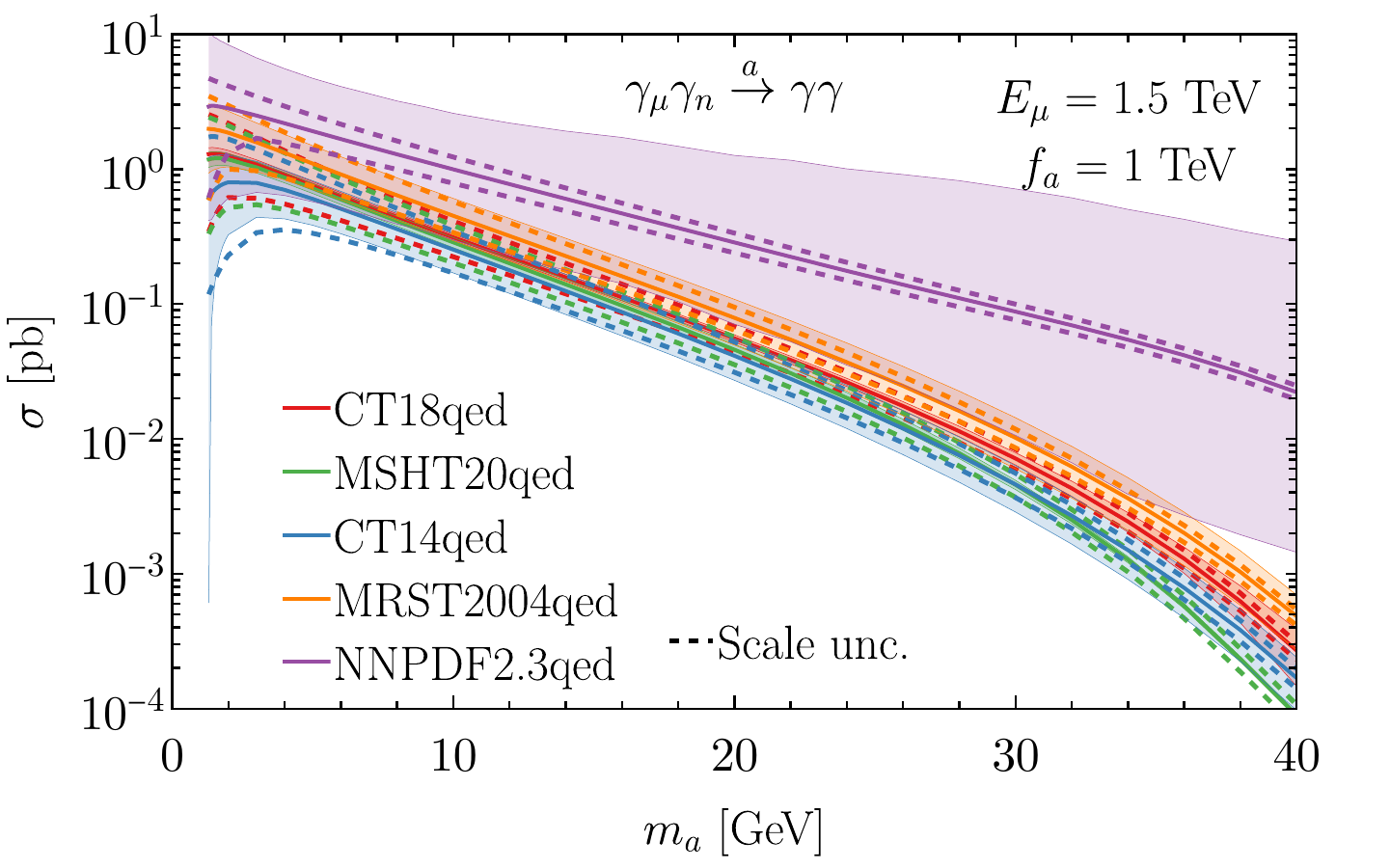}
		\includegraphics[width=0.48\textwidth]{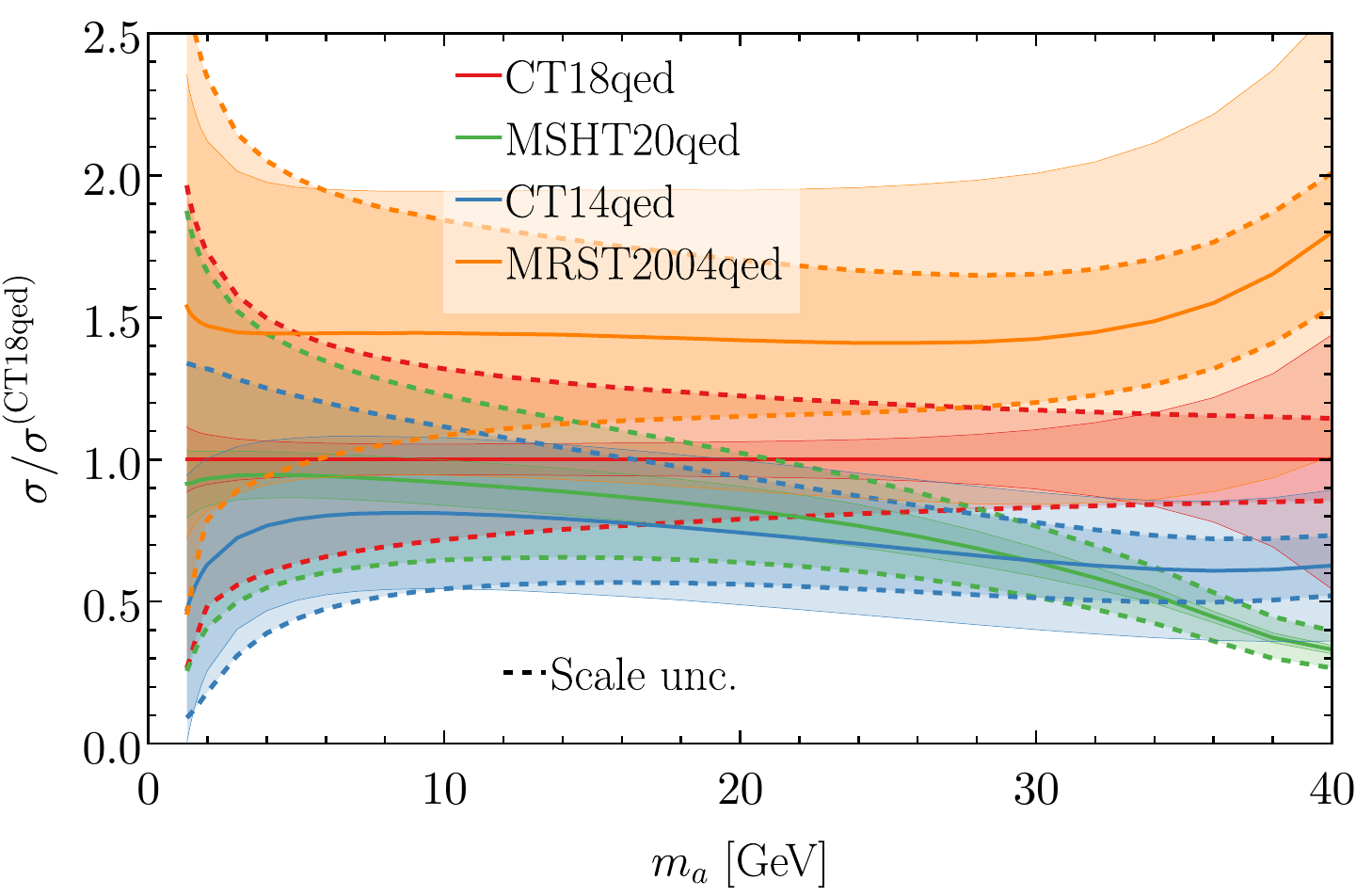} 
		\includegraphics[width=0.51\textwidth]{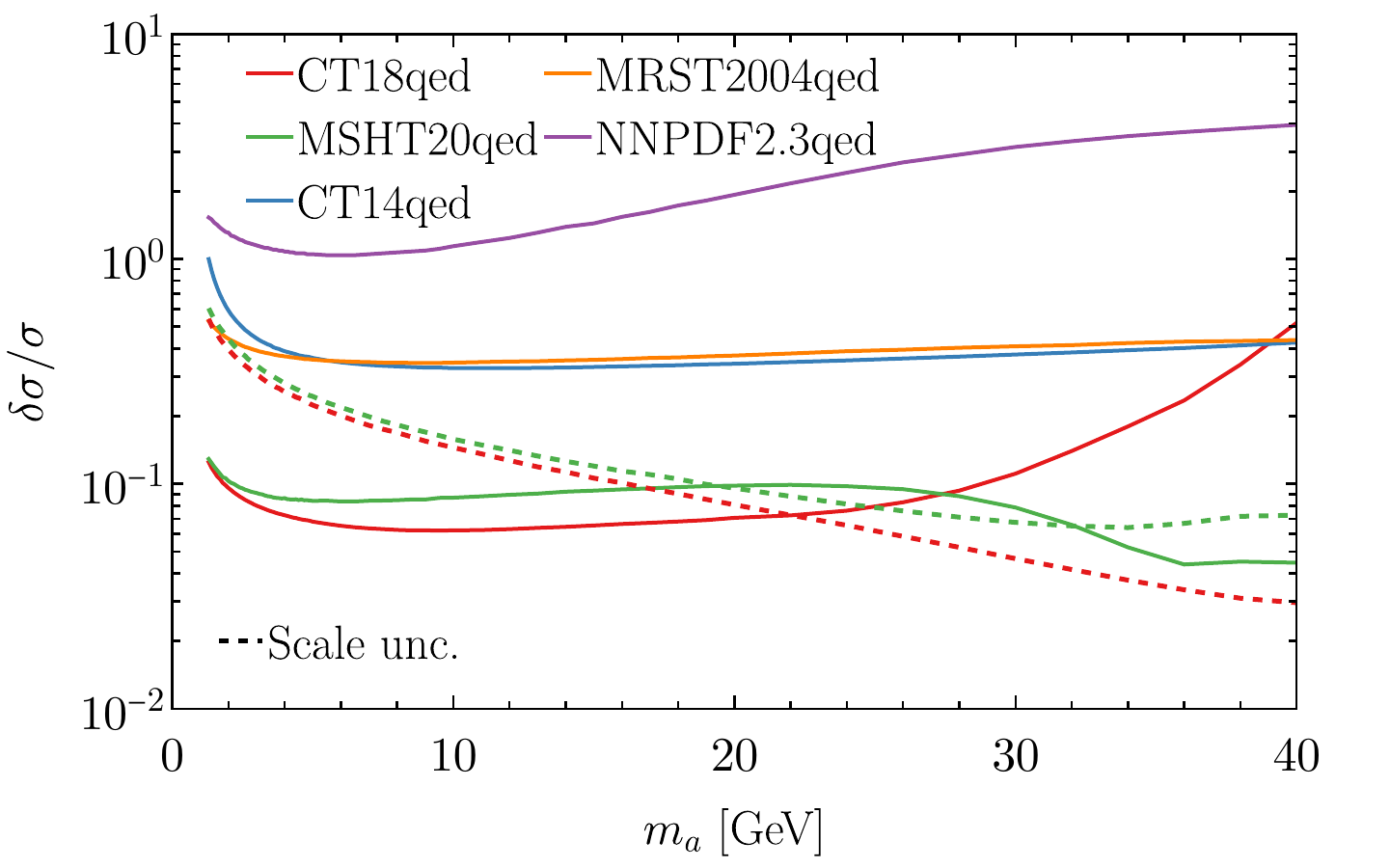}   
		\caption{Axion-like particle production cross sections in the muon-neutron scattering at a $E_\mu=1.5~\TeV$ muon beam dump experiment. The thin and dashed bands indicate the corresponding PDF (68\% CL) and scale uncertainties. The upper left panel compares the single $s$-channel with complete $s,t,u$-channel calculations, while the rest include all the $s,t,u$ channels.
		}
		\label{fig:Axion}
	\end{figure}
	
	In Fig.~\ref{fig:Axion}, we present the ALP production cross sections at a high-energy muon beam dump experiment, with the beam energy $E_\mu=1.5~\TeV$, which corresponds to a 3 TeV muon collider~\cite{Delahaye:2019omf,MuonCollider:2022xlm}. 
	The corresponding center-of-mass collision energy is 
	\begin{eqnarray}
		\sqrt{s}\simeq\sqrt{2E_\mu m_N}\simeq53~\GeV,
	\end{eqnarray}
	where $m_N$ is the mass of a nucleon.
	The calculation is done at the leading order with MadGraph~\cite{Alwall:2014hca} interfacing with the UFO~\cite{Degrande:2011ua} model file~\cite{Brivio:2017ije} generated with FeynRules~\cite{Alloul:2013bka}.
	We take $f_a=1~\TeV$ as a benchmark and show the ALP production cross section's dependence on the ALP mass from GeV up to $m_a=40~\GeV$, which mainly probes the photon PDF in the large $x$ region. The photon radiated off the neutron target corresponds to the neutron's photon PDF, while the muon's photon is taken from the improved Weizsaecker-Williams approximation~\cite{Frixione:1993yw}.
	The central factorization and renormalization scales are taken as the axion mass, $\mu_F=\mu_R=m_a$, while the scale uncertainty is quantified with the 9-point variation. The axion decay width is self-consistently determined solely through the $a\to\gamma\gamma$ channel as
	\begin{equation}
		\Gamma_a=\Gamma(a\to\gamma\gamma)=\frac{1}{4\pi}\frac{m_a^3}{f_a^2}.
	\end{equation}
	
	We first compare the calculation with only including $s$-channel diagram and the one with the complete $s,t,u$ ones in the upper left panel of Fig.~\ref{fig:Axion}. We see that in the whole $m_a$ region, the result with $s,t,u$ channels coincides with the one with the $s$-channel only very well. This can be understood from the resonance effect, with the negligible contribution from the $t,u$-channel diagrams. Similar behavior has been observed in the electron-ion collider (EIC) scenario~\cite{Liu:2021lan}.
	However, some potential additional decay channels, \emph{e.g.}, into invisible dark matter, $a\to\chi\bar{\chi}$, can enlarge the axion decay width $\Gamma_a$, which will signify the destructive interference between the $s$ and $t,u$ channels in Fig.~\ref{fig:feynAxion} and reduce the production cross section as a consequence.
	
	The PDF and scale uncertainties are presented in Fig.~\ref{fig:Axion}, which shows that the scale uncertainty dominates at low $m_a$ region, while PDF uncertainty takes over at large $m_a$.
	The comparison among various neutron's photon PDF results is presented in the rest plots of Fig.~\ref{fig:Axion}.
	We see that with an increase of ALP mass, the ALP production cross section drops very quickly, from about 1 pb down to $10^{-4}$ pb. The scale uncertainty decreases from 60\% down to 2\%, mainly driven by the decrease of the PDF variation as a result of a smaller high-scale strong coupling.  
	
	On the lower left panel, we normalize the cross sections to the CT18qed prediction and compare the corresponding ratios as well as the uncertainties. We see that the MSHT20qed result is in good agreement with the CT18qed result at a low $m_a$ when $m_a\lesssim10~\GeV$, while gradually gets a significantly smaller when $m_a\gtrsim20~\GeV$, even up to 60\% when $m_a=40~\GeV$, mainly driven by its much smaller photon PDF at large $x$, as shown in Figs.~\ref{fig:MSHT20}-\ref{fig:n2p}.
	As a reference, the CT18qed PDF uncertainty stabilizes at a few percent when $m_a<30~\GeV$, which increases quickly up to 50\% when $m_a=40~\GeV$. In comparison, the MSHT20qed uncertainty stays at a 10\% level and even decreases a little when ALP mass approaches $m_a=40~\GeV$, which indicates its different large-$x$ photon PDF variation, as discussed in Sec.~\ref{sec:compare}. 
	
	We also include the first generation of photon PDFs in this comparison. We see that the MRST2004qed prediction gets 50\% larger than the CT18 benchmark, with a much larger PDF uncertainty $(\sim40\%)$, mainly propagated from the large $x$ PDFs, as shown in Fig.~\ref{fig:MSHT20}. The CT14qed prediction is about $60\%\sim80\%$ of the CT18qed one, with about 40\% variation with respect to its central one. In comparison, both the central prediction and uncertainty band of the NNPDF2.3qed PDF can go much larger than the other PDF sets.
	
	\section{Conclusions}
	\label{sec:conclude}
	In this work, as a follow-up to our CT18qed photon PDF study~\cite{Xie:2021equ}, we complete it by calculating the neutron's photon content with the latest LUXqed formalism~\cite{Manohar:2016nzj, Manohar:2017eqh}, which determines the photon PDF with the precisely-measured structure functions. Similar to the proton case, the neutron's photon PDF is determined in two methodologies, \emph{i.e.}, CT18lux and CT18qed. In the CT18lux approach, the photon PDF is directly calculated at any scale by applying the LUXqed master formula, Eq.~(\ref{eq:LUXqed}). In comparison, the CT18qed photon PDF is initialized at the starting scale, $\mu_0$, with the LUXqed, and evolves to a higher scale with the DGLAP evolution, which has reached the NNLO QCD and NLO QED accuracy. 
	
	To determine our CT18 neutron's elastic photon PDF, we adopt the neutron's electromagnetic form factors extracted from the global neutron data~\cite{Ye:2017gyb} instead of from the simple Galster parameterization~\cite{Galster:1971kv, Kelly:2004hm} used in MMHT2015qed~\cite{Harland-Lang:2019pla} and MHST20qed~\cite{Cridge:2021pxm} PDF sets.
	Moreover, instead of integrating the elastic form factor up to the initial scale $\mu_0$ and evolve to high scale as MMHT2015qed~\cite{Harland-Lang:2019pla} and MHST20qed~\cite{Cridge:2021pxm}, we take the full LUXqed formalism to determine the elastic component at all scale, which ends up a significant larger elastic photon PDF as shown in Fig.~\ref{fig:el100GeV}. At different scales, the CT18 elastic photon more or less remains the same, with a minor difference coming from the scale evolution of the QED coupling constant.
	
	The inelastic photon PDF, with the CT18lux methodology, is determined using the LUXqed formalism, which integrates the structure functions (SFs) over all the $(x,Q^2)$ kinematic regions, as shown in Fig.~\ref{fig:breakup}. In the perturbative QCD region with $Q^2>Q^2_{\rm PDF}$, we relate the neutron's quark-gluon PDFs to the proton ones with isospin symmetry approximation. The low-$Q^2$ structure functions are taken from the experimental measurements and the corresponding fits. In the low-$Q^2$ continuum region, the HERMES GD-11P and GD-11D fits~\cite{HERMES:2011yno} as well as the R1998 fit~\cite{E143:1998nvx} of the longitudinal-to-transverse cross-section ratio $R_{L/T}=\sigma_L/\sigma_T$ are adopted.
	The resonance structure functions are determined through the CLAS~\cite{CLAS:2003iiq} or Christy-Bosted (CB) fit~\cite{Christy:2007ve, Bosted:2007xd, Christy:2021abc} in the resonance region. We take a smooth matching of the SFs in the transition regions between the resonance and low-$Q^2$ HERMES continuum, as well as the one between the low-$Q^2$ HERMES and high-$Q^2$ pQCD continuum. Various low-$Q^2$ nonperturbative resources, which induce the variation of the inelastic photon are carefully examined, including the CLAS/CB resonance fits, the HERMES, R1998, pQCD SFs in the small-$x$ and small-$Q^2$ region (sxQ2), higher twist (HT) and target-mass (TM) corrections, PDF matching (Q2PDF), as well as the missing higher order (MHO) uncertainty. It turns out that the resonance SFs (CLAS) can introduce the largest variation at a low scale in the moderate-$x$ range, while the HERMES extrapolation (sxQ2) takes over in the low-$x$ region, both of which dies out at a high scale, while the TM corrections induce a large uncertainty at a large scale and large $x$.
	
	With the CT18qed methodology, the neutron's inelastic photon is initialized at a low scale $\mu_0=1.3~\GeV$ and evolves to high sales with the NNLO QCD and NLO QED DGLAP equation.
	With an iterative approach (Fig.~\ref{fig:flowchart}), the momentum sum rule and the isospin symmetry violation (ISV) are self-consistently determined, which turn out to have a negligible effect on the final photon PDF, as shown in Fig.~\ref{fig:qed2lux}.
	Similar to the CT18lux, the variation of low-$Q^2$ nonperturbative resources is examined, which turns out that the CT18qed photon accumulates advantages of the large-scale suppression of the resonance SF effect as well as the small TM corrections to the low-scale PDFs.
	
	The comparison of neutron's photon in the second-generation QED PDFs, \emph{i.e.}, CT18lux, CT18qed, MSHT20qed~\cite{Cridge:2021pxm} is performed in Fig.~\ref{fig:MSHT20}. In comparison with the CT18lux, the CT18qed photon PDF gets enhancement at small-$x$, as a result of the DGLAP evolution, which equivalently integrates out larger LO SFs, with respect to the high-order one in the LUXqed formalism. The larger high-$x$ CT18lux photon is induced by the larger $\msbar$ matching term induced by a smaller high-$Q^2$ perturbative $F_2$ than the low-$Q^2$ non-perturbative one at the initialization scale in CT18qed. 
	In comparison with the error band, the CT8lux one gets slightly larger error bands, due to a worse control of the TM corrections. For this reason, we advocate the CT18qed as a primary use for phenomenological calculations in related studies.
	
	The MSHT20qed gets a good agreement with the CT18qed in the moderate-$x$ region $(10^{-3}\lesssim x\lesssim10^{-1})$, while smaller in both small- and large-$x$ regions. The smaller MSHT20qed small-$x$ photon is due to its smaller charge-weighted singlet, while the smaller large-$x$ photon is induced by both the noticeably smaller charge-weighted singlet and its stationary approximation of Eq.~(\ref{eq:MSHT20}). The photon PDF uncertainty of MSHT20qed is comparable to CT18qed one, with some moderate difference in the small-$x$ region driven by the charged-weighted singlet while the large-$x$ region driven by the different low-$Q^2$ non-perturbative SF variations.
	We also compare our PDFs with the first generation QED PDFs, \emph{i.e.}, MRST2004qed~\cite{Martin:2004dh}, CT14qed~\cite{Schmidt:2015zda}, and NNPDF2.3qed~\cite{Ball:2013hta} in Fig.~\ref{fig:1stgen}, which shows a significant improvement in precision, thanks to the LUXqed formalism.
	
	We also investigate the phenomenological implications of our neutron's photon PDFs with two photon-initiated scattering processes: the neutrino-nucleus $W$-boson production and the axion-like particle production in a high-energy muon beam-dump experiment. In both scenarios, the cross sections are very sensitive to the neutron's photon PDF at large $x$. With respect to the first generation of QED PDFs, the second generation ones, both CT18qed and MHST20qed, get a significant theoretical improvement in the PDF uncertainty. In comparison with the CT18qed predictions, the MSHT20qed gets smaller cross sections for the WBP near the energy threshold and for large-mass ALP production at a muon beam dump experiment, due to the corresponding large-$x$ PDF behaviors. 
	
	As a companion of this article, the LHAPDF6~\cite{Buckley:2014ana} grids of the CT18qed PDFs, including the elastic, inelastic components as well as the sum as a total, will be provided through the CTEQ-TEA GitLab repository, \url{https://cteq-tea.gitlab.io/project/00pdfs/}. 
	The construction of each eigenvector set is discussed in detail in Sec.~\ref{sec:compare}.
	The final PDF uncertainty should be combined in terms of Eq.~(\ref{eq:unc}). We have re-scaled the $n_{\textrm{low-}Q^2}$ low-$Q^2$ nonperturbative error sets with a factor of 1.645 to obtain the 90\% CL, respecting the CT Hessian error criterion~\cite{Hou:2019efy,Stump:2001gu,Pumplin:2001ct}, for the convenience of combination.
	Different from the default CT18qed proton QED PDF set which takes the $\mu_0=3~\GeV$ to minimize the non-perturbative PDF uncertainty~\cite{Xie:2021equ}, we take the $\mu_0=1.3~\GeV$ for the neutron one as default, which get better control of the low-$Q^2$ resource, largely by suppressing the mismatch effect between the HERMES and pQCD continuum regions. 
	
	Finally, this and previous work~\cite{Xie:2021equ} focus on the photon content of free nucleons. The nuclear corrections play a role in various places, such as the neutron's electromagnetic form factors~\cite{Ye:2017gyb} and the resonance structure functions~\cite{CLAS:2003iiq,Christy:2007ve,Bosted:2007xd,Christy:2021abc}. In our treatment, the nuclear uncertainty on the neutron's low-$Q^2$ SFs in the HERMES continuum region~\cite{HERMES:2011yno} is folded in the HERMES and R1998 error sets. Moreover, the nuclear effects (including shadowing~\cite{Armesto:2006ph,Kopeliovich:2012kw} or EMC~\cite{EuropeanMuon:1992pyr}), as well as the electric charge, are expected to have a non-negligible impact on the neutron's inelastic and elastic photon inside a heavy nucleus, such as the lead $^{208}_{~82}$Pb, relevant to the LHC and EIC heavy-ion collisions. This leaves room for future improvement, which can be done with a joint effort with the nuclear PDF fitting group, such as the nCTEQ collaboration.
	
	\acknowledgments
	We thank Thomas Cridge for the help with the clarification of MSHT20qed results, and our CTEQ-TEA colleagues for useful discussions. The work of KX was supported by the U.S. Department of Energy under grant No. DE-SC0007914, the U.S. National Science Foundation under Grants No. PHY-2112829, No. PHY-2013791, and No. PHY-2310497, and also in part by the PITT PACC. The work of BZ~was supported by the Simons Foundation. The work of TJH~was supported by the U.S. Department of Energy under contract DE-AC02-06CH11357. The work of KX was performed partly at the Aspen Center for Physics, which is supported by the U.S. National Science Foundation under Grant No. PHY-1607611 and No. PHY-2210452. This work used resources of high-performance computing clusters from SMU M2/M3, MSU HPCC, as well as Pitt CRC.

\bibliographystyle{utphys}
	\bibliography{ref_Xie}
\end{document}